\documentclass[11pt]{article}
\usepackage{graphicx}
\usepackage{amssymb,amsmath,amsfonts}

\def\Re{{\cal R \mskip-4mu \lower.1ex \hbox{\it e}\,}}
\def\Im{{\cal I \mskip-5mu \lower.1ex \hbox{\it m}\,}}
\def\ie{{\it i.e.}}
\def\eg{{\it e.g.}}
\def\etc{{\it etc}}
\def\etal{{\it et al.}}

\def\sub#1{_{\lower.25ex\hbox{$\scriptstyle#1$}}}
\def\tev{\,{\ifmmode\mathrm {TeV}\else TeV\fi}}
\def\gev{\,{\ifmmode\mathrm {GeV}\else GeV\fi}}
\def\mev{\,{\ifmmode\mathrm {MeV}\else MeV\fi}}
\def\mpl{\ifmmode M_{pl}\else $M_{pl}$\fi}
\def\mpl{\ifmmode \overline M_{Pl}\else $\bar M_{Pl}$\fi}
\def\to{\rightarrow}

\def\subw{_{\rm w}}
\def\mh{\ifmmode m\sbl H \else $m\sbl H$\fi}
\def\mch{\ifmmode m_{H^\pm} \else $m_{H^\pm}$\fi}
\def\mt{\ifmmode m_t\else $m_t$\fi}
\def\mc{\ifmmode m_c\else $m_c$\fi}
\def\mz{\ifmmode M_Z\else $M_Z$\fi}
\def\mw{\ifmmode M_W\else $M_W$\fi}
\def\mws{\ifmmode M_W^2 \else $M_W^2$\fi}
\def\mhs{\ifmmode m_H^2 \else $m_H^2$\fi}   
\def\mzs{\ifmmode M_Z^2 \else $M_Z^2$\fi}
\def\mts{\ifmmode m_t^2 \else $m_t^2$\fi}
\def\mcs{\ifmmode m_c^2 \else $m_c^2$\fi}
\def\mchs{\ifmmode m_{H^\pm}^2 \else $m_{H^\pm}^2$\fi}
\def\ztwo{\ifmmode Z_2\else $Z_2$\fi}
\def\zone{\ifmmode Z_1\else $Z_1$\fi}
\def\mtwo{\ifmmode M_2\else $M_2$\fi}
\def\mone{\ifmmode M_1\else $M_1$\fi}
\def\tb{\ifmmode \tan\beta \else $\tan\beta$\fi}
\def\xw{\ifmmode x\subw\else $x\subw$\fi}
\def\ch{\ifmmode H^\pm \else $H^\pm$\fi}
\def\lum{\ifmmode {\cal L}\else ${\cal L}$\fi}
\def\inpb{\,{\ifmmode {\mathrm {pb}}^{-1}\else ${\mathrm {pb}}^{-1}$\fi}}
\def\infb{\,{\ifmmode {\mathrm {fb}}^{-1}\else ${\mathrm {fb}}^{-1}$\fi}}
\def\epem{\ifmmode e^+e^-\else $e^+e^-$\fi}
\def\ppb{\ifmmode \bar pp\else $\bar pp$\fi}
\def\bsg{\ifmmode B\to X_s\gamma\else $B\to X_s\gamma$\fi}
\def\bsll{\ifmmode B\to X_s\ell^+\ell^-\else $B\to X_s\ell^+\ell^-$\fi}
\def\bstt{\ifmmode B\to X_s\tau^+\tau^-\else $B\to X_s\tau^+\tau^-$\fi}
\def\lamt{\ifmmode \tilde\lambda\else $\tilde\lambda$\fi}
\def\shat{\ifmmode \hat s\else $\hat s$\fi}
\def\that{\ifmmode \hat t\else $\hat t$\fi}
\def\uhat{\ifmmode \hat u\else $\hat u$\fi}

\newskip\zatskip \zatskip=0pt plus0pt minus0pt
\def\matth{\mathsurround=0pt}
\def\lsim{\mathrel{\mathpalette\atversim<}}
\def\gsim{\mathrel{\mathpalette\atversim>}}
\def\atversim#1#2{\lower0.7ex\vbox{\baselineskip\zatskip\lineskip\zatskip
  \lineskiplimit 0pt\ialign{$\matth#1\hfil##\hfil$\crcr#2\crcr\sim\crcr}}}

\def\grtsim{\,\,\rlap{\raise 3pt\hbox{$>$}}{\lower 3pt\hbox{$\sim$}}\,\,}
\def\lsim{\,\,\rlap{\raise 3pt\hbox{$<$}}{\lower 3pt\hbox{$\sim$}}\,\,}

\renewcommand{\thefootnote}{\fnsymbol{footnote}}

\begin{document} \begin{titlepage}
\rightline{\vbox{\halign{&#\hfil\cr
%&DRAFT\cr
&MIT-CTP-3988\cr
&SLAC-PUB-13388\cr
%&March 2006\cr
}}}
\begin{center}
\thispagestyle{empty} \flushbottom {{\Large\bf Supersymmetry Without Prejudice 
\footnote{Work supported in part
by the Department of Energy, Contract DE-AC02-76SF00515}}}
\medskip
\end{center}

\centerline{Carola F. Berger{\footnote{e-mail:cfberger@mit.edu}}}
\vspace{6pt}
\centerline{\it  Center for Theoretical Physics,
Massachusetts Institute of Technology, Cambridge, MA 02139, USA}
\vspace{12pt}
\centerline{James S. Gainer{\footnote{e-mail:jgainer@slac.stanford.edu}}, 
JoAnne L. Hewett{\footnote{e-mail:hewett@slac.stanford.edu}}, 
Thomas G. Rizzo{\footnote{e-mail:rizzo@slac.stanford.edu}}}
\vspace{6pt} 
\centerline{\it SLAC National Accelerator Laboratory, 2575 Sand Hill Rd., Menlo Park, CA, 94025, USA}

\vspace*{0.3cm}

\begin{abstract}
We begin an exploration of the physics associated with the general CP-conserving MSSM with Minimal Flavor Violation, the pMSSM. The 19 soft SUSY 
breaking parameters in this scenario are chosen so as to satisfy all existing experimental and theoretical constraints assuming that the WIMP is a conventional 
thermal relic, 
\ie, the lightest neutralino. We scan this parameter space twice using both flat and log priors for the soft SUSY breaking mass parameters and compare the 
results which yield similar conclusions. Detailed constraints from both LEP and the Tevatron searches play a particularly important role in obtaining 
our final model samples. We find that the pMSSM leads to a much broader set of predictions for the properties of the SUSY partners as well as for a number of 
experimental observables than those found in any of the conventional SUSY breaking scenarios such as mSUGRA. This set of models can easily lead to atypical 
expectations for SUSY signals at the LHC. 
\end{abstract}

%\begin{center}

%\end{center}

\renewcommand{\thefootnote}{\arabic{footnote}} \end{titlepage} 

%
%
%
%%%%%%%%%%%%%%%%%%%%%%%%%%%%%%%---- Put text here

\section{Introduction}

The LHC will soon begin operations in earnest and will thereafter begin a detailed study of the Terascale. Amongst the many prospective  
new physics signals that may be realized at the LHC, the possible observation of an excess of events with missing energy is  
awaited with great anticipation. Given the strong collection of evidence for the existence of dark matter \cite{Bertone:2004pz}   
and the naturalness of the WIMP paradigm \cite{Primack:1988zm}, it is no surprise that many beyond the Standard Model (SM) scenarios predict the appearance of 
missing energy signatures at the LHC \cite{missinglhc}. Perhaps the most well explored of these 
scenarios is R-Parity conserving, N=1 Supersymmetry (SUSY) where the WIMP is identified as the lightest supersymmetric 
particle (LSP), which most commonly corresponds to the lightest neutralino or  gravitino. The simplest manifestation of this picture is 
the Minimal Supersymmetric Standard Model (MSSM), where the gauge group of the SM is maintained and the 
matter content of the SM is augmented only by an additional Higgs doublet. 

While the MSSM has many nice features, perhaps its least understood aspect is the nature of the mechanism which breaks the Supersymmetry.  
There is an ever growing list of possible candidate scenarios, including mSUGRA \cite{msugrab}, GMSB \cite{gmsbb}, 
AMSB \cite{amsbb}, and gaugino mediated supersymmetry breaking \cite{ggmsbb}.  Each of these 
mechanisms strongly influences both the mass spectra and decay patterns of the various sparticles and hence it is highly 
non-trivial to make model-independent statements about the properties of the LSP or about SUSY signatures at colliders such as the LHC.  This puts 
forth the question of
whether there is some way to study the MSSM in a broad fashion while simultaneously adopting as few simplifying 
assumptions as possible.  The purpose of this paper is to address this issue.  
In trying to answer this question, one is immediately faced with the daunting task of dealing 
with the over 120 parameters, arising mainly in the soft-breaking sector, which describe SUSY breaking in the MSSM \cite{mssmrev}.  Clearly the 
phenomenological study of any model with such a 
large number of parameters is totally hopeless. Obviously {\it some} assumptions need to be made in order to reduce this large number 
of independent parameters to a viable subset and there are many possible avenues one could follow.  
Most studies assume a specific SUSY breaking mechanism, employing theoretical assumptions at the GUT scale, and 
thus reduce the number of independent model parameters to 4 or 5.
In this investigation, we will not make any assumptions at the high scale.
We will, however, restrict ourselves to the CP-conserving MSSM (\ie, no new phases) with minimal flavor violation (MFV)\cite{mfv}. 
Moreover, to simplify matters further and help soften the impact of the experimental constraints arising from the flavor sector, 
we will require that the first two generations of sfermions be degenerate. This leaves us with 19 
independent, real, weak-scale, SUSY Lagrangian parameters to consider. These include the gaugino masses $M_{1,2,3}$, the  
Higgsino mixing parameter $\mu$, the ratio of the Higgs vevs $\tan \beta$, the mass of the pseudoscalar 
Higgs boson $m_A$, and the 10 squared masses of the sfermions (5 for the assumed degenerate first two generations and a separate 5 
for the third generation). Finally, due to the small Yukawa couplings for the first two generations, independent 
$A$-terms are only phenomenologically relevant for the third generation and will be included in our study for the $b,t$ and $\tau$ sectors. This set of 19 
parameters leaves us with what has been called the phenomenological MSSM (pMSSM) in the literature{\cite{Djouadi:2002ze}} and is the scenario that we will 
study in some detail below. We note that for this study we will make the further assumption that the LSP is the neutralino and is a conventional thermal relic 
within the standard cosmology with a radiation dominated era.

The primary goal of the first phase of this study is to obtain a large set of MSSM parameter points 
that can be used for future studies of SUSY signatures, \eg, signals at the LHC or in cosmic rays.  To accomplish this, 
we will perform a scan over this 19-dimensional parameter space and subject the resulting specific sets of parameter 
choices (which we denote as models) to a large and well-known (and some perhaps, not so well-known) set of experimental and 
theoretical constraints{\cite {Freitas:2008uk}}. Furthermore, we will perform two independent scans which will employ different priors for choosing 
the above mass parameters; this is important since some bias may be introduced into the analysis by how one choose parameter space points. By choosing more 
than one prior and comparing results this sensitivity can be estimated. We will show that while both scans yield qualitatively similar results, quantitative 
differences will be observed in some cases.  
  
We will perform our comparison to the experimental data with some care to identify sets of parameter space points which are at present phenomenologically viable. 
As we will see below, applying these constraints can be complicated and is not 
always as straightforward as employed in some other analyses.  This is particularly true in the case where experimental 
bounds are predicated on specific SUSY breaking mechanisms. As we are not a priori 
interested in any particular region of parameter space (\eg, we are enforcing the WMAP dark matter 
density{\cite {Komatsu:2008hk}} observations only as an upper bound) or in obtaining a set of points that provide a `best fit' to the data,  
we will not need to make use of Markov-chain Monte Carlo techniques{\cite {MCMC,orig}}.{\footnote {The MCMC approach asks a different 
question than we do here, namely `What regions of the parameter space provide the best fits to the data?', by using, for example, a chi-squared analysis. This
requires one to assume precise knowledge of the probability distributions for both experimental results as well as theoretical predictions. For example,
given the uncertainties associated with the g-2 of the muon, is it justified to treat the error associated with the apparent difference with the SM prediction  
as Gaussian?  These types of questions introduce prejudice into the analysis, which we try to avoid, and so we have instead adopted the scanning approach.}}  
We will then 
perform a detailed study of the properties of these surviving 
models and, in particular, highlight those aspects which are not generally encountered when only a given set of SUSY breaking 
scenarios is considered. This allows us to get a perspective on  the wide range of predictions for various observables that 
can arise in the more general pMSSM framework in comparison to those found in specific SUSY breaking scenarios. 

In the next Section, we will discuss our scanning techniques, the ranges of the soft breaking parameters that we employ, as well as the values of the SM input 
parameters that we use. In Section 3, we will detail the various theoretical and experimental constraints at some length and will explain exactly how they are 
employed in our analysis. Section 4 presents our analysis and a survey of the results, while Section 5 contains an overview discussion of the lessons learned from 
this analysis. Our conclusions can be found in Section 6.

\section{Scans, Parameters Values and Ranges}

To perform our multi-dimensional parameter scan, we need to choose suitable
ranges for the soft-breaking parameters 
that we employ and determine how the values of these parameters
are picked within these assumed ranges. In the analysis presented 
here, we will make two independent parameter scans. In the first case, we randomly generate $10^7$ sets of parameters, 
assuming flat priors, \ie, we assume that the parameter values are chosen {\it uniformly} throughout their allowed ranges.
With an eye toward the experimental constraints to be discussed in the next Section, for the flat prior scan we employ the following ranges:
\begin{eqnarray}
100 \gev \leq m_{\tilde f} \leq 1\tev \,, \nonumber\\
50\gev \leq |M_{1,2},\mu|\leq 1 \tev\,, \nonumber \\ 
100 \gev \leq M_3\leq 1 \tev\,, \nonumber \\ 
|A_{b,t,\tau}| \leq 1 \tev\,, \\
1 \leq \tan \beta \leq 50\,, \nonumber \\ 
43.5\gev \leq m_A \leq 1 \tev\,. \nonumber  
\end{eqnarray}
Note that absolute value signs are present in these quoted ranges as we will allow the soft-breaking parameters to 
have arbitrary sign. Given that the soft parameters $M_i,A_i$ and $\mu$ can all appear with arbitrary signs (\ie, phases) but that only six relative signs are 
physical{\cite{Kraml:2007pr}} we choose $M_3$ to be positive. This scan will generate SUSY spartners with light to moderate 
masses.

For our second scan, we randomly generate $2\times 10^6$ sets of parameters assuming log priors for the mass parameters. We then modify the allowed soft parameter 
ranges as follows: 
\begin{eqnarray}
100 \gev \leq m_{\tilde f} \leq 3\tev \,, \nonumber\\
10\gev \leq |M_{1,2},\mu|\leq 3 \tev\,, \nonumber \\ 
100 \gev \leq M_3\leq 3 \tev\,, \nonumber \\ 
10\gev\leq |A_{b,t,\tau}| \leq 3 \tev\,, \\
1 \leq \tan \beta \leq 60\,, \nonumber \\ 
43.5\gev \leq m_A \leq 3 \tev\,. \nonumber  
\end{eqnarray}
Here, $\tilde f = \tilde Q_L, \tilde Q_3, \tilde L_1, \tilde L_3, \tilde u_1,
\tilde d_1, \tilde u_3, \tilde d_3, \tilde e_1,$ and $ \tilde e_3$.
This expanded range of parameters will allow us 
access to both very light as well as some heavy sparticle states that may only be produced at the SLHC. The goal of this second scan is to make 
contrasts and comparisons to the flat prior study in order to determine the dependence of the model properties on the scan assumptions. The possible 
bias introduced by choosing a particular prior in analyses such as ours is an important problem that has been often discussed in the literature{\cite {big}}. The only way 
it can be properly addressed is to take more than one assumed prior, compare the results and look for differences. For the case of our two priors we will see 
that they yield qualitatively similar results, but the detailed predictions in the two cases will be found to be quantitatively different in several 
aspects.

%We will also comment on several smaller scans made in the breaking parameter ranges %that make a very light LSP more likely.  

Note that the above mass parameters are assumed to be evaluated at the SUSY scale, \ie,  
by convention the geometric mean of the two stop masses $\lsim 1$ 
TeV.  The physical spectra for the sparticles themselves are generated in all cases using the code SuSpect2.34{\cite {Djouadi:2002ze}}. 
Once a set of parameters is chosen, we subject the resulting MSSM model to the entire set of constraints discussed in the next Section; those models which 
survive will be analyzed in further detail below.  Of course, as models are passed through each successive experimental `filter,' the number of surviving 
models will be further and further reduced.  

Another important aspect of such a broad study is the set of input values for the SM parameters, \eg, $\alpha_s(M_Z),~m_t, ~M_W$, \etc., 
~that we employ. As is well-known, some of the predictions for the various observables we analyze below can be sensitive 
to these particular numerical choices. While we have chosen these numerical values with some care, we are not advocating any one particular set of values for 
these parameters. We provide their values here only for completeness since it is important for the interested reader to know exactly how our analysis was 
performed.  Table \ref{smparam} lists the values of these SM parameters used in our analysis.  
While we only quote central values here, many of the relevant errors are included when 
evaluating the set of observables we discuss in the next Section. We follow this approach, rather than including the SM input values in our 
scan, since they are essentially known quantities unlike the pMSSM parameters.

\begin{table}
\centering
\begin{tabular}{|c|c|} \hline\hline
Parameter  & Value   \\ \hline
$\alpha(M_Z)$ & 127.918~~~[Ref.\cite{Amsler:2008zz}] \\
$\alpha_s(M_Z)$ & 0.1198~~~[Ref.\cite{Baikov:2008jh}] \\
$M_Z$ & 91.1875 GeV~~~[Ref.{\cite {LEPEWWG}}] \\
$\Gamma_Z$ & 2.4952 GeV~~~[Ref.{\cite{Amsler:2008zz,LEPEWWG}}] \\
$\sin^2\theta_w|_{\rm on-shell}$ & 0.22264~~~[Ref.{\cite {LEPEWWG}}]\\
$M_W$ & 80.398 GeV~~~[Ref.{\cite {LEPEWWG}}]\\
$\Gamma_W$ & 2.140 GeV~~~[Ref.{\cite {LEPEWWG}}]\\
$m_s$(1 GeV) & 128 MeV~~~[Ref.\cite{Amsler:2008zz}]\\
$m_c^{\rm pole}$ & 1666 MeV~~~[Ref.\cite{Baikov:2008jh}]\\
$m_b$ & 4.164 GeV~~~[Ref.\cite{Baikov:2008jh}]\\
$m_b^{\rm pole}$ & 4.80 GeV~~~[Ref.\cite{Baikov:2008jh}]\\
$m_t^{\rm pole}$ & 172.6 GeV~~~[Ref.\cite{Group:2008nq}]\\
$V_{us}$ & 0.2255~~~[Ref.\cite{Amsler:2008zz}]\\
$V_{cb}$ & $41.6\times 10^{-3}$[Ref.\cite{Amsler:2008zz}]\\
$V_{ub}$ & $4.31\times 10^{-3}$[Ref.\cite{Amsler:2008zz}]\\
$V_{ub}/V_{cb}$ & 0.104~~~[Ref.\cite{Amsler:2008zz}]\\
$m_{B_d}$ & 5.279 GeV~~~[Ref.\cite{Amsler:2008zz}]\\
$f_{B_d}$ & 216 MeV~~~[Ref.\cite{Amsler:2008zz}]\\
$\tau_{B_d}$ & 1.643 ps~~~[Ref.\cite{Amsler:2008zz}]\\
$f_{B_s}$ & 230 MeV~~~[Ref.\cite{Onogi}]\\
$\tau_{B_S}$ & 1.47 ps~~~[Ref.\cite {HFAG}]\\
\hline\hline
\end{tabular}
\caption{Central values of the SM input parameters used in our analysis.}
\label{smparam}
\end{table}

\section{Theoretical and Experimental Constraints}

Apart from the restrictions imposed by the ranges chosen above for the 19 pMSSM soft breaking parameters, there are a large 
number of both theoretical and experimental constraints that need to be considered in order to obtain phenomenologically viable models. Many of 
these are familiar while others will be somewhat less so to the arbitrary reader. The direct application of subsets of these numerous 
constraints and the specific ranges which are considered viable varies widely in 
the literature from one analysis to another{\cite {big}}; hence, we will discuss them here in some in detail. 

The first set of constraints we impose are theoretical and rather standard and are applied while generating the sparticle spectrum with the SuSpect code: the 
sparticle spectrum must be tachyon free and cannot lead to color or charge breaking minima in the scalar potential. 
We also require that electroweak symmetry breaking be consistent and that the Higgs potential 
be bounded from below. Note, however, since we do {\it not} impose grand unification (as we are ignorant of the ultraviolet 
completion of the MSSM if we do not implement a specific SUSY breaking scenario), we do not require that the RGE evolution of the couplings be 
well-behaved up to the GUT scale. Because we are only concerned with the physics between the weak and TeV scales and the overall effects of RGE  
evolution in this range will be rather modest, it is unlikely that a Landau pole or other irregularities can be generated in this narrow mass window. 

At this point in our analysis, we will make the further assumption that the WIMP be a conventional thermal relic so that the LSP can be identified as 
the lightest neutralino; the possibility of the LSP being the gravitino or the axino will be ignored in the rest of the present 
analysis{\cite {SUSYDM}}.   The scenario where the gravitino is the 
LSP{\cite{Feng:2004mt}} will be discussed in a future analysis within a different astrophysical framework.  We note that the case where the sneutrino is the LSP 
can be easily eliminated in the pMSSM by combining several of the constraints discussed below. The possibility that the LSP is a non-thermal relic can also be 
entertained and may lead to somewhat different results. 

The code  micrOMEGAs2.21{\cite {MICROMEGAS}} takes the MSSM spectrum output from SuSpect and allows us 
to implement the constraints arising from a number of precision and flavor measurements: precision electroweak constraints via 
$\Delta \rho$, the rare decays $b\to s\gamma$ and $B_s \to \mu^+\mu^-$, as well as the $g-2$ of the muon. 
We take the $95\%$ CL allowed experimental range for $\Delta \rho$ directly from the analysis presented in the 
2008 PDG{\cite{Amsler:2008zz}}: $-0.0007 \leq \Delta \rho \leq 0.0026$. 
For $b\to s\gamma$ we use the 
combined experimental result from HFAG{\cite {HFAG}}, $B_{b\to s\gamma}=(3.52 \pm 0.25)\cdot 10^{-4}$, as well as the recent SM theoretical 
predictions as given by both Misiak \etal {\cite{Misiak:2006zs}} and Becher and Neubert{\cite{Becher:2006pu}}. 
Combining both the experimental and theoretical errors, we require, rather conservatively, 
that the predicted branching fraction lie in the range $B_{b\to s\gamma} = (2.5-4.1)\cdot 10^{-4}$. For the decay $B_s \to \mu^+\mu^-$, we employ 
the recently reported  combined limit on the branching fraction obtained by CDF and D0{\cite{toback}}, $B_{B_s\to\mu\mu}\leq 4.5 \cdot 10^{-8}$ at $95\%$ CL. Note 
that our philosophy will be to apply some experimental 
constraints rather loosely, as they can always be tightened in obtaining a final sub-sample. This will also allow us to easily study the 
sensitivity of the various observables to the particular values of the soft breaking parameters as well as determine the favored ranges of the various 
observables themselves.  

As is well-known, measurements of the anomalous magnetic moment of the muon, $(g-2)_\mu$, differ{\cite{Bennett:2006fi}} from the recently updated prediction 
of the SM by more than $\sim 3\sigma$: this difference is found to be $\Delta(g-2)_\mu=(30.2\pm 8.8)\cdot 10^{-10}$ in
one recent analysis{\cite{Passera:2008hj}} and $(29.5\pm 7.9)\cdot 10^{-10}$ in another{\cite{DeRafael:2008iu}}, with both analyses 
using low energy $e^+e^-$ data as input. It has been argued that this deviation might be a 
signal for new physics beyond the SM, \eg, light Supersymmetry. In order to explore this 
possibility more fully within the general pMSSM context and also to allow for the scenario 
that the SM value may essentially end up being 
correct once the theoretical, experimental and input errors are more fully understood and reduced, we will implement the loose constraint that 
$(-10 \leq \Delta(g-2)_\mu \leq 40)\cdot 10^{-10}$ in our analysis. This will also allow us to observe how the MSSM predictions are distributed within this range 
for the set of surviving models.

In addition to these constraints which are essentially built into the micrOMEGAs2.21 code, we employ several others that arise from heavy flavor measurements: ($i$) the 
branching fraction for $B\to \tau \nu$; 
combining the results from HFAG{\cite {HFAG}} and those presented at ICHEP08{\cite {Chang}} we will require this branching fraction to lie in the range 
$B_{B\to\tau\nu}=(55-227)\cdot 10^{-6}$. Here we make use of the theoretical analysis by Isidori and Paradisi{\cite {gino}} as well as Erikson, 
Mamoudi and Stal{\cite {ems}} which include the SUSY loop contributions to the bottom quark Yukawa coupling.  
%($ii$) The constraint on the SUSY contributions to ratio of the $Z$ partial widths %$R_b=\Gamma(Z\to b\bar b)/\Gamma(Z\to hadrons)$, which agrees 
%well with the SM prediction, will be employed. Here 
%we will require that $\Delta R_b=(R_b)_{EXP}-(R_b)_{SM}$ be in the range $(-82$ to %$176)\cdot 10^{-5}$ arising the experimental results from  
%the LEPEWWG{\cite{Amsler:2008zz}} and the theoretical calculation as given by Boulware %and Finnell[REF]. 
%The relevant Higgs and sparticle couplings are determined via SuSpect. 
($ii$) With the assumption of MFV{\cite {mfv}} and degenerate first 
and second generation sfermions, the resulting constraints from meson-antimeson mixing on the squark mass spectra are relatively modest{\cite {mesonmix}}. 
We performed a detailed numerical study which demonstrated that these conditions can be
generally satisfied if the ratio of first/second and third generation squark soft breaking masses (of a given charge and helicity) 
differ from unity by no more than a factor of $\sim 5$.  We will impose this restriction on the mass spectra of the models we consider.  We also apply similar 
restrictions in the slepton sector. 

It is well-known that restrictions on the MSSM parameter space arise from employing the LSP as a form of dark matter. Two major constraints are found in 
this case: the first arises from the 5 year WMAP measurement{\cite {Komatsu:2008hk}} of the relic density, 
which we employ only as an upper limit on the LSP contribution, \ie, $\Omega h^2|_{LSP} \leq 0.1210$. Here we simply acknowledge the possibility that even 
within the MSSM and the thermal relic framework, dark matter may have multiple components with the LSP being just one possible contributor.{\footnote {We note 
that within our analysis framework, {\it given the order in which our constraints are applied}, the use of the WMAP constraint removes $\sim 50\%$ of the 
models which have survived up to that point in the case of flat priors.}} 
The second constraint arises from direct detection searches{\cite {dmsearch}} for dark matter via both spin-independent and spin-dependent LSP 
elastic scattering off nuclei in the laboratory. Over essentially all of the parameter space, the spin-independent constraints are presently found to be dominant. 
We include the cross section bounds arising from XENON10{\cite {XENON10}}, CDMS{\cite {CDMS}}, CRESST I{\cite {CRESST}} 
and DAMA{\cite {DAMA}} data. Given the uncertainties from low energy physics (\eg,  nuclear form factors) 
in the determination of these cross sections, we allow for a factor of 4 uncertainty in these calculations. This numerical factor 
was obtained by varying the low energy input parameters independently over their allowed ranges for several benchmark points. Thus we will allow 
models to predict cross sections as much as 4 times larger than the usually quoted experimental bounds. We employ 
micrOMEGAs2.21{\cite {MICROMEGAS}} in evaluating both $\Omega h^2$ as well as the dark matter scattering cross sections. We note that a `low' prediction 
for the thermal relic density in some of our models will lead to an effective direct search cross section which is smaller than what one would obtain if 
the WMAP bound were saturated. 

We now turn to the constraints imposed by LEP data. First, we require that there be no new charged sparticles or charged Higgs bosons with 
masses below $M_Z/2$ due to the lack of evidence for such states in both
direct or indirect measurements at the $Z$ pole;
we also impose this same constraint on the lightest neutral Higgs boson. Based on data from LEPII{\cite {lepstable}}, we further require that there be no 
new {\it stable} charged particles of any kind with masses below 100 GeV. We note that such a situation may occur if, for example, the nLSP is charged and 
highly degenerate with the LSP, which is a common occurrence in the AMSB scenario. Additionally, we constrain any new contributions to the invisible width of the 
$Z$ boson to be $\leq 2$ MeV{\cite {LEPEWWG}}. 
This will restrict the masses and couplings of light neutralinos that have significant Higgsino content and thus may appear in the decay $Z\to \chi_1^0\chi_1^0$. 
While it is unlikely that decays such as $Z\to \chi_2^0\chi_1^0$ will be kinematically accessible, in such cases the $\chi_2^0$ will 
more than likely decay into visible final states.  We isolate such parameter space model points in the analysis below and scrutinize them 
further{\footnote {In the one model that falls into this category we find that $\chi_2^0$ is essentially 
stable but its contribution to the invisible width of the $Z$ is tiny.}} 
(using the SUSY-HIT1.1{\cite {SUSYHIT}} to analyze the $\chi_2^0$ decays). 
Furthermore, we note that for the range of sfermion mass  
soft breaking parameters we consider, $Z$ decay to pairs of sneutrinos is not kinematically allowed so that this final state cannot contribute in 
any way to the purely invisible width.

Data from LEPII provides further direct search constraints on the sparticles, although they are oftentimes sensitive to detailed assumptions about the model; we 
provide only an outline of these many bounds here. 
ALEPH{\cite {ALEPH}} has placed a lower limit on the light squark masses, assuming that 
the gluino is more massive than the squarks, via their decay to a jet+LSP, \ie, jet + missing energy. Provided that the mass difference between the squark 
and the LSP, $\Delta m$, is $\geq 10$ GeV to avoid very 
soft jets, they obtain a lower limit of 92 GeV on the squark masses. Given our soft breaking parameter spectrum ranges above, we employ this 
constraint directly, including the $\Delta m$ cut.  Models in our scan with $\Delta m <10$ GeV are hence not affected by this search limit.  A comparable 
search{\cite {bbb}} for light sbottoms using the same sort of decay pattern results 
in a lower bound of 95 GeV on the 
mass of this sparticle. Corresponding lower bounds have been placed{\cite {LEPSUSY}} on the masses of right-handed sleptons undergoing decays to leptons plus 
missing energy of $m\gsim 100(95,90)$ GeV for the selectron(smuon,stau).  This limit is only applicable if the slepton masses are at least a few percent larger 
than that of the LSP, otherwise the final state leptons will be too soft.  As in the case of squark production, this mass splitting requirement is critical and 
is included in our analysis. Strictly speaking, 
these bounds are only applicable to the right-handed sleptons, however, they 
may also be applied to left-handed sleptons provided the corresponding sneutrino $t-$channel exchange contribution is not very important. We make 
this assumption in our analysis. An analogous situation applies to chargino production. If the LSP-chargino mass splitting is $\Delta m >2$ GeV, a direct lower 
limit of 103 GeV on the chargino mass is obtained from LEPII data.  However, if this splitting is $\Delta m <2$ GeV, the 
bound degrades to 95 GeV, {\it provided} that also $\Delta m >50$ MeV, otherwise the chargino would appear as a stable particle in the 
detector and would then be excluded by the stable particle searches discussed above. In the case where the lightest chargino is dominantly Wino, this limit 
is found to be applicable only when the electron sneutrino is more massive than 160 GeV. There are also two searches for light stops to consider: 
if $\tilde t_1$ is too light to decay into $Wb\chi_1^0$, then 
the search for the decay $\tilde t_1 \to c\chi_1^0$ provides a lower bound on the stop mass of 97 GeV. If $\tilde t_1$ is heavy enough to decay into  
$\ell b\tilde \nu$, then the corresponding lower bound is 95 GeV. 

As provided in detail by the LEP Higgs Working Group{\cite {LEPHIGGS}},
there are five sets of constraints on the MSSM Higgs sector imposed by LEPII data. 
We employ the SUSY-HIT routine to analyze these, recalling that the uncertainty on the calculated mass of the lightest Higgs boson is approximately 
3 GeV{\cite {uncertain}} as determined by SuSpect.
The first pair of these constraints 
applies to the products $g_{ZZh}^2\times B(h\to b\bar b,\tau^+\tau^-)$, where $g_{ZZh}$ represents the $ZZh$ coupling,
relative to their SM values as a function of the light Higgs mass.  A second pair  
of constraints applies to the corresponding couplings $g_{ZhA}^2\times B(hA\to b\bar b b\bar b,~b\bar b\tau^+\tau^-)$ as a function of the sum of the $h$ and $A$ masses. 
The last of these LEPII Higgs bounds 
applies to the mass of the charged Higgs boson as a function of its branching fraction into the $\tau \nu$ and $c\bar s$ final states. We include all of 
these experimental constraints in detail in our results presented below.

The final set of constraints we incorporate arises from Tevatron data. We first consider the restrictions imposed on the squark and gluino sectors 
arising from the null result of the multijet plus missing energy search performed by D0{\cite {domet}}. In our study, we generalize 
their analysis to render it model independent.  For each of our models that have survived the set of constraints thus far, we compute the NLO SUSY cross 
sections for squark and gluino production using PROSPINO2.0{\cite {PROSPINO}}. Once produced, these sparticles are decayed via SDECAY/HDECAY to 
obtain the relevant decay chains and branching fractions and are then passed to PYTHIA6.4{\cite {PYTHIA}} to include
hadronization and fragmentation. We then use PGS4{\cite {PGS}}  
to simulate the D0 detector and to impose the analysis kinematic cuts; we tune PGS4 by reproducing the results and efficiencies for the three benchmark mSUGRA points 
chosen by D0 in their published multijet study. Given an integrated luminosity of 2.1 $fb^{-1}$, we find that the $95\%$ CL upper limit 
on the number of signal events from 
combining all of the production channels is 8.34 using the method of Feldman and Cousins{\cite {fc}}.  
%To match the D0 analysis we perform the analogous study 
%at the PGS4 level for the entire set of models that have survived all the previous cuts. 
If our calculation for a model leads to an 
event rate larger than this value, that model is removed from the remaining sample. 

Analogously, we employ constraints from the CDF search for trileptons plus missing energy{\cite {cdftrilepton}}, 
which we also generalize to the full pMSSM. Our procedure is essentially the same 
as in the jets plus plus missing energy analysis described above, except that we now employ a CDF tuning for PGS4 which we obtain by reproducing the CDF 
benchmark point results.  
Here we use the leading order cross section together with a universal K-factor of 1.3 to mimic the full NLO cross section. We used PROSPINO to check that this 
K-factor is a good approximation for the range of gaugino masses relevant to the Tevatron.
In  this case we only make use of the `3 tight lepton' analysis from CDF as it is the cleanest and easiest to implement with PGS4. The $95\%$ CL upper bound on the 
possible SUSY signal in this channel is then 4.65 events assuming a  luminosity of 2.02 $fb^{-1}$ as used in the CDF analysis. Again, parameter sets leading to 
larger event rates are dropped from the remainder of our analysis. 

Over the narrow mass range $90 \leq m_A \leq 100$ GeV, searches for SUSY Higgs signals by both CDF and D0{\cite {tevhiggs}} provide the approximate 
constraint $\tan \beta \geq 1.2m_A-70$.  We include this restriction in our analysis. 

Finally, both CDF{\cite {cdfstable}} and D0{\cite {dostable}} have placed lower limits on the masses of heavy stable charged particles. In our analysis we employ the 
stronger D0 constraint which we take to be of the form $m_{\chi^+}\geq 206 |U_{1w}|^2 +171 |U_{1h}|^2$ GeV at $95\%$ CL in the case of charginos.  Here, the matrix $U$ 
determines the Wino/Higgsino content of the lightest chargino and is used to interpolate between the separate Wino and Higgsino results as quoted by D0. 
As we will see below, this is an {\it extremely} powerful constraint for a large number of our models since we find that chargino-LSP 
mass degeneracies are common in our model sample, particularly when the LSP is nearly pure Wino or Higgsino or a combination of these two cases. 

We note in passing that both CDF and D0 also have a set of analyses aimed at searches for light stops and bottoms{\cite {stops}} with various 
assumptions about the SUSY mass spectrum and possible decay channels. Most of these searches are only applicable 
if these sparticles are lighter than the top quark. Since these are a difficult set of analyses to generalize to the pMSSM we will constrain our model 
set to the subset of those which contain stops and sbottoms with masses larger than $m_t$. We separately store those models satisfying all of our other 
constraints above but contain 
lighter stops and sbottoms (with masses above the LEPII limits) for a further later detailed study. However, we will find that set of models in this category are found 
to be extremely rare as will be obvious from the stop and sbottom mass distributions presented in the next Section. 

A final comment on the Tevatron SUSY searches is in order. As can be seen above, there are only two generic, flavor-independent, high $E_T$ searches for SUSY 
particles at the Tevatron. Given our large parameter space we will see that these searches are relatively easy to evade. Some fraction of the remaining parameter 
space may, 
however, be closed if further searches of these kind can be performed at the Tevatron. An obvious example of this is the usual `cascade' decay picture which is a hallmark 
of the LHC SUSY searches where, \eg, squarks decay via electroweak gauginos leading to a multi-jet, multi-lepton plus missing $E_T$ final state. Such a decay signature  
may certainly occur for many of our models and it thus remains only a question of the signal rates vs. backgrounds and the development the most efficient cuts to 
determine how effective such a search might be at the 
Tevatron. We encourage our Tevatron colleagues to consider performing such a search and to report their findings in as model independent a manner as possible.

\section{Analysis and Results}

We now describe the results from our analysis for both the cases of flat and logarithmic priors.

\subsection{Flat Priors}

In the case of flat priors, we randomly generated $10^7$ parameter space points (\ie, models).  After subjecting these models to all of the constraints detailed above, we found that only $\sim 68.5\cdot 10^3$ satisfy all of the restrictions.  Here, we discuss the properties of these surviving models in some detail.  The first characteristic one would like to know is the nature of the sparticle spectra in these 
various models. Since we cannot analyze so many complete spectra here, we will instead examine a number of specific properties of the models which then will provide an overall qualitative feel for their general behavior. 

\begin{figure}[htbp]
\centerline{
\includegraphics[width=9.0cm,angle=0]{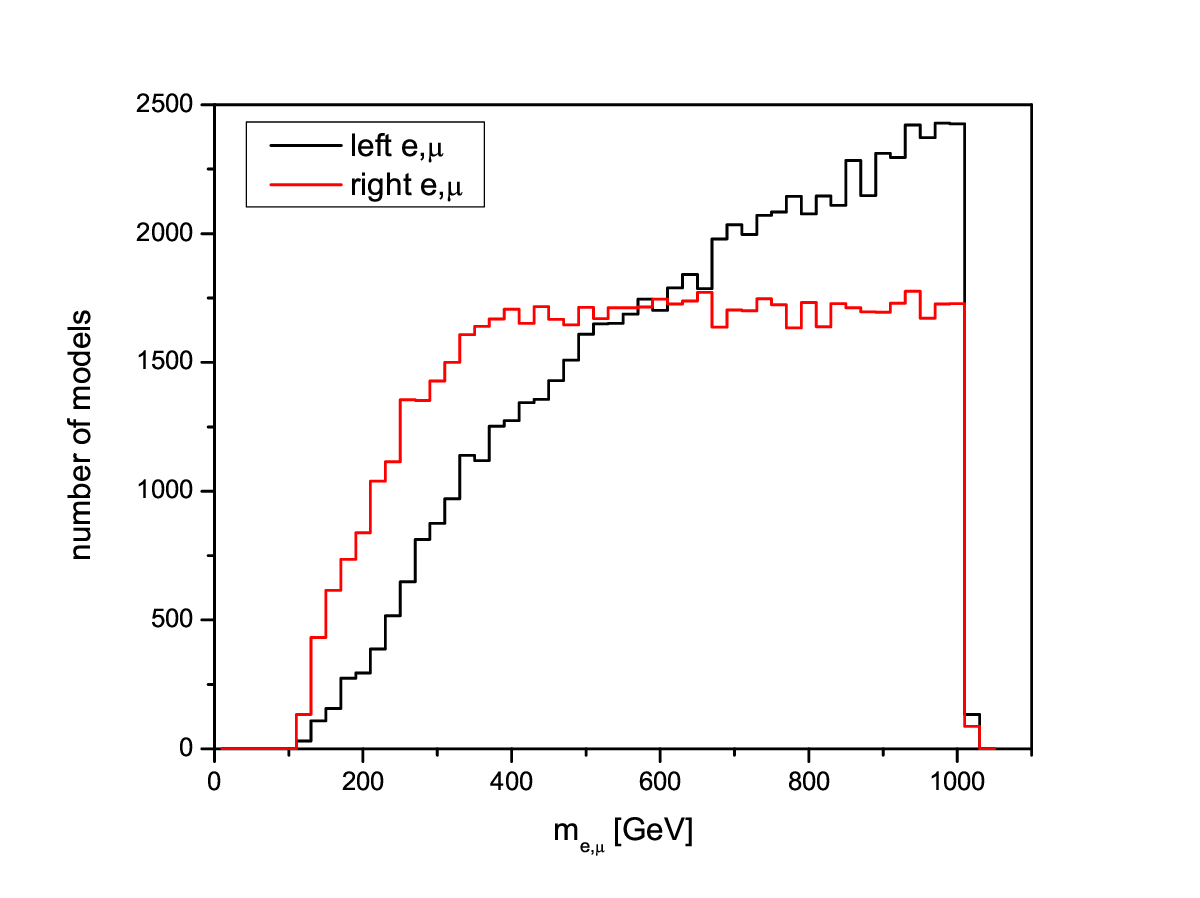}
\hspace*{0.1cm}
\includegraphics[width=9.0cm,angle=0]{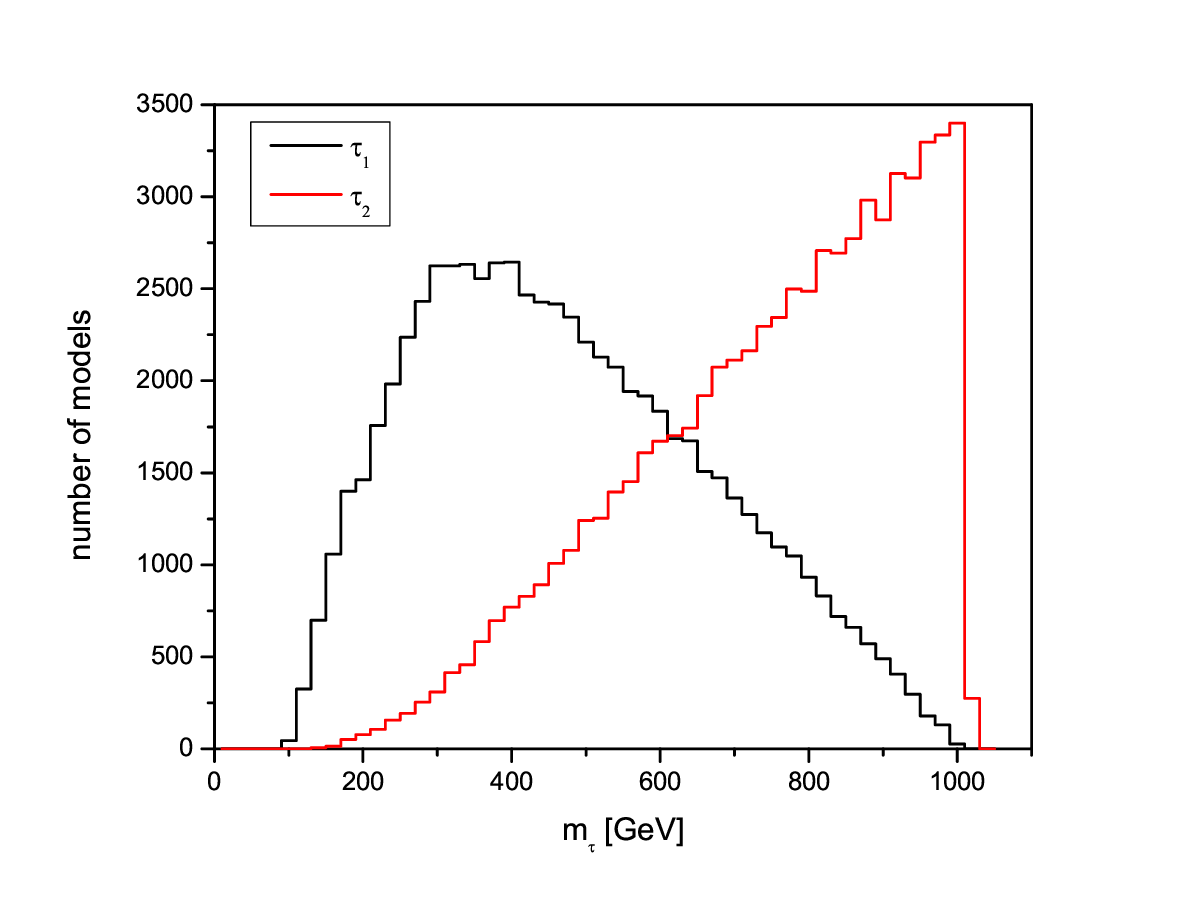}}
\vspace*{0.1cm}
\centerline{
\includegraphics[width=9.0cm,angle=0]{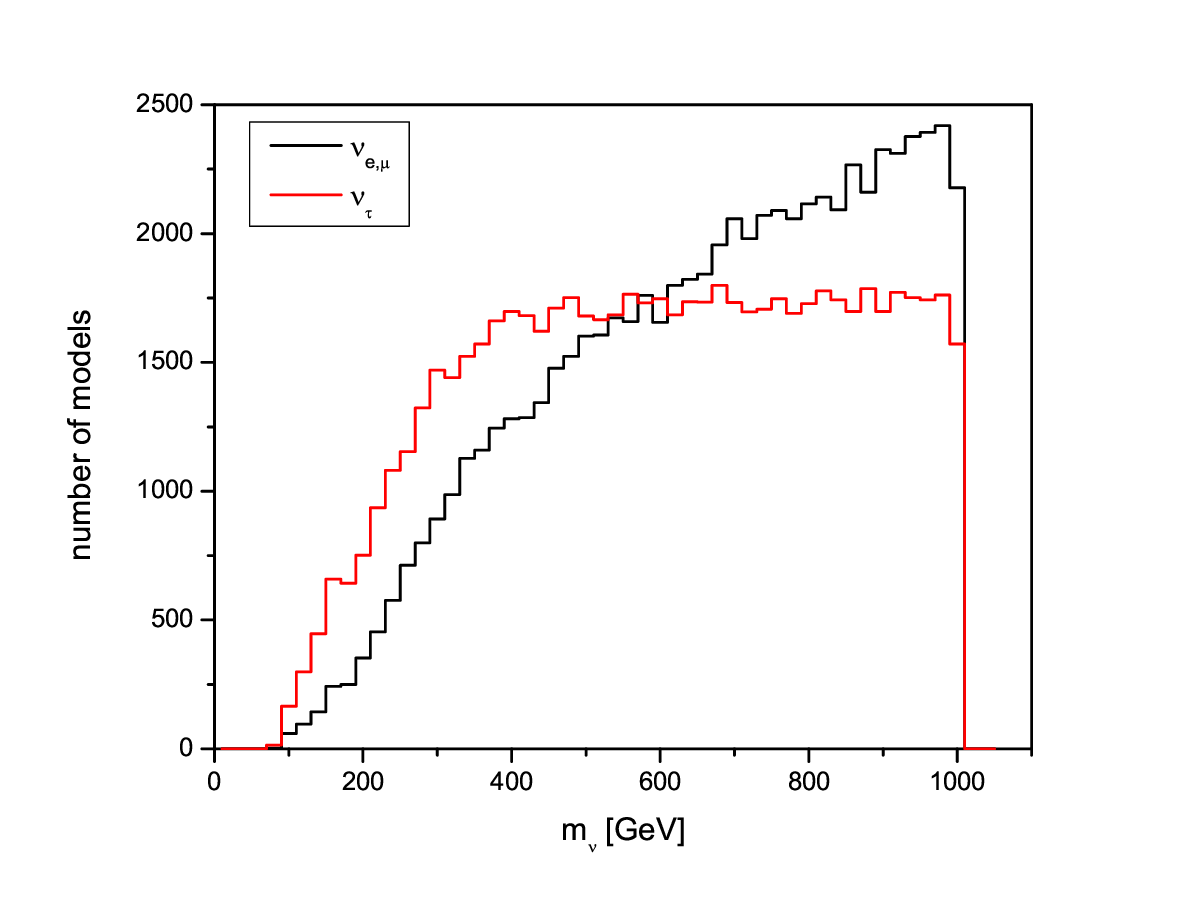}}
\caption{Distribution of slepton masses for the set of flat prior models satisfying all of our constraints: selectrons/smuons(top left), staus(top right) and 
sneutrinos(bottom).}
\label{fig1}
\end{figure}

\begin{figure}[htbp]
\centerline{
\includegraphics[width=9.0cm,angle=0]{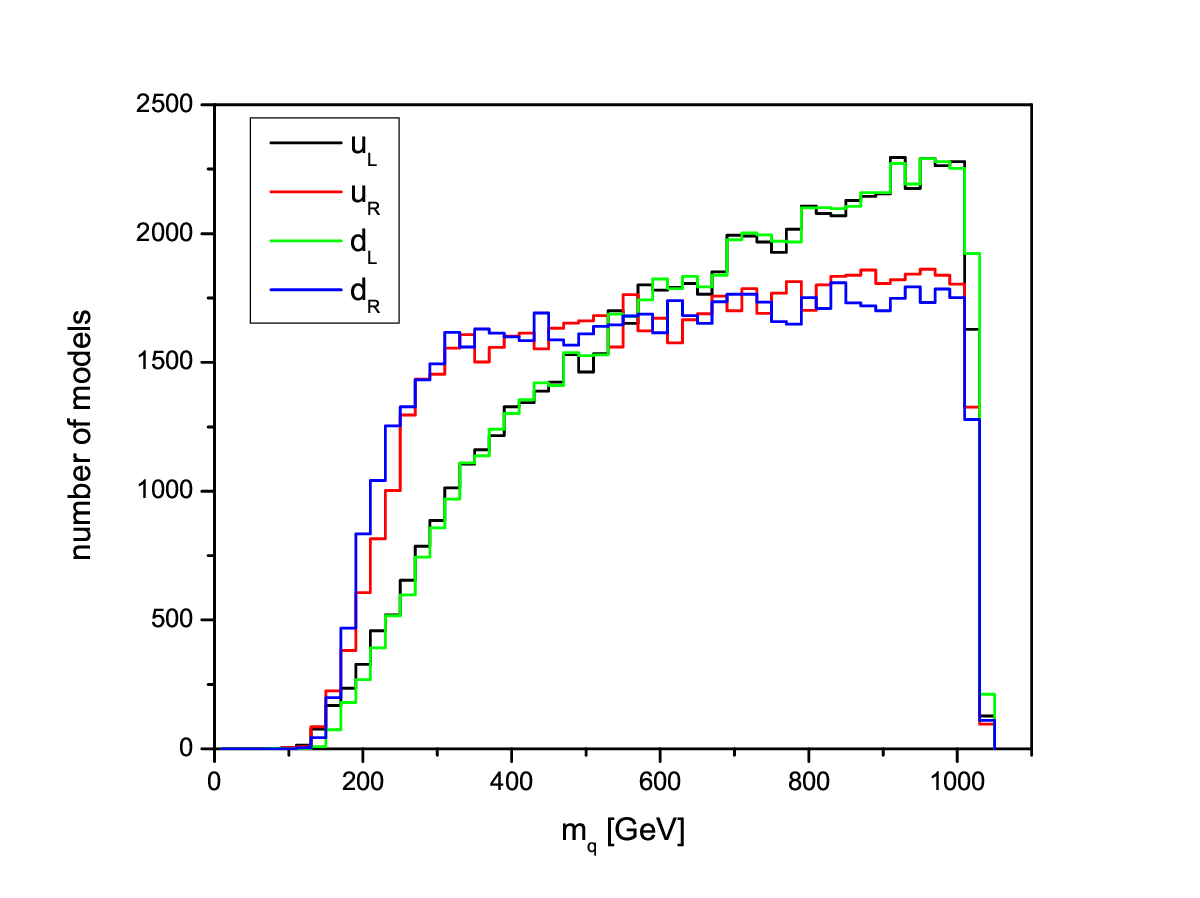}
\hspace*{0.1cm}
\includegraphics[width=9.0cm,angle=0]{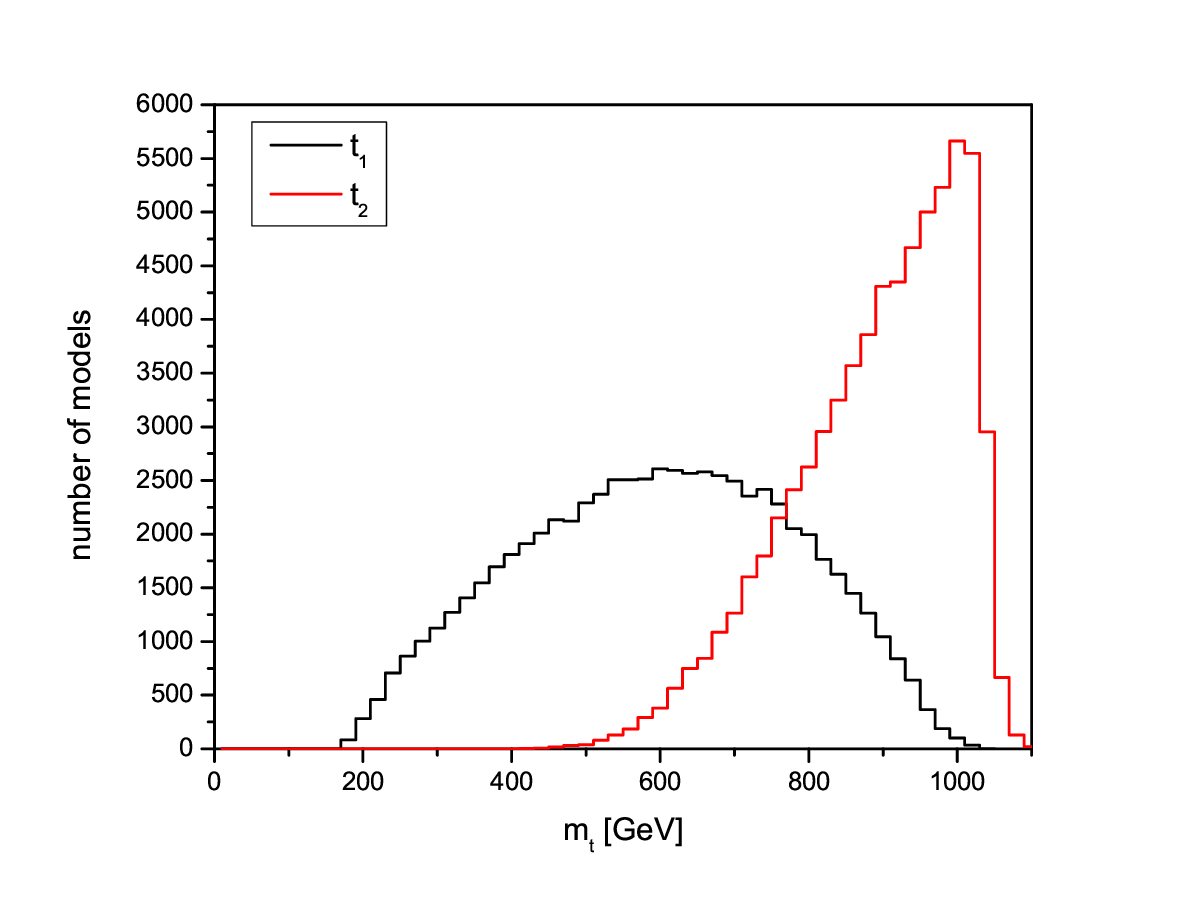}}
\vspace*{0.1cm}
\centerline{
\includegraphics[width=9.0cm,angle=0]{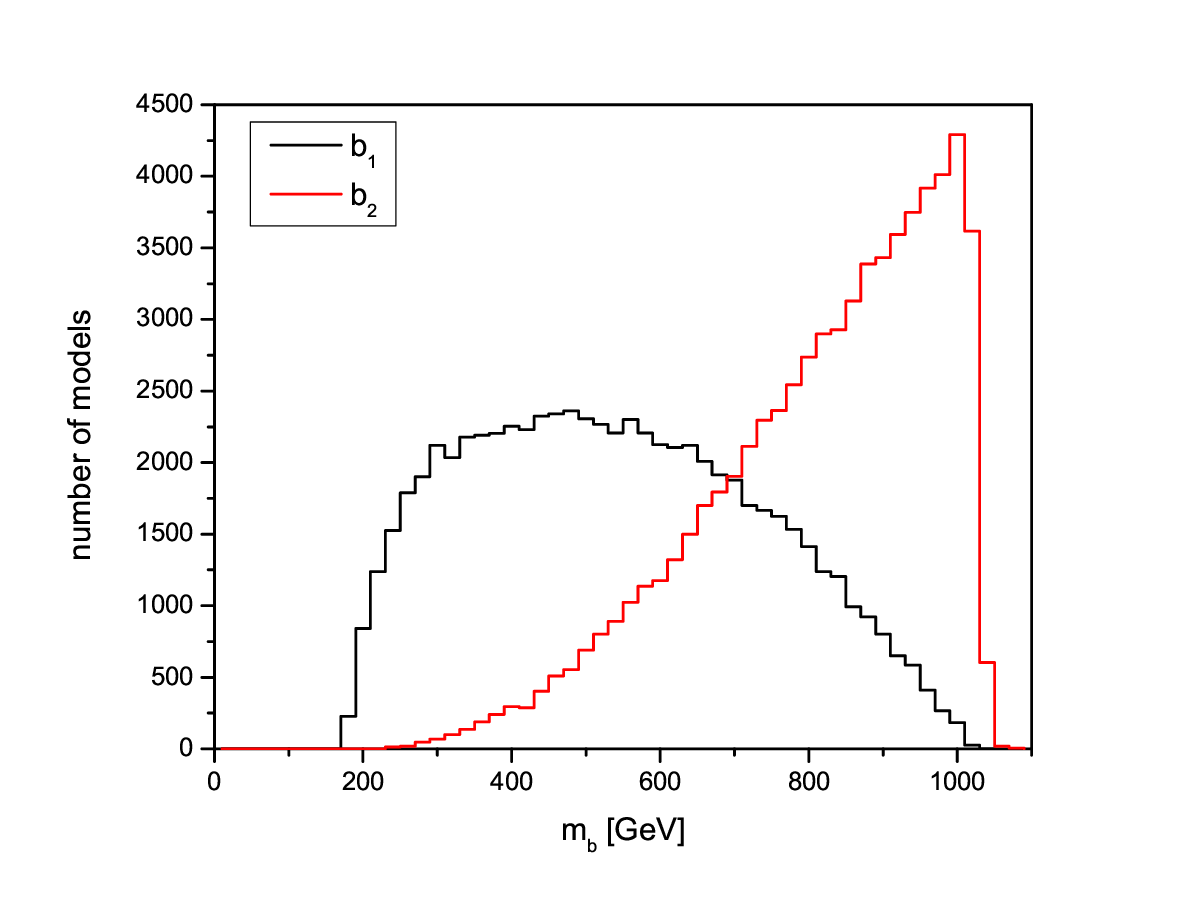}}
\caption{Distribution of squark masses for the set of flat prior models satisfying all of our constraints: first/second generation squarks are in the upper left panel, stops in the upper right panel and sbottoms are in the lower panel.}
\label{fig2}
\end{figure}

\begin{figure}[htbp]
\centerline{
\includegraphics[width=13.0cm,angle=0]{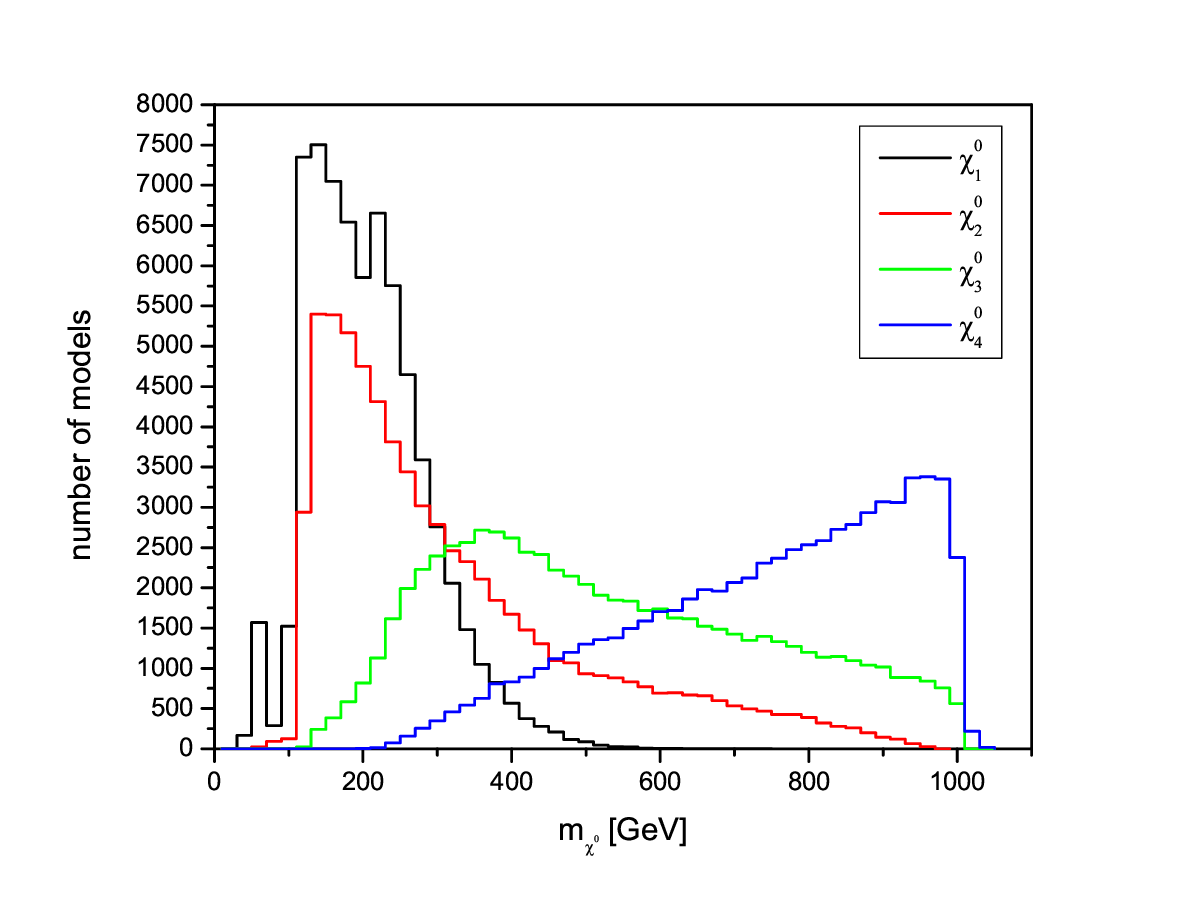}}
\vspace*{0.1cm}
\centerline{
\includegraphics[width=13.0cm,angle=0]{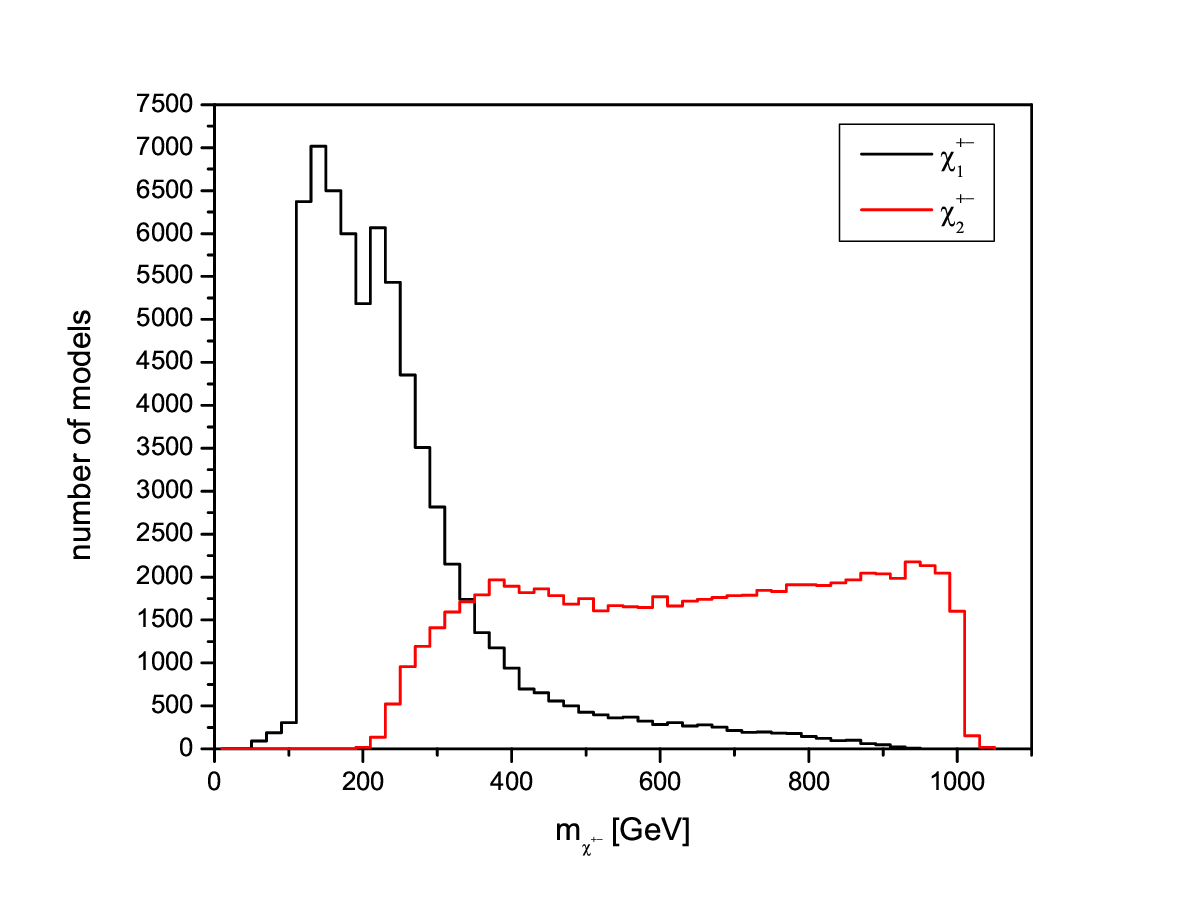}}
\caption{Distribution of neutralino and chargino masses for the set of flat prior models satisfying all of our constraints.}
\label{fig3}
\end{figure}

\begin{figure}[htbp]
\centerline{
\includegraphics[width=13.0cm,angle=0]{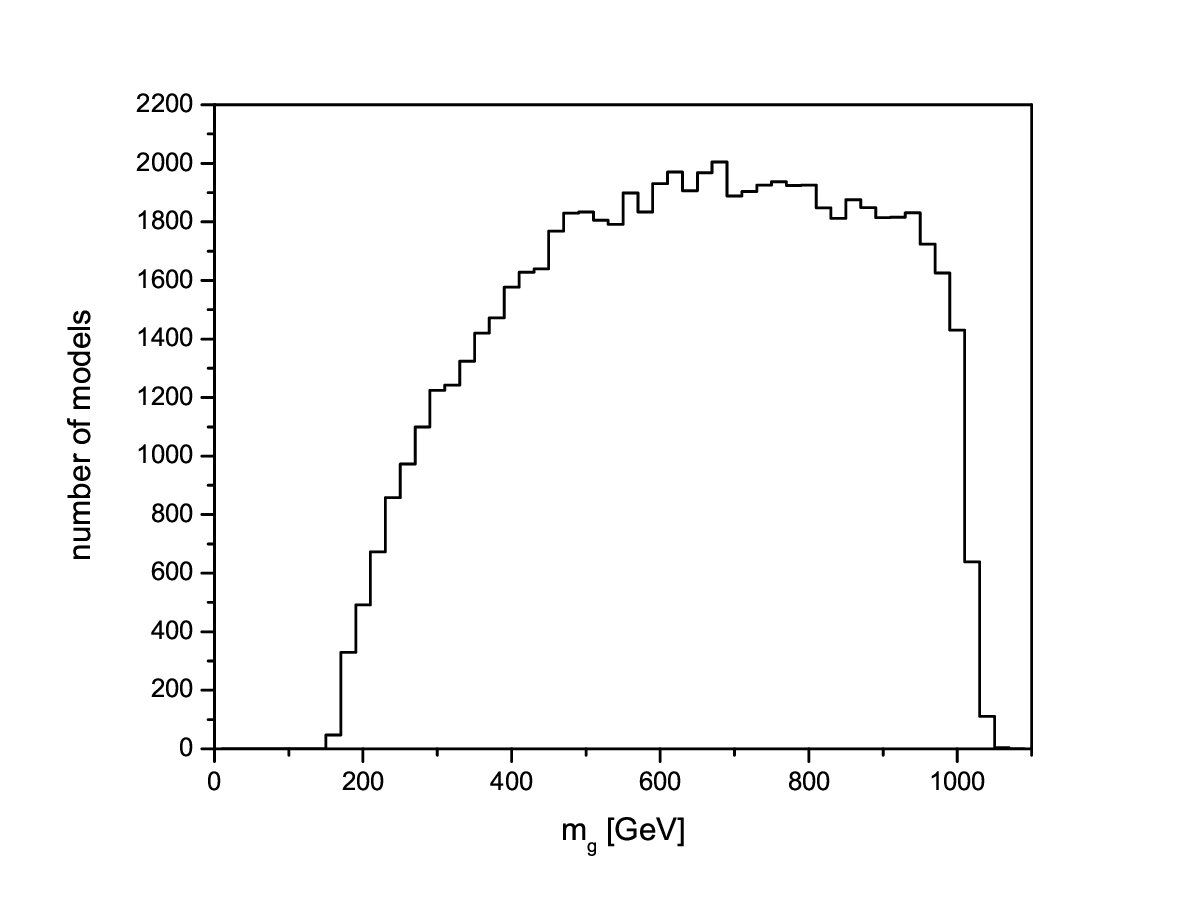}}
\vspace*{0.2cm}
\centerline{
\includegraphics[width=13.0cm,angle=0]{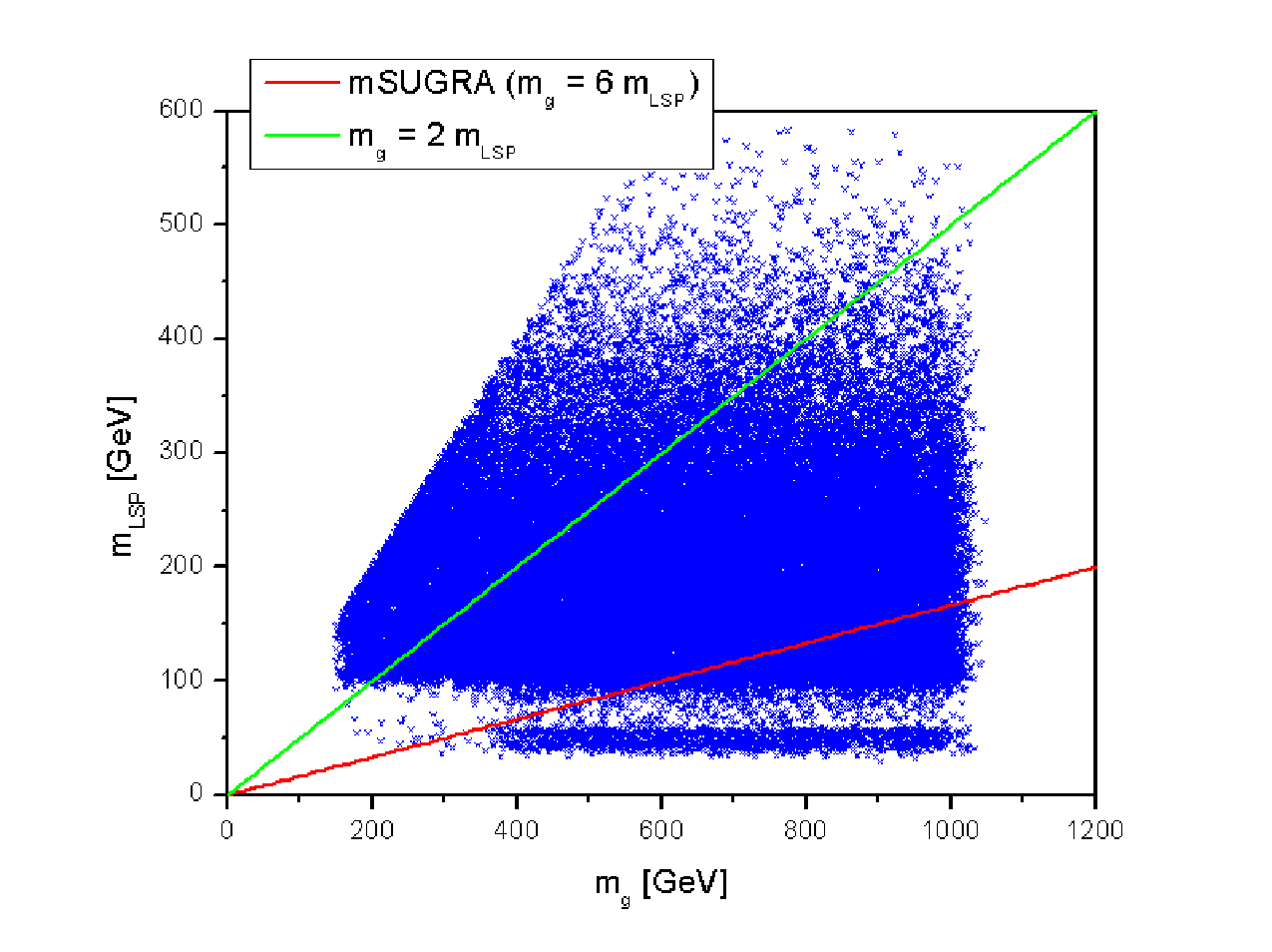}}
\caption{Distribution of gluino masses and a comparison of the gluino and LSP masses for the set of flat prior models satisfying all of our constraints.}
\label{fig4}
\end{figure}

\begin{figure}[htbp]
\centerline{
\includegraphics[width=13.0cm,angle=0]{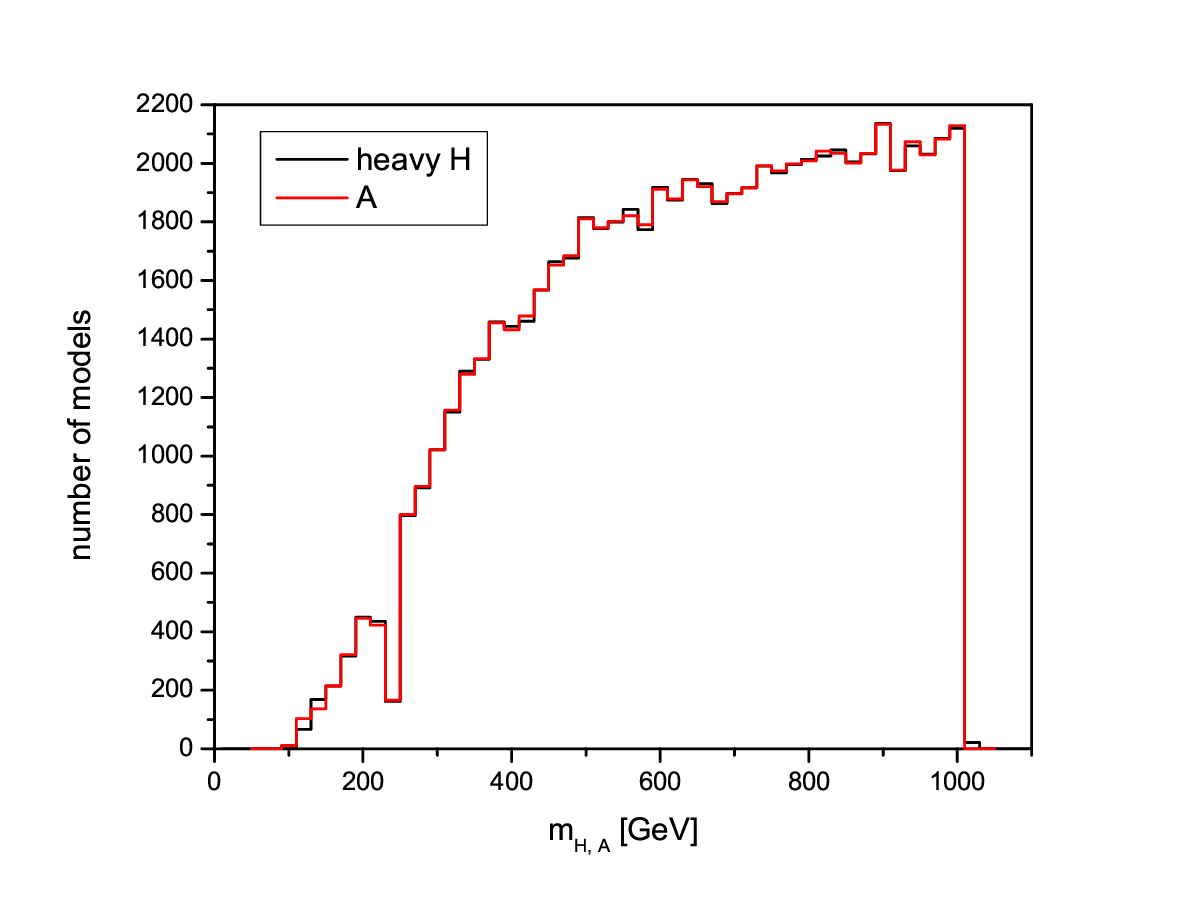}}
\vspace*{0.1cm}
\centerline{
\includegraphics[width=13.0cm,angle=0]{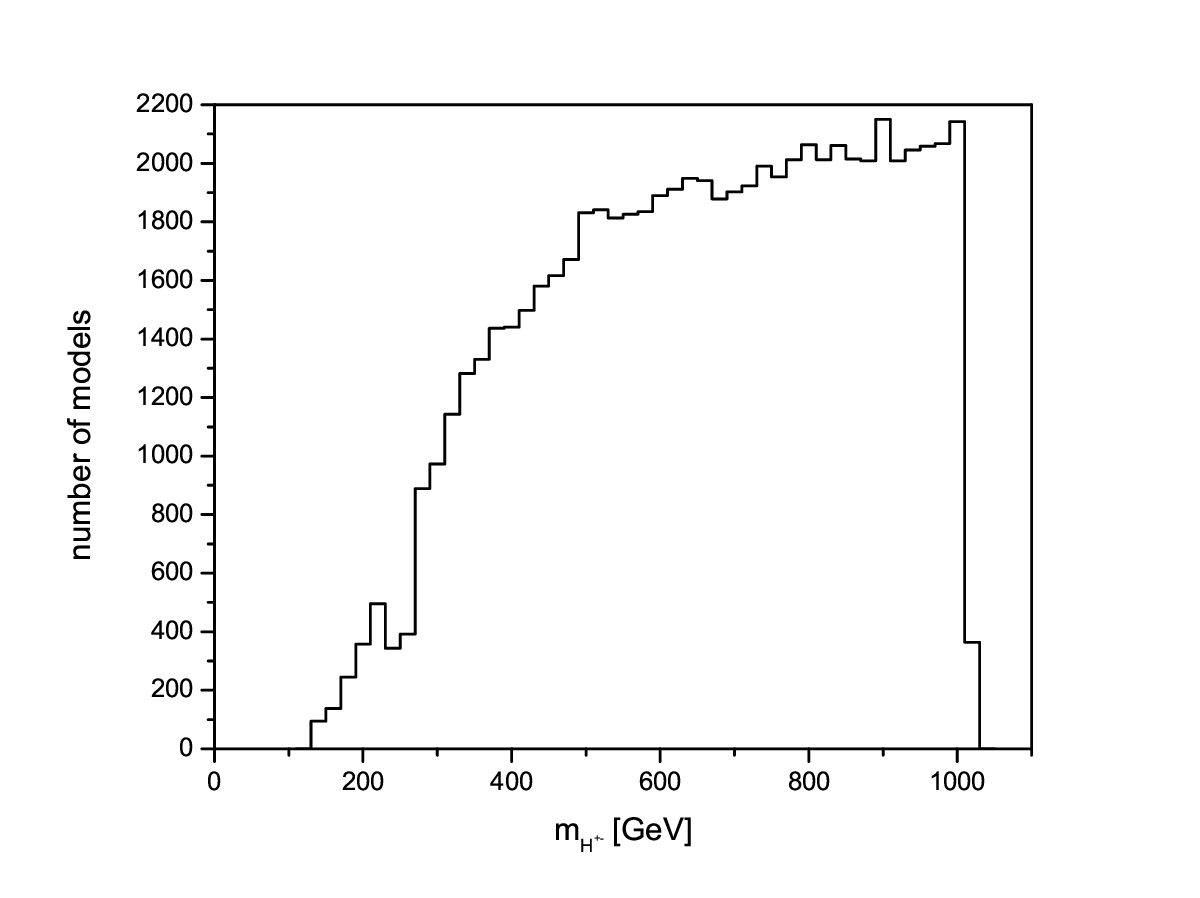}}
\caption{Distribution of heavy and charged Higgs masses for the set of flat prior models satisfying all of our constraints.}
\label{fig5}
\end{figure}

\begin{figure}[htbp]
\centerline{
\includegraphics[width=14.0cm,angle=0]{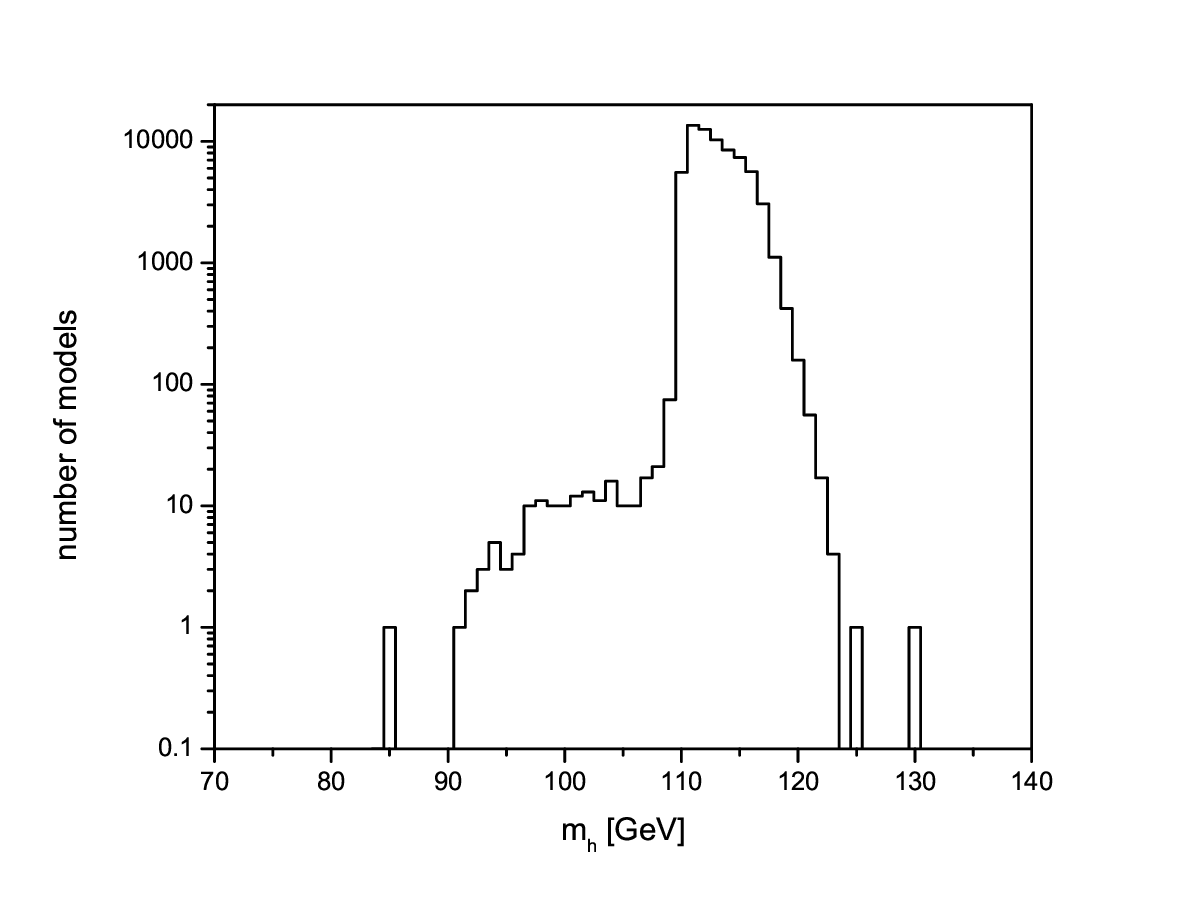}}
\caption{Distribution of the light Higgs mass on a log scale for the set of flat prior models satisfying all of our constraints.}
\label{fig6}
\end{figure}

Figures~\ref{fig1}~--~\ref{fig6} show histograms of the distributions of the various sparticle masses, as well as for the charged and neutral Higgs boson masses, in our set of surviving models . In Fig.~\ref{fig1}, we display the distributions of the selectron/smuon, stau and sneutrino masses. We observe several features: 
($i$) right-handed selectrons/smuons tend 
to be lighter than their corresponding left-handed sleptons, ($ii$) the lighter stau tends to be the lightest charged slepton, \ie, there is an inclination for it to be lighter than selectrons 
or smuons with either helicity label; this is most likely due to stau mixing effects, ($iii$) tau sneutrinos have a proclivity to be lighter 
than electron/muon sneutrinos (which is not surprising, recalling the connection between the masses of the left-handed charged sleptons and their associated sneutrinos), 
and ($iv$) a reasonable fraction of these models, $\gsim 10\%$, would predict that some type of slepton 
is kinematically accessible at a 500 GeV $e^+e^-$ linear collider. 

We present the squark mass distributions in
Fig.~\ref{fig2}, where we see that the lightest stops and sbottoms are naturally heavier than the top mass (and thus the constraint on stop/sbottom masses imposed above is found to 
have little effect). The upper left panel reveals particularly interesting results
for the first/second generation (\ie, `light') squarks.
Here we see that there are a significant number of models with first/second generation  squarks masses below $\sim 300$ GeV, \ie, 
well below the limits  
usually quoted from Tevatron squark searches{\cite {domet}}.  It is important to remember that all the models 
shown here have been subjected to and have {\it passed} the Tevatron search constraints from D0. However, the Tevatron searches are
targeted at mSUGRA models, which have a very restricted type of mass spectra. There are several ways that light squarks may survive 
the D0 analysis and not be excluded: ($i$) if a left-handed squark is light it will tend not to decay immediately to missing energy but instead undergo a cascade (assuming the electroweak gauginos are also light). In this 
case the D0 search strategy can easily fail as the search criterion are invalid. 
($ii$) If a right-handed squark is relatively light, it will more than likely decay into a jet plus the LSP. However if the mass difference between the squark and the LSP is 
relatively small (something that does {\it not} normally happen in mSUGRA) the resulting events will contain rather soft jets and
will not be affected by the kinematic cuts imposed by D0.
($iii$) Depending on the gluino mass, the production rate of the SUSY signal can be suppressed so that there are not enough events to reach the 
statistical boundary for an observable signal. 
This result is important as relatively light squarks could exist and may have been missed by the Tevatron searches.  In fact, 
there appears to be a reasonable window for light right-handed squarks to be kinematically accessible at a 500 GeV linear collider. Hunters of squarks at the LHC must keep this 
possibility in mind when constructing detailed search strategies. 

Figure~\ref{fig3} displays the mass distributions for both neutralinos and charginos. While the LSP tends to be quite light, we see it is also common for both 
$\chi_1^\pm$ and $\chi_2^0$ to {\it also} to quite light with a mass almost always below $\sim 250-300$ GeV. As we will see below, the almost pure 
Higgsino LSP scenario is reasonably common in our model set and the rough degeneracy of these three states then follows nearly automatically. 
The almost pure Wino LSP scenario is also quite common amongst our models and it also predicts a near LSP-$\chi_1^\pm$ degeneracy. Both of these scenarios can 
lead to long-lived charged particle signatures at the LHC. 
The mass distribution for gluinos, and its correlation with the LSP mass, is given in Fig.~\ref{fig4}. Here, again, we see that many models allow for a rather light 
gluino, something we would not have expected based on mSUGRA due to 
the results of the D0 jet plus missing energy search. However, most of these light gluinos escape the D0 search due 
to, \eg,  the smaller mass splitting with the LSP as discussed above in the case of squarks; this leads to 
softer jets that avoid the D0 search requirements{\cite {jay}}. 

In Fig.~\ref{fig5} we see that the $A, H$ and $H^\pm$ masses tend to track each other even for rather low mass values. A few percent of our models have 
heavy Higgs bosons which are reasonably light and are thus in the non-decoupling scenario. We note that there is  
a small depletion in the heavy Higgs counts when $m_{H,A,H^+} \simeq 2 m_h$. This is due to a small problem with the calculation of the $H\to 2h$ decay width near 
threshold{\cite {discuss}} in 
the version of SUSY-HIT that we employ. In order to avoid this problem, we have removed a small 
region of the phase space, leading to the depletion observed here in the $H,A$ and $H^+$ distributions.
Figure~\ref{fig6} shows the distribution for the lightest Higgs mass (recall there is a 3 GeV uncertainty in the SuSpect calculation). Here 
we see that ($i$) it is difficult to generate masses much above 125 GeV (this is most likely related to our chosen parameter range that stop masses 
be below $\sim 1$ TeV), 
and ($ii$) a small set of models predict light Higgs bosons with masses as low as $85-90$ GeV. 
Clearly in such models the Higgs couplings are either quite different than 
in the SM with, \eg, reduced ZZh couplings with a lighter $m_A$ (such as in the non-decoupling region), 
and/or the light Higgs decays into LSP pairs with a large 
branching fraction (while simultaneously avoiding the constraint on $Z$ decay into invisible particles).  
In fact, we find that both situations do occur in our set of models. We note in passing that we encounter a second SUSH-HIT issue in the case of very 
light Higgs in the mass range between 80 and 90 GeV. Here, an anomalously large partial width for the $\gamma \gamma$ mode is returned by the code yielding 
an invalid result. These few cases appear in both the flat and log prior samples and are dropped from our final model sample.  


\begin{figure}[htbp]
\centerline{
\includegraphics[width=13.0cm,angle=0]{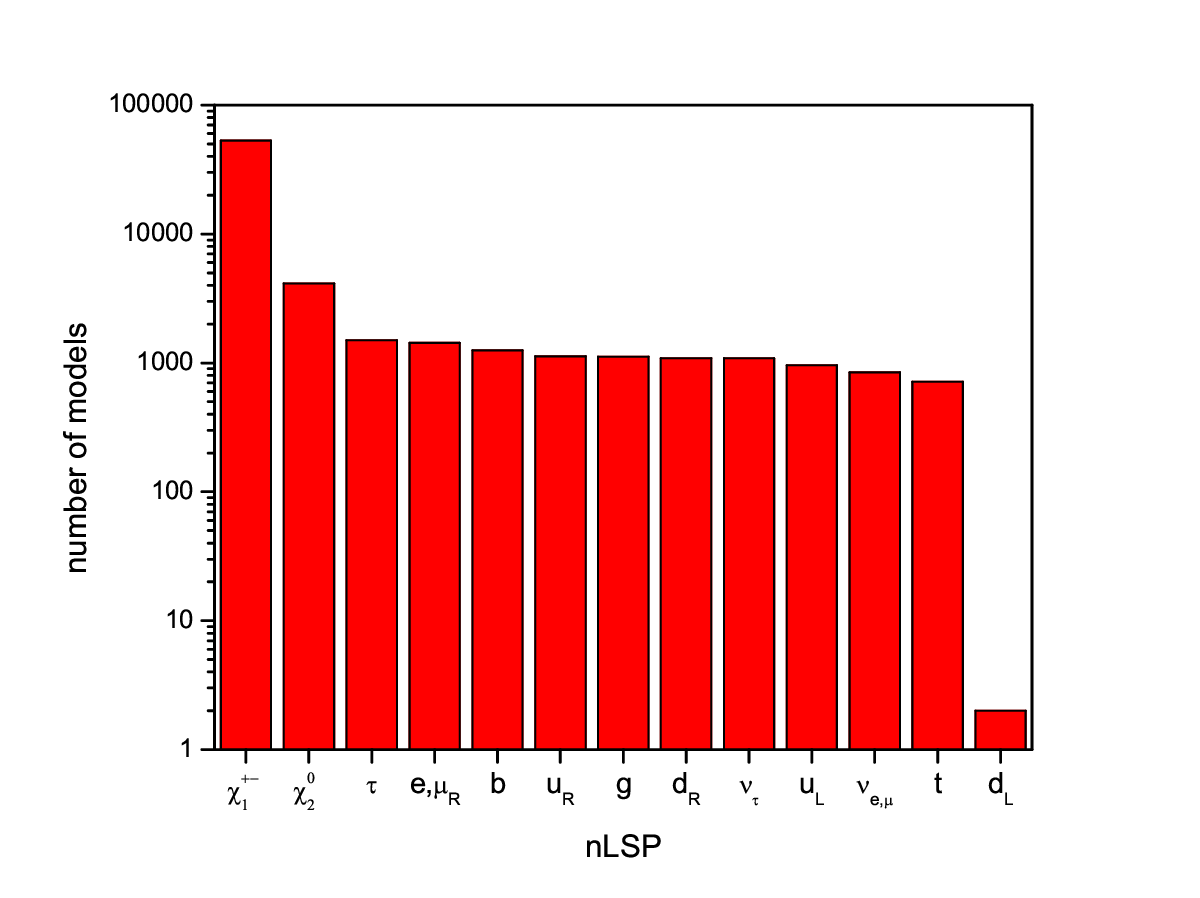}}
\vspace*{0.2cm}
\centerline{
\includegraphics[width=15.0cm,angle=0]{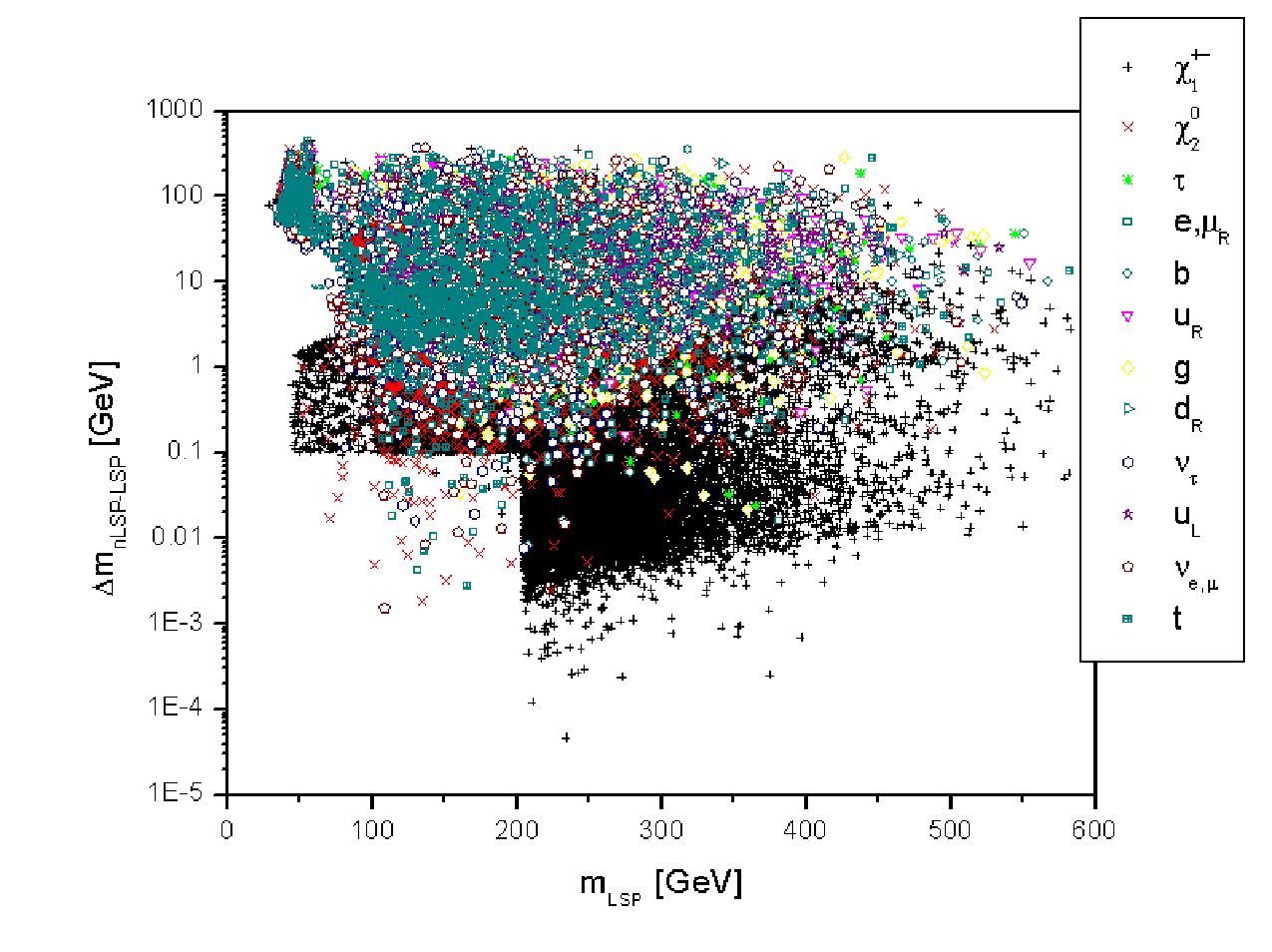}}
\caption{(Top) Identity of the nLSP. (Bottom) nLSP-LSP mass splitting as a function of the LSP mass, with the identity of the nLSP as labeled. Both assume flat priors.}
\label{fig7}
\end{figure}

The properties of the nLSP are examined in Fig.~\ref{fig7}, which shows the various possible identities of the nLSP in our surviving flat prior model 
set, as well as the mass splitting between the LSP and the nLSP as a function of the LSP mass. These two properties essentially
determine the collider phenomenology of a model. 
From this figure, we see that the lightest 
chargino plays the role of the nLSP much of the time but many of the other sparticles can also be the nLSP at the level of a few percent. The second most common nLSP 
is found to be $\chi_2^0$ occurring in $\sim 6-7\%$ of the cases; these are mainly pure Higgsino LSP scenarios. Other identities of the nLSP may lead to unusual signatures 
at the LHC. In the lower panel of this figure we see not only the nLSP identity but also the nLSP-LSP mass splitting, $\Delta m$, as a function of 
the LSP mass itself. Here we see that there is a strong concentration of models with small values of $\Delta m$ in our sample; in hindsight this is not too surprising 
since, as we shall see below, the LSP in our model set is commonly either pure Wino or Higgsino or a small admixture of these two cases. 
The bulk of the mostly empty square region which appears  
on the lower left-hand side of the figure has been removed by the stable chargino search at the Tevatron; only non-chargino nLSPs remain in this region{\footnote 
{Note that sleptons with masses in this range have cross sections which are too small to be excluded by the Tevatron search.}}. Again, we emphasize that the LSP in the models with small 
values of $\Delta m$ tend much of the time   
to be close to a Wino or Higgsino (or combination thereof) electroweak eigenstate as we will see below. We note that a stable
heavy charged particle search at the LHC should exclude or discover the models with heavier Chargino nLSPs and small values of $\Delta m$.
There are many interesting 
features worth noting in the lower panel of this figure.  For example, we see that 
there are a cluster of models with light charginos, \ie, masses less than 100 GeV, but with small ($\lsim 1$ GeV) values of $\Delta m$. Such particles, though 
sufficiently short-lived, would only leave soft tracks in a LEP detector and would likely to have been missed. We also see a set of models with a very small $\Delta m$ where $\chi_2^0$ is the 
nLSP but with masses such that $Z\to \chi_2^0\chi_1^0$ is kinematically forbidden. Again, such final states would be nearly impossible to 
observe and would have been missed at LEP.  This figure also reveals 
cases with light squark nLSPs and small values of $\Delta m$.  
Furthermore, we observe that models in this figure with the lightest LSPs and also have larger values of 
$\Delta m \sim 10-300$ GeV, can have almost any SUSY particle as the nLSP. The small gap in the region $1\lsim\Delta m\lsim 30$ for LSP masses 
$\lsim 100$ GeV may be easily explained. One possibility is that in the many models where we encounter an 
almost pure Wino or Higgsino LSP with a small mass difference with a chargino nLSP, the mass of such a chargino must be in excess of $\sim 100$ GeV due to LEP 
constraints.  Bino LSPs, on the otherhand, are not subject to such constraints and can be much lighter 
though are found to be somewhat less common in our sample..

Based on these results there are many interesting possible scenarios that can arise at the LHC. 
Here, we give two examples involving small mass splittings between the nLSP and LSP. In the first case we consider, we
find that there are a class of models with a squark nLSP with very light masses $\sim 80-120$ GeV and mass splittings from the LSP of only a few GeV or less. In the 
low end of this squark mass range, such events would have appeared as a two-photon background at LEP, whereas for the full mass range the relative softness of the 
jets in the final state 
would imply that the squarks would have been missed by the D0 multijet search.  
Such sparticles would be notoriously difficult to detect at the LHC; it is perhaps possible that using 
event samples with additional gluon ISR would be useful{\cite {jay}} but would require further study. A second interesting scenario is the case 
where $\tilde \chi_2^0$ is the nLSP having a small mass splitting with the LSP. The properties of $\tilde \chi_2^0$
neutralinos are usually determined via the cascade decays of heavier colored sparticles produced 
with large cross sections or when they are produced in association with the lightest chargino via the trilepton mode discussed above.  The most common case 
where $\tilde \chi_2^0$ is the nLSP 
in our model sample is for both it and the LSP to be dominantly Higgsino-like, further implying that the lightest chargino would also be nearby in mass. In this situation,
due to the Higgsino-like nature, 
we would not expect significant trilepton production and the appearance of
$\chi_2^0$ in cascade decays would be most unusual. If  $\tilde \chi_2^0$ is produced in a cascade decay (by, say, stop pair 
production 
followed by the decay $\tilde t_1 \to \tilde \chi_2^0$), the $\tilde \chi_2^0$ decay to the LSP would be controlled by the size of the allowed phase space which may be 
well below 1 GeV in some cases leading to a possible secondary vertex. For some mass splittings, decays through a virtual $Z$ or $h$ may dominate leading to an 
additional pair of (b)jets or leptons. In other cases, decays into a photon would be dominant leading to a final state with additional non-pointing photons.

Since the nLSP can be almost any SUSY particle, and the corresponding mass splittings can be small in all cases, essentially all of the conventional SUSY long-lived 
particle scenarios can be captured in our model set. For example, long-lived stops or staus (as in GMSB\cite{gmsbb}), 
gluinos (as in Split SUSY\cite{SSS}) as well as charginos (as in AMSB\cite{amsbb}) 
all occur in our sample as do other possibilities including long-lived selectrons, sneutrinos and sbottoms. 
Of course, long-lived $\tilde \chi_1^0$ neutralinos can occur in GMSB, where they are the nLSP, whereas here long-lived neutralinos can only be the 
$\tilde \chi_2^0$. As an aside, it is interesting to note that the set of mass spectra obtained above for the various sparticles, as well as the large 
number of models with small nLSP-LSP mass splittings, are qualitatively similar to the corresponding ones  
in a set of models given in Ref. {\cite {AKTW}}.  These models were found to be difficult to distinguish at the LHC and were subsequently further analyzed in 
detail in the ILC context{\cite {feature}}. As we will see below, this similarity will remain valid in the case of our models generated via log priors as well. 
This present analysis thus demonstrates that there is nothing special (or qualitatively wrong) about that particular\cite{AKTW} model set. 

\begin{figure}[htbp]
\centerline{
\includegraphics[width=13.0cm,angle=0]{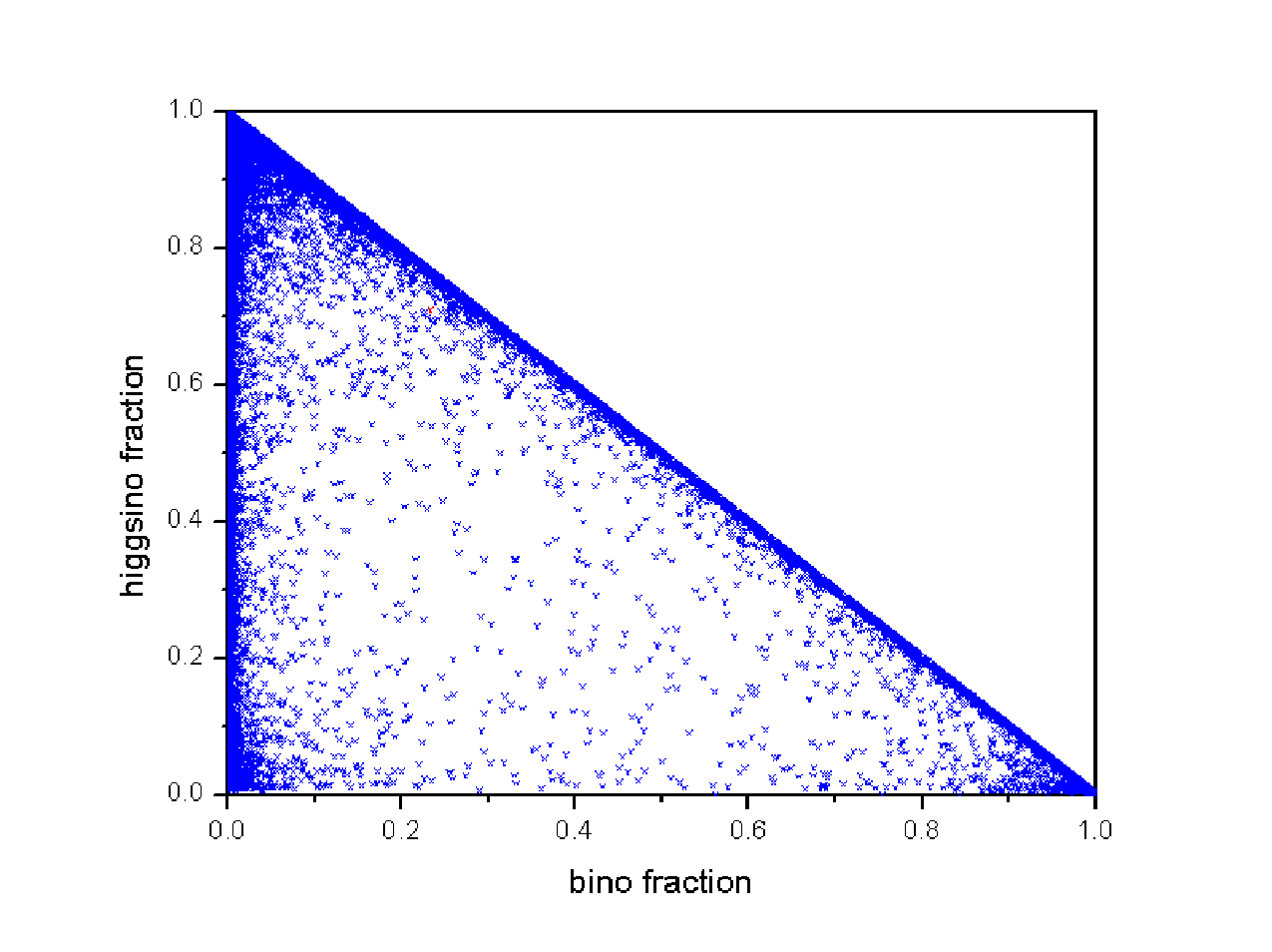}}
\vspace*{0.1cm}
\centerline{
\includegraphics[width=13.0cm,angle=0]{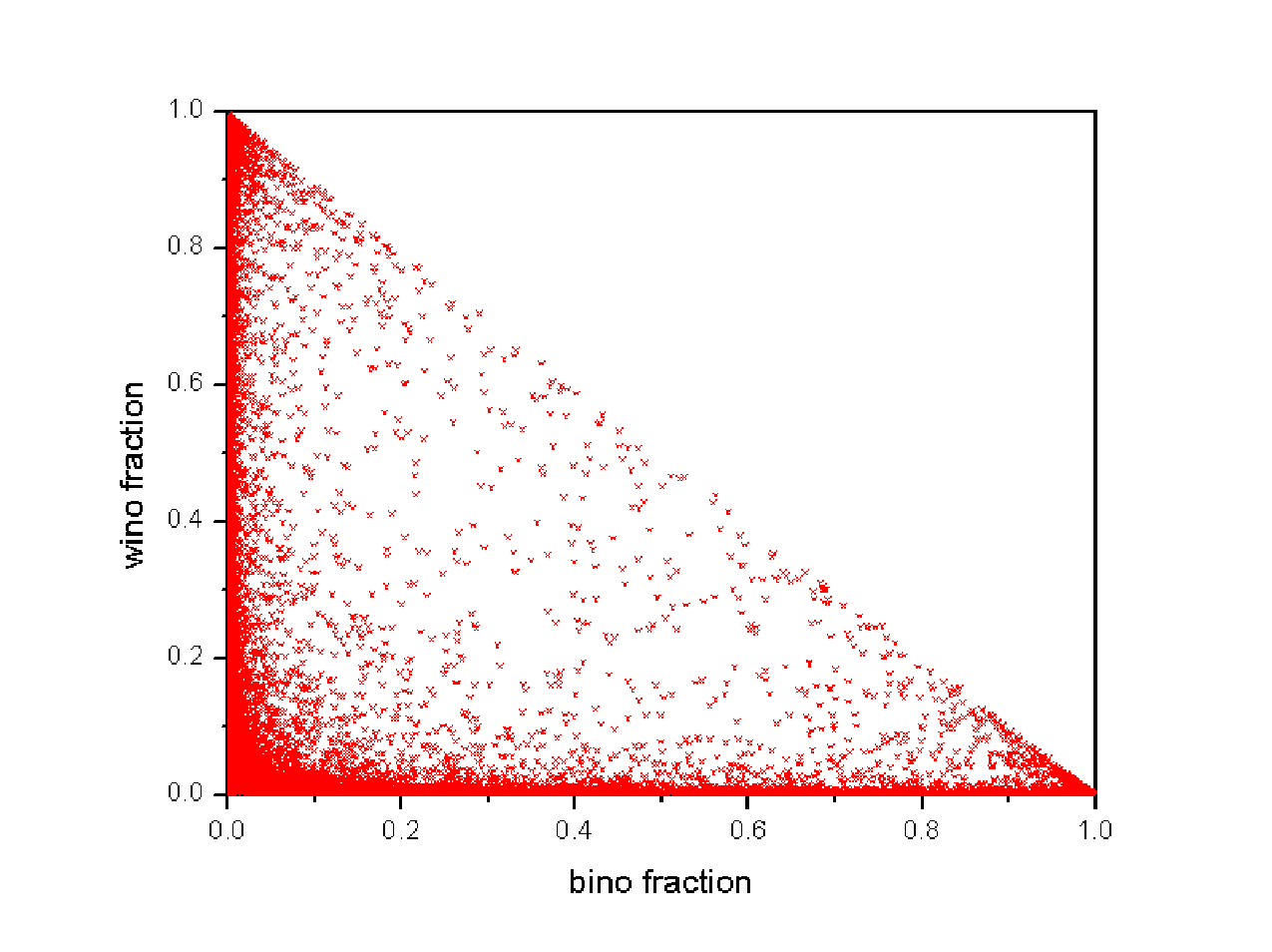}}
\caption{Wino/Higgsino/Bino content of the LSP in the case of flat priors.}
\label{fig8}
\end{figure}

Figure~\ref{fig8} shows the detailed Wino/Higgsino/Bino content of the LSP in our set of models. Here we see that, as advertised, the LSPs in our sample tend to be 
either a rather pure electroweak eigenstate or a mixture of only two of these states. This occurs quite naturally when one of the parameters in the neutralino 
mass matrix takes on a value which is much smaller than the others. 
LSPs with large weak eigenstate admixtures, that would populate the central regions of 
these figures, are seen to be relatively rare in our model sample.  In addition, both Wino-like and Higgsino-like LSPs are seen to be somewhat (\ie, weakly) dominant 
over the Bino-like case familiar from mSUGRA although all three of these possibilities are very well represented in our model sample.

\begin{figure}[htbp]
\centerline{
\includegraphics[width=13.0cm,angle=0]{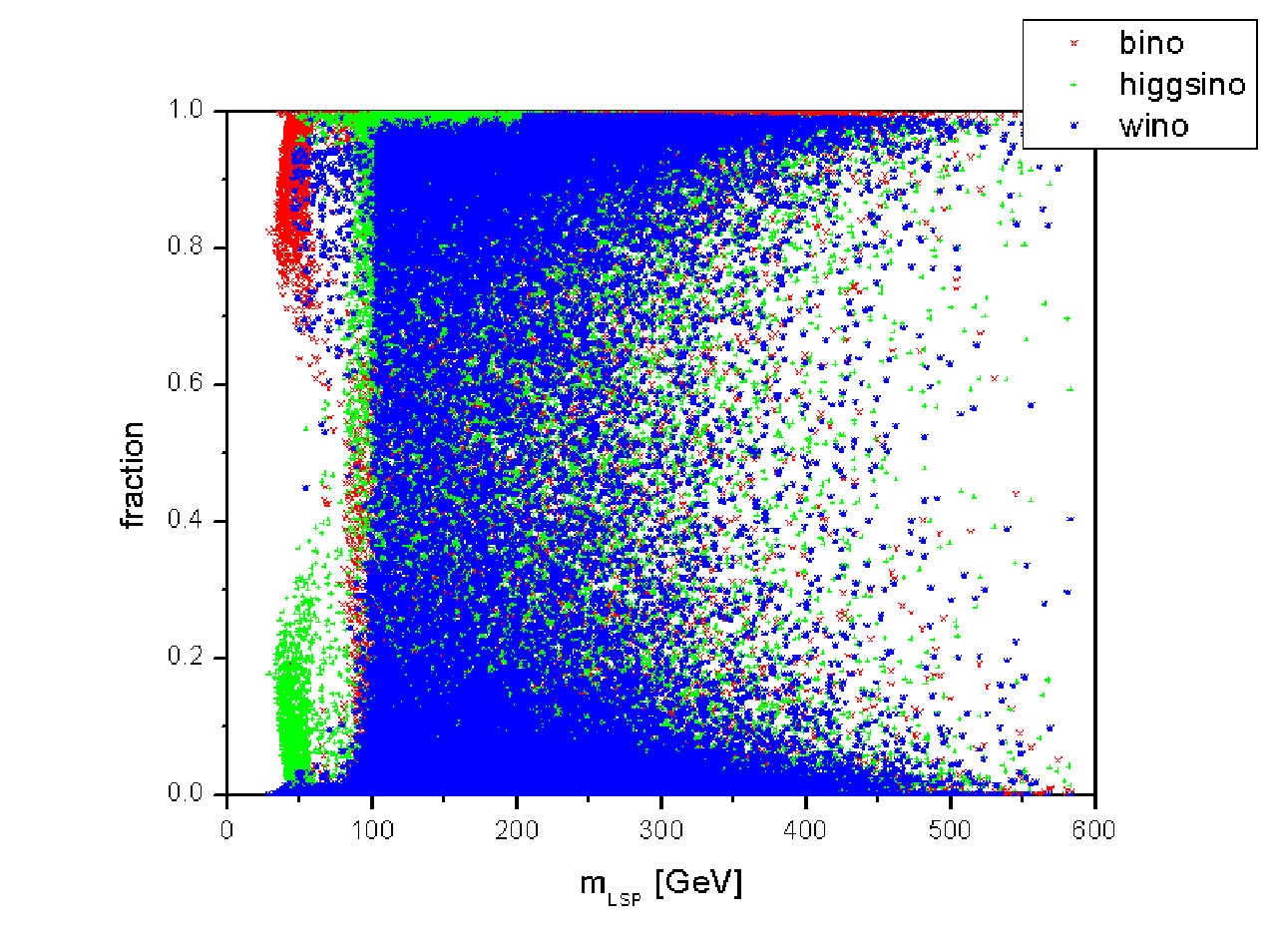}}
\vspace*{0.1cm}
\centerline{
\includegraphics[width=13.0cm,angle=0]{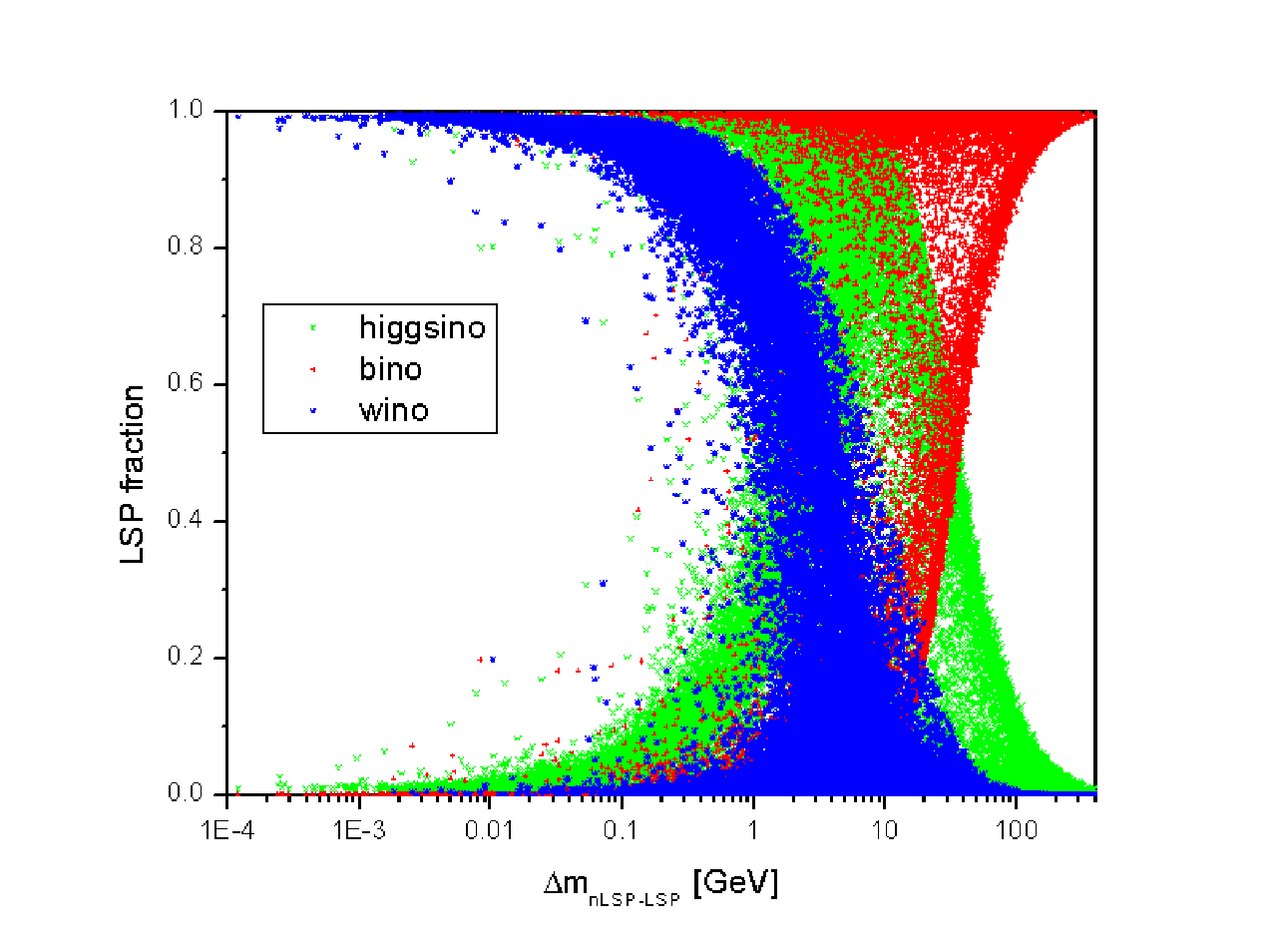}}
\caption{Fractional Wino/Higgsino/Bino content of the LSP in the case of flat priors as functions of the LSP mass(top) and the nLSP-LSP mass splitting(bottom). 
Note that every model has three entries in both of these figures.}
\label{fig8p}
\end{figure}

Complementary information on the Wino/Higgsino/Bino content of the LSP is provided in Figure~\ref{fig8p}. Here we see several things: ($i$) light LSPs tend to have 
a large Bino content with a possible moderate amount of admixing with the Higgsino, and ($ii$) for tiny mass splittings with the LSP the wino content of the 
LSP is quite high, while for large splittings the Bino content apparently dominates. The Higgsino content is large in the mass splitting region between these two 
extreme cases.

\begin{figure}[htbp]
\centerline{
\includegraphics[width=6.0cm,angle=-90]{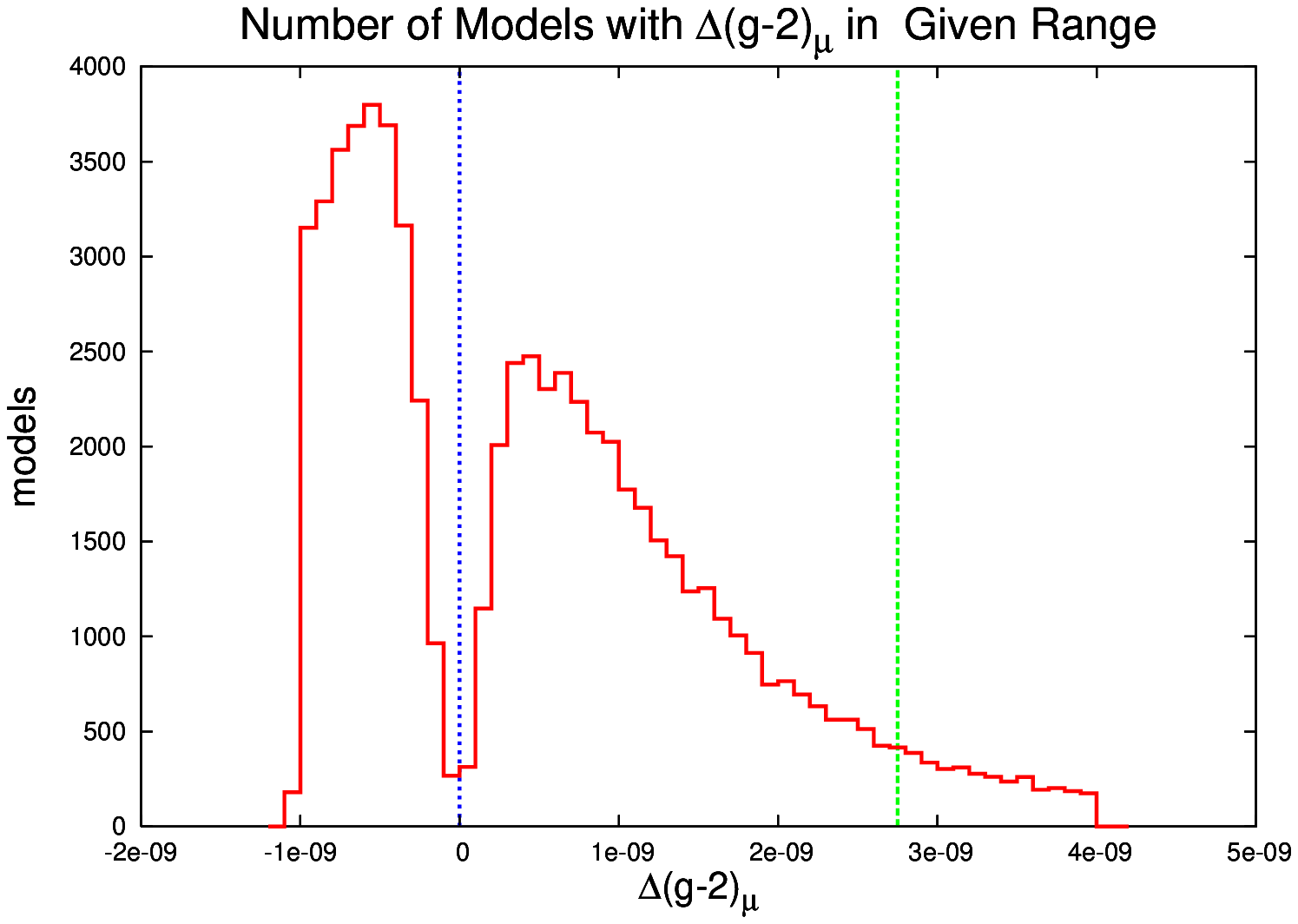}
\hspace*{0.1cm}
\includegraphics[width=6.0cm,angle=-90]{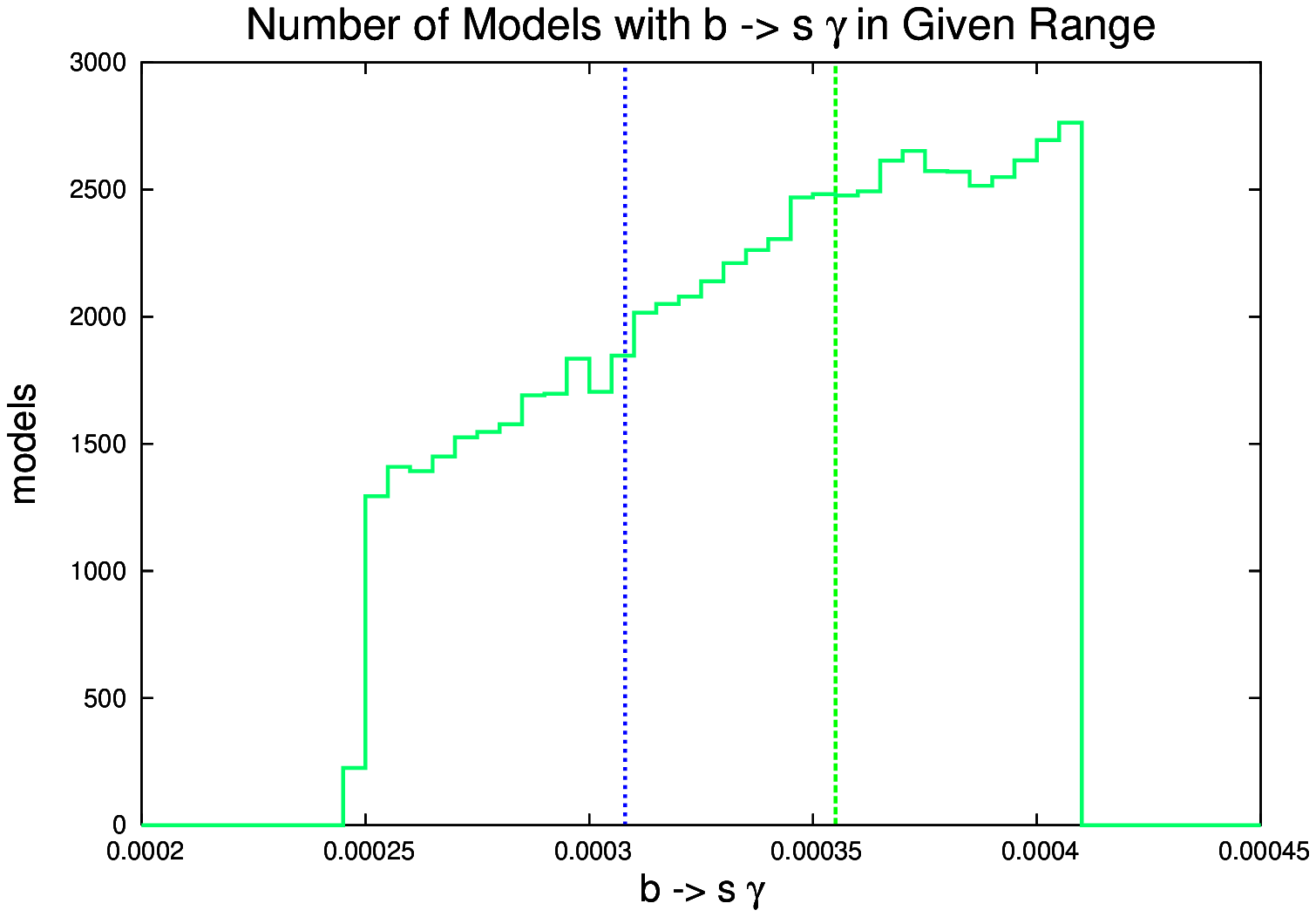}}
\vspace*{0.1cm}
\centerline{
\includegraphics[width=6.0cm,angle=-90]{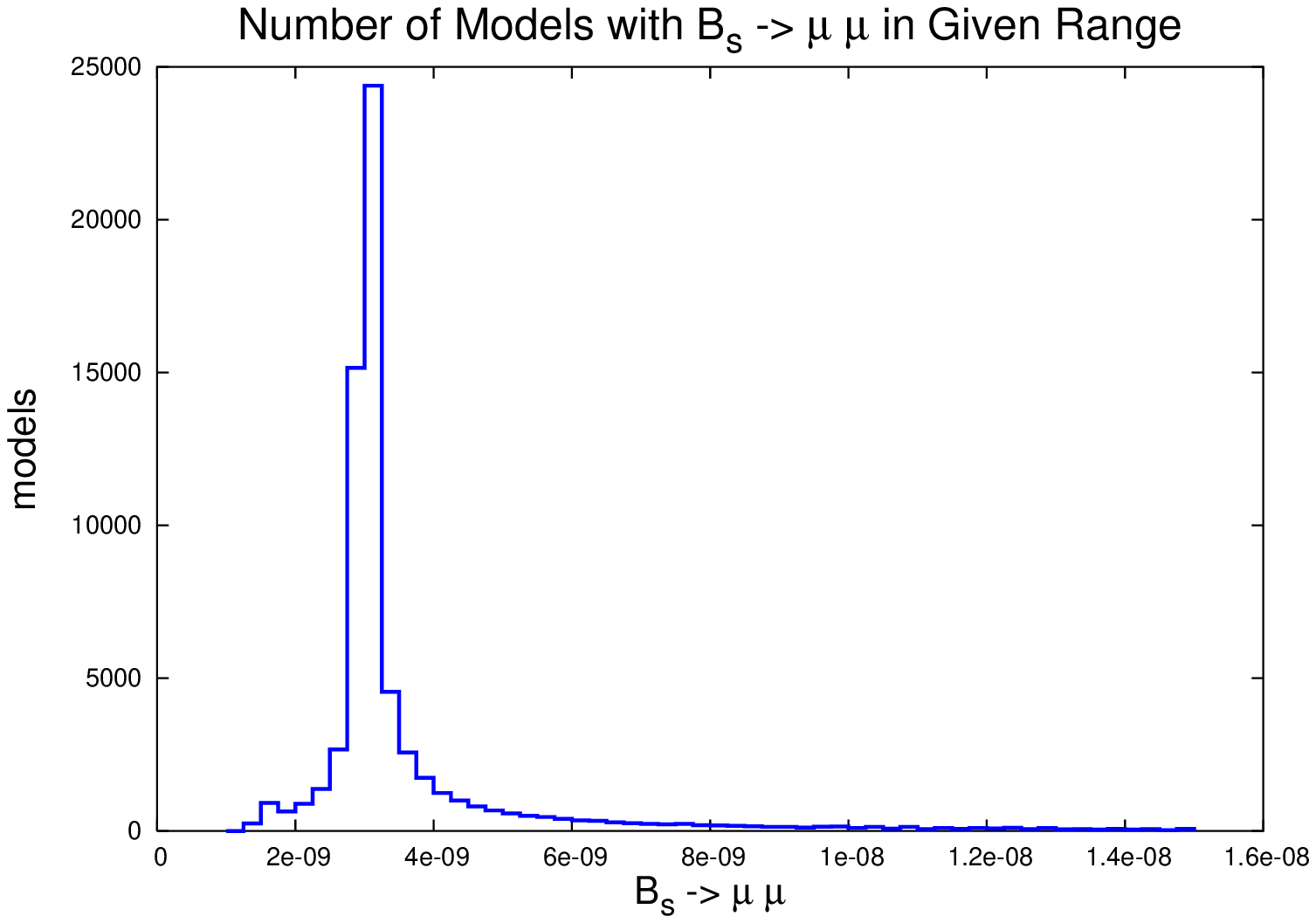}
\hspace*{0.1cm}
\includegraphics[width=6.0cm,angle=-90]{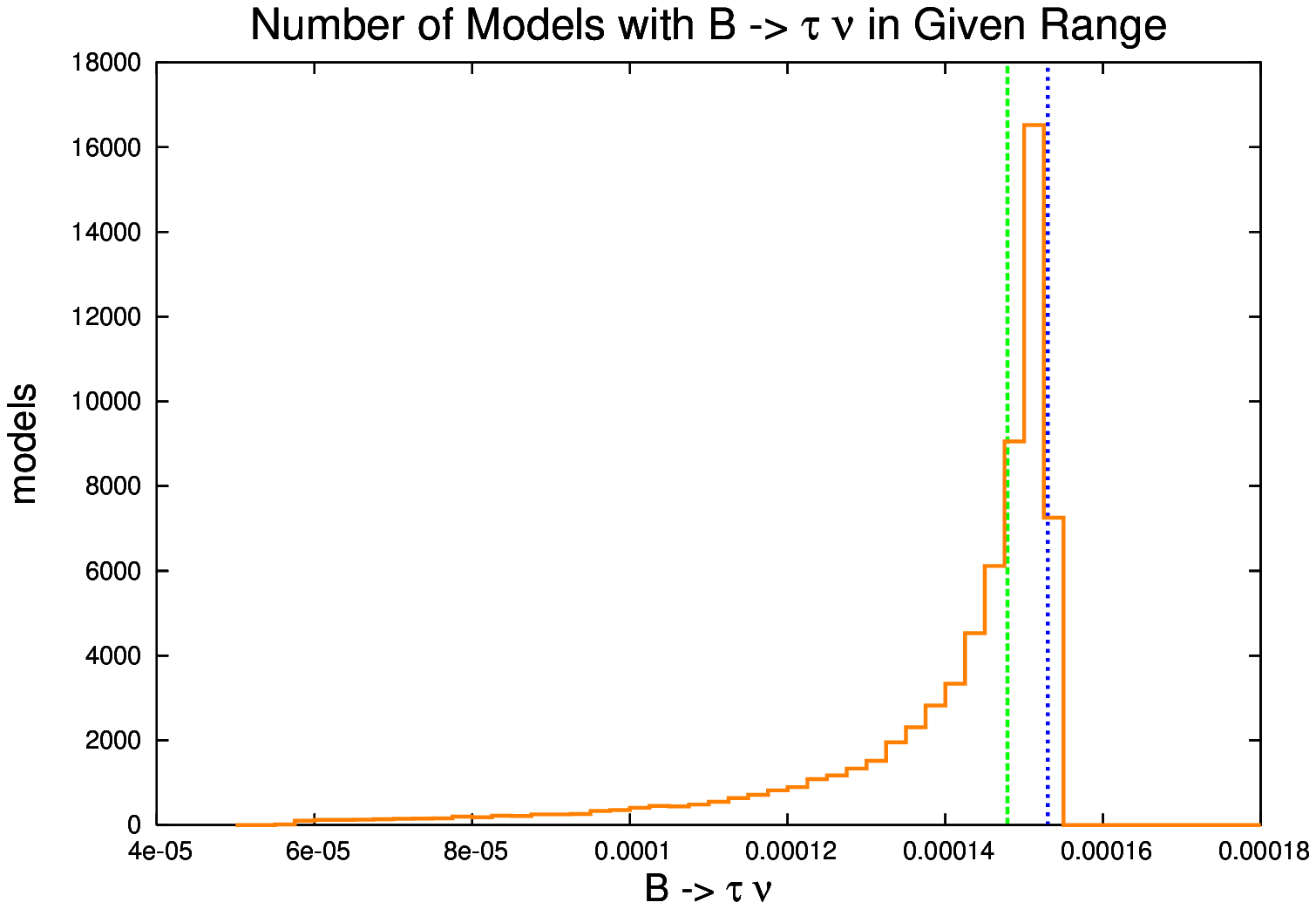}}
\centerline{
\includegraphics[width=6.0cm,angle=-90]{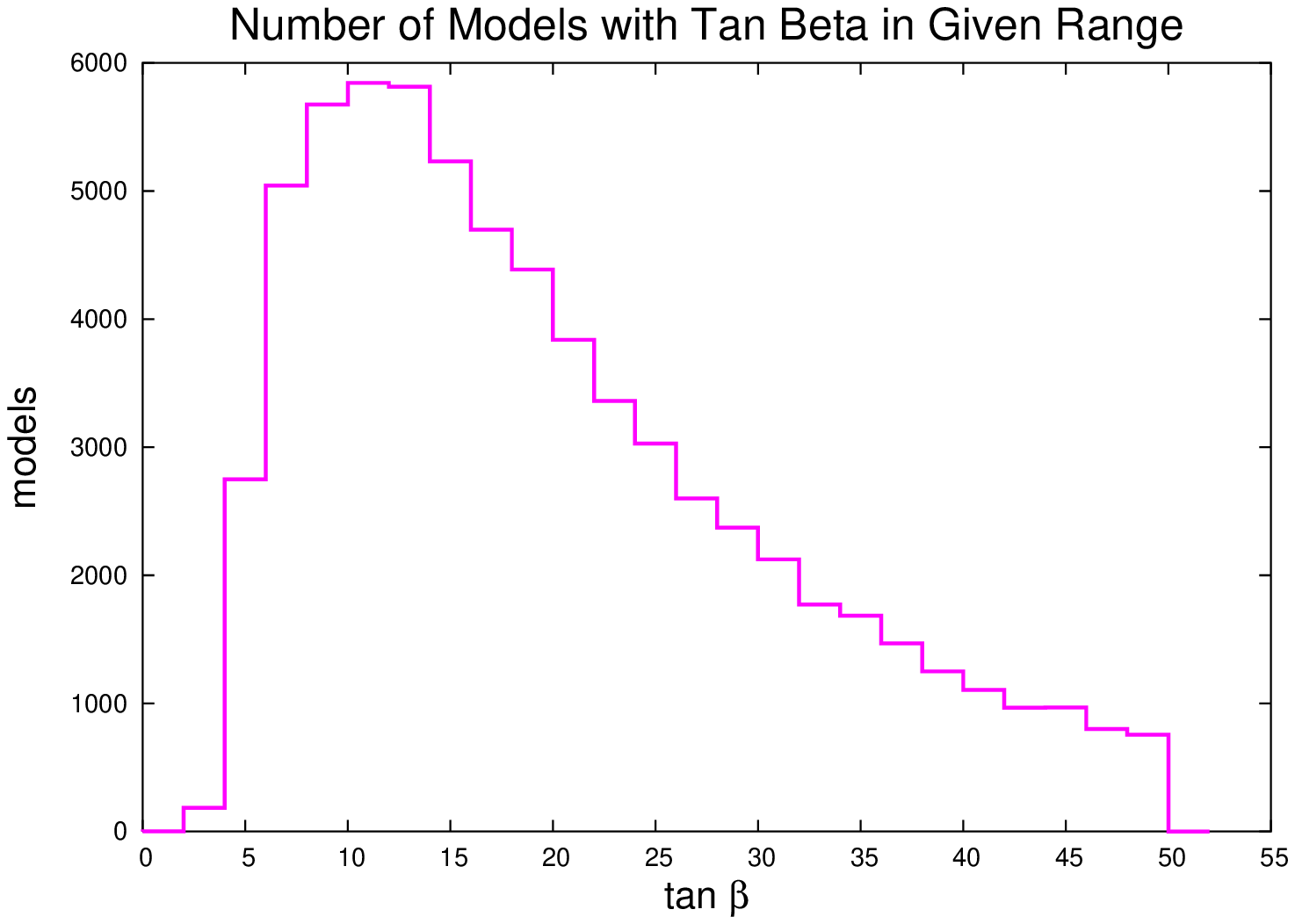}
\hspace*{0.1cm}
\includegraphics[width=6.0cm,angle=-90]{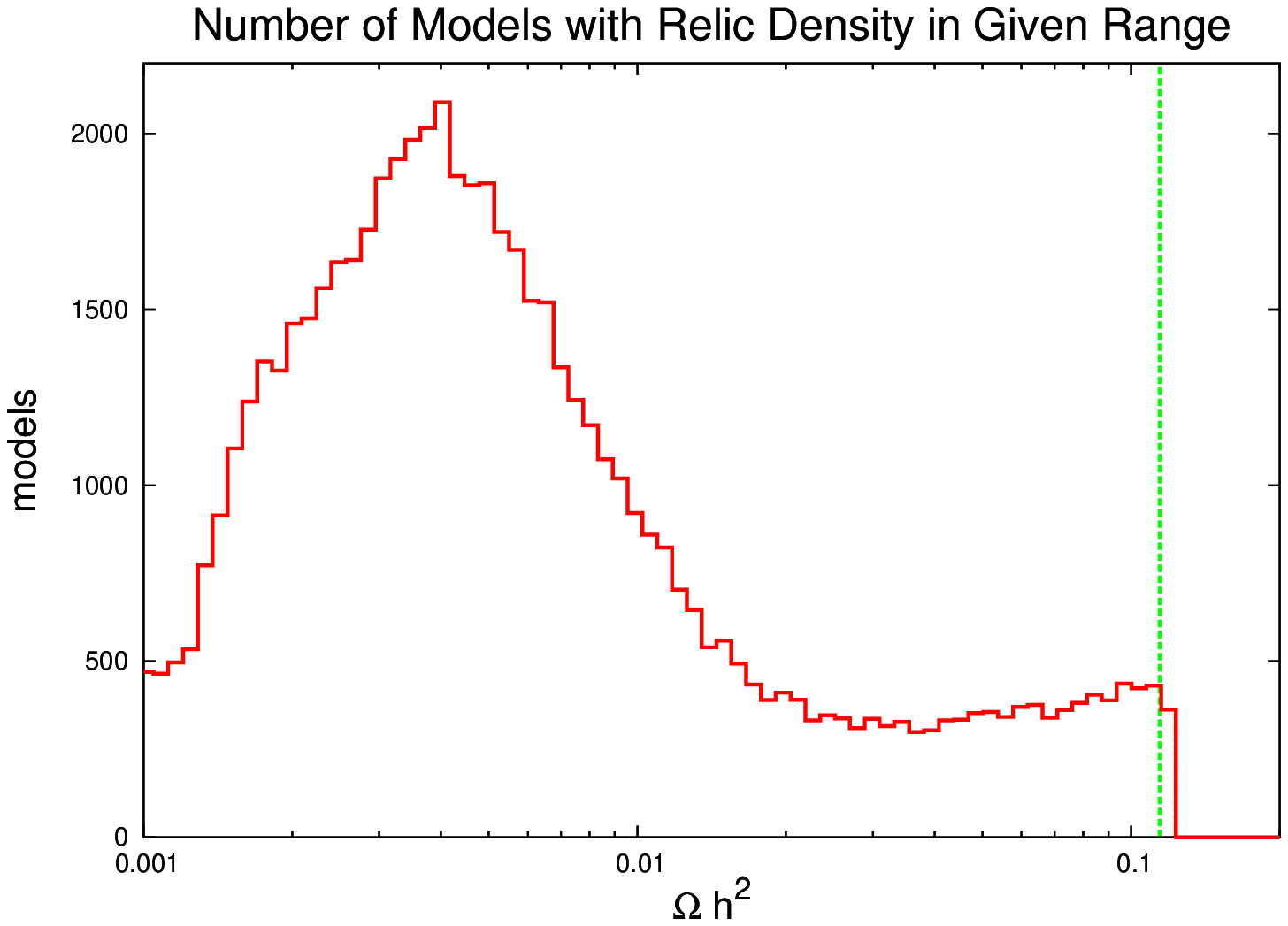}}
\caption{Distributions of predictions for several observables as well as $\tan \beta$ for our model sample subject to the constraints discussed in the text. 
The blue and green dashed lines show the SM predictions as well as the current central values obtained by experiment, respectively. }
\label{fig9}
\end{figure}

Let us now turn to the pMSSM postdictions, \ie, the distribution of values from our model set for the experimental quantities used as constraints in obtaining that model set. 
Figure~\ref{fig9} displays these postdictions for several of the experimental observables, as well as for the values of $\tan \beta$, from our surviving model 
sample. Note that this particular distribution peaks at values near $\tan \beta \simeq 12$ and favors smaller values of this quantity. As expected, the distribution for $\delta (g-2)_\mu$ is 
approximately bimodal, depending upon the sign of the Higgsino mixing parameter $\mu$. Here, we see that most models do not predict a large enough shift in this observable to fully `explain' 
the central value of 
the observed discrepancy with the SM. We also see that while there is a slight preference for somewhat larger values of $B(b\to s\gamma)$ the predicted values for 
$B(B_s\to \mu^+\mu^-)$ lie significantly below the current Tevatron bound and hover around the SM prediction for this decay. 
Generally, $B(B\to \tau \nu)$ is not found to be a great discriminator amongst our model set. Note that the distribution for the predicted value of the relic density $\Omega h^2$ 
is peaked at rather small values implying that in most of our models the dark matter must be dominantly non-LSP in origin. However we also see that 
there is a long tail leading to a significant number of models with $\Omega h^2$ values close to the WMAP upper limit.

\begin{figure}[htbp]
\centerline{
\includegraphics[width=6.0cm,angle=-90]{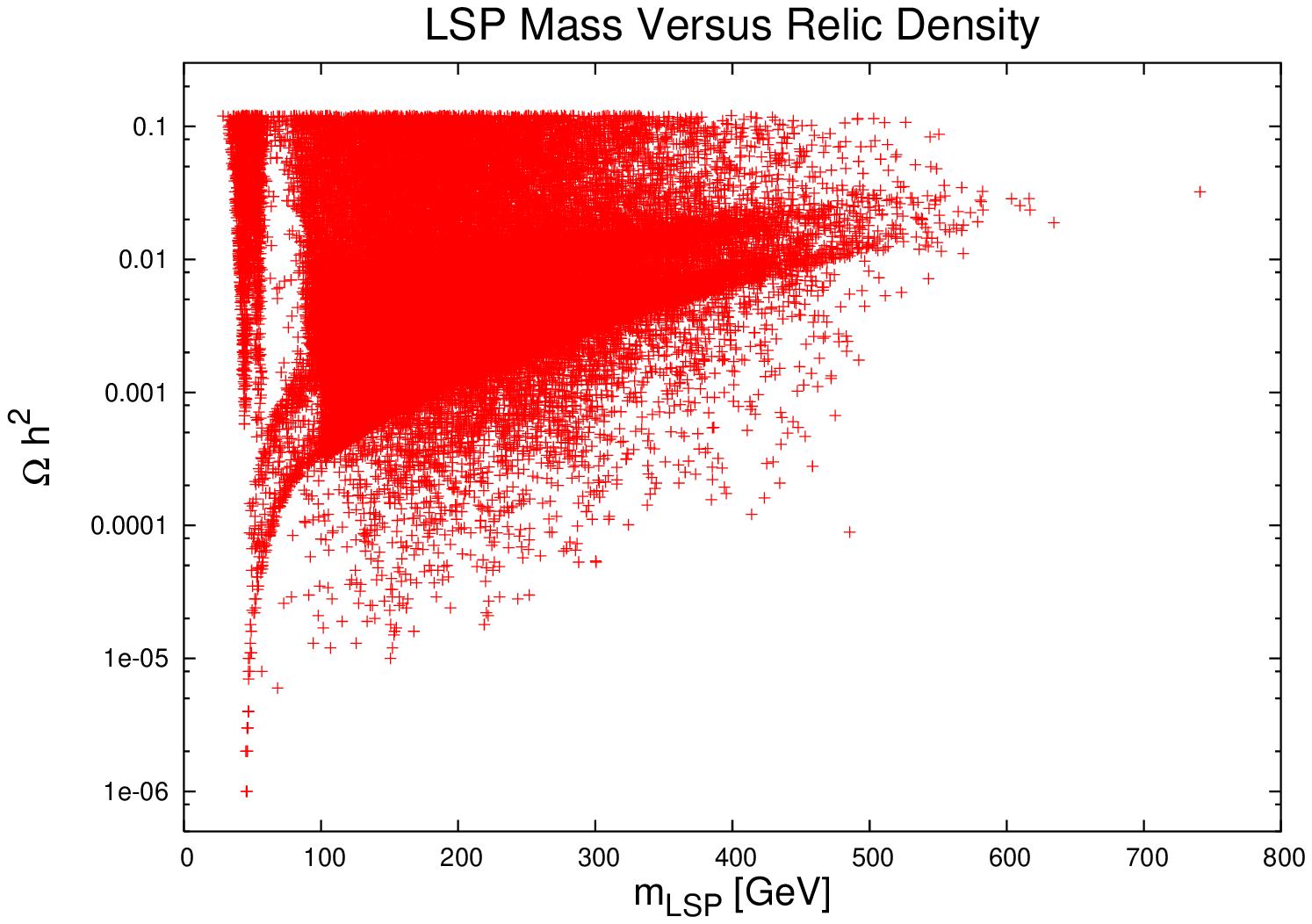}
\hspace*{0.1cm}
\includegraphics[width=6.0cm,angle=-90]{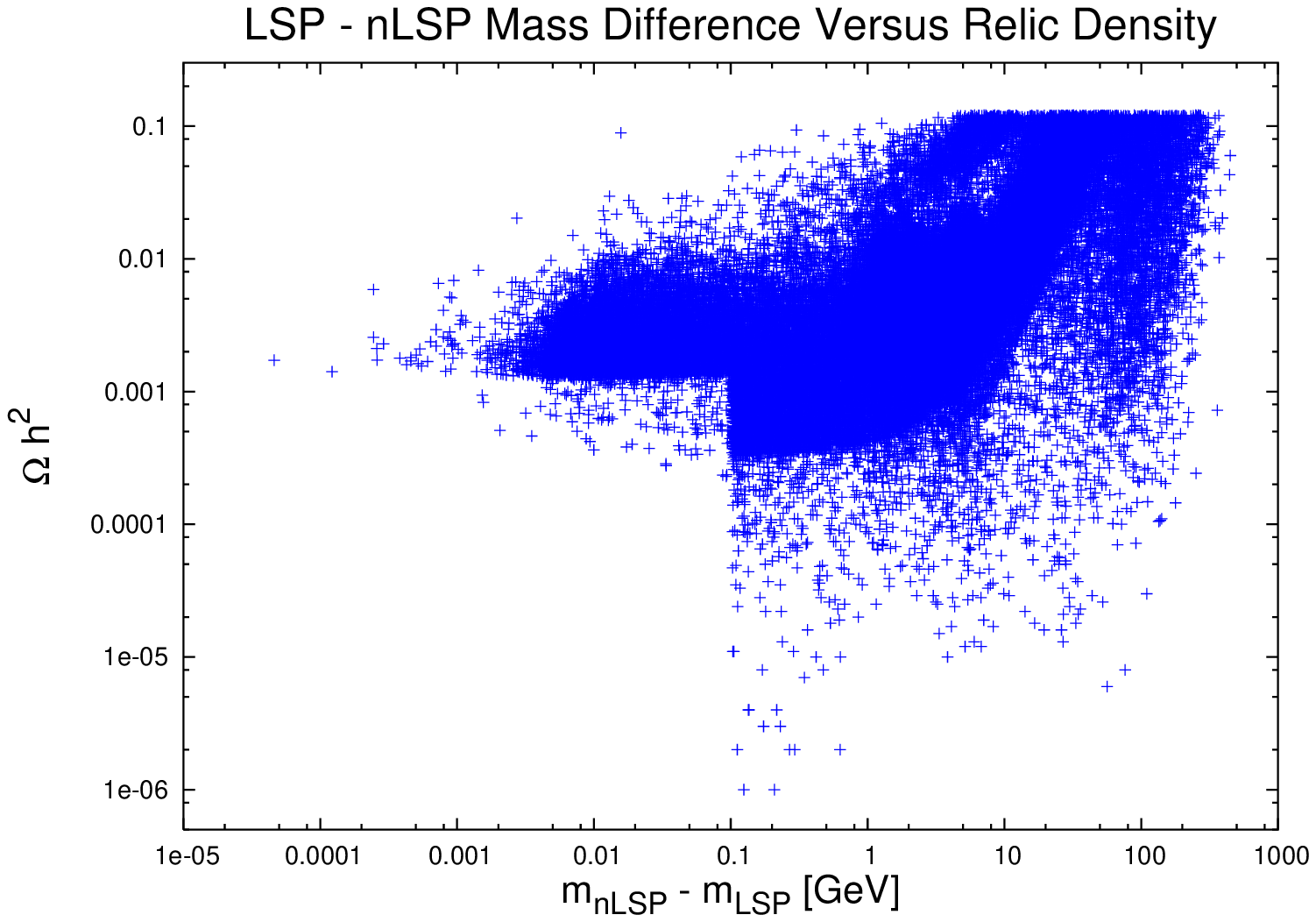}}
\vspace*{0.1cm}
\centerline{
\includegraphics[width=6.0cm,angle=-90]{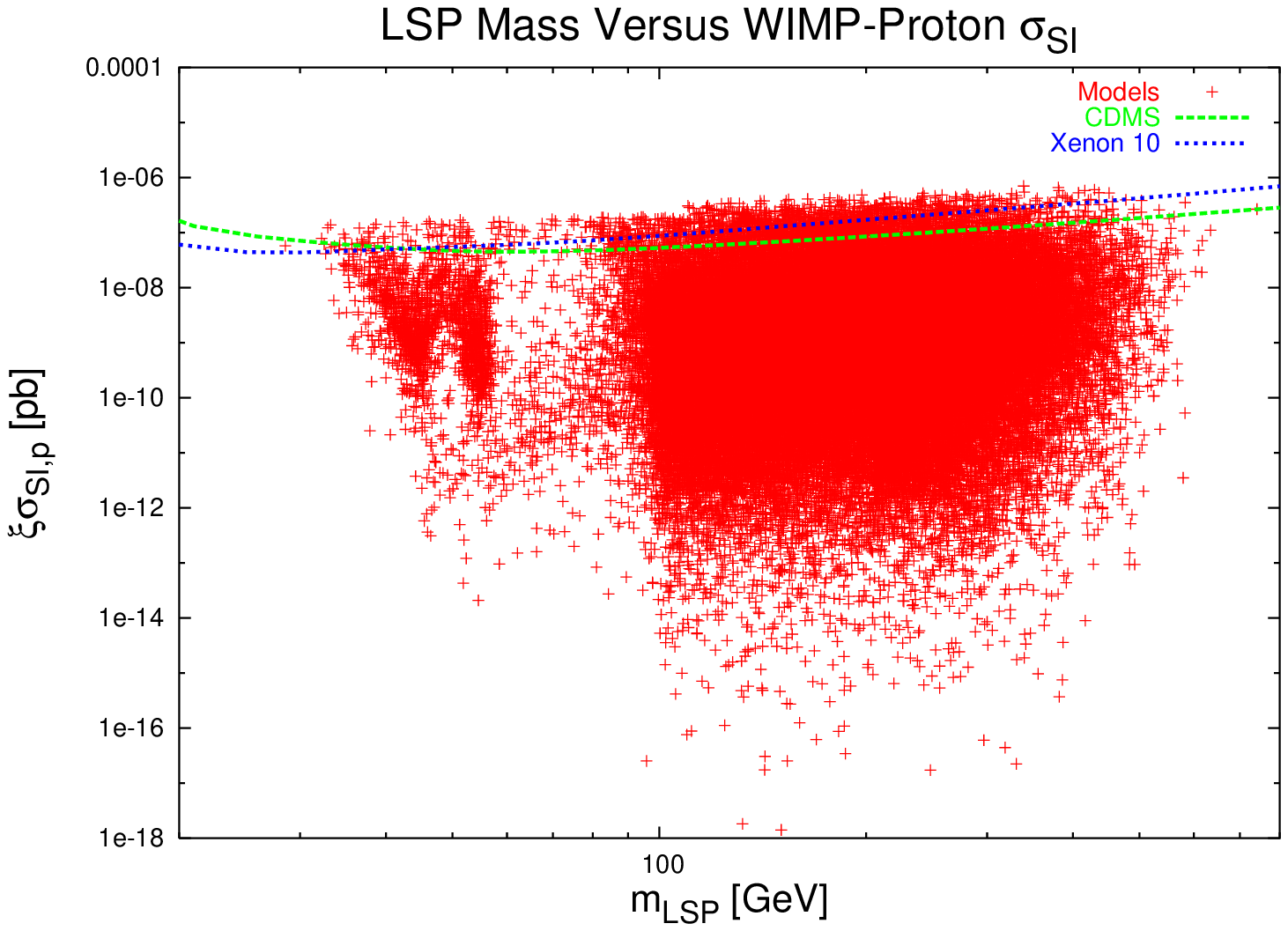}
\hspace*{0.1cm}
\includegraphics[width=6.0cm,angle=-90]{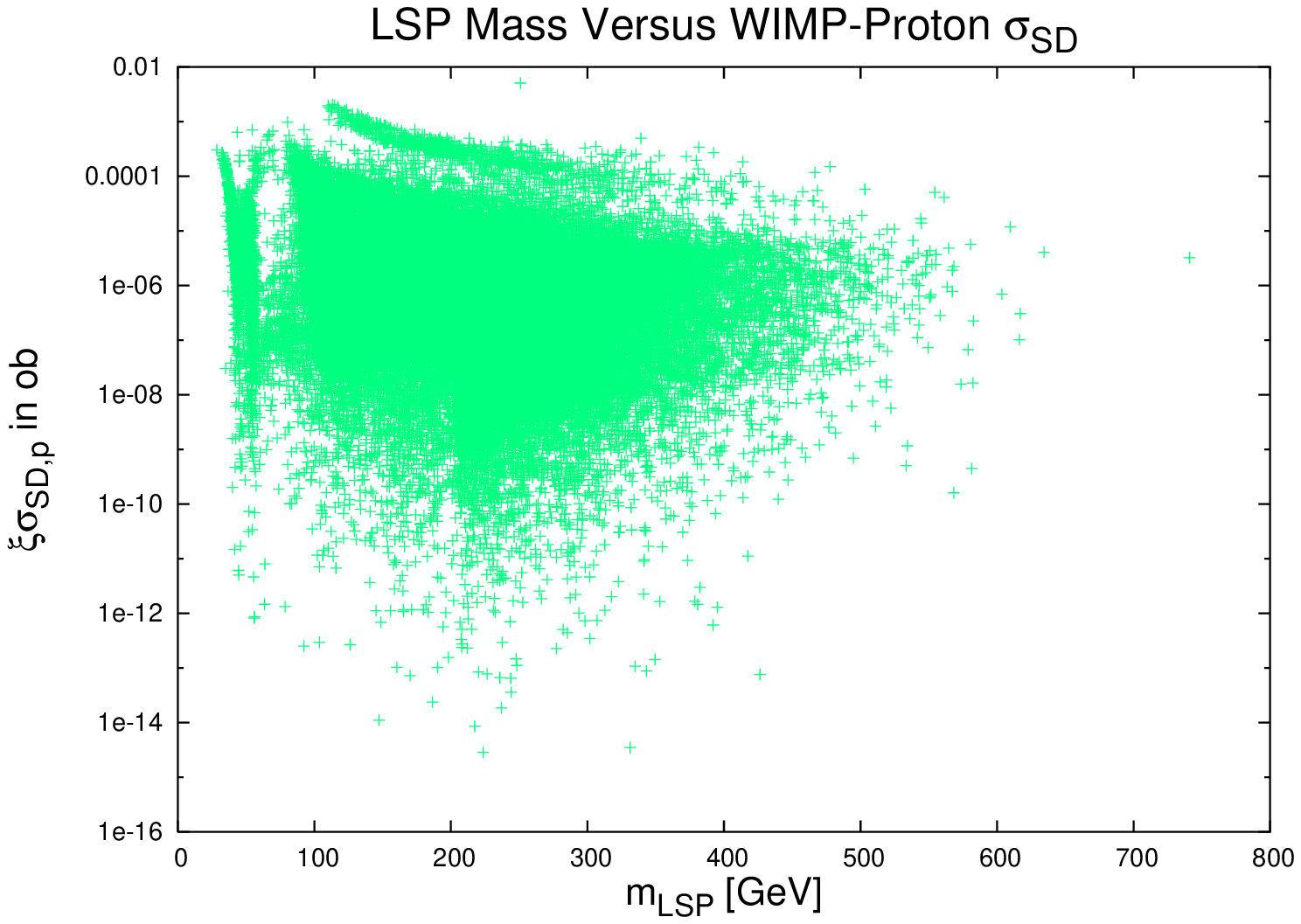}}
\centerline{
\includegraphics[width=6.0cm,angle=-90]{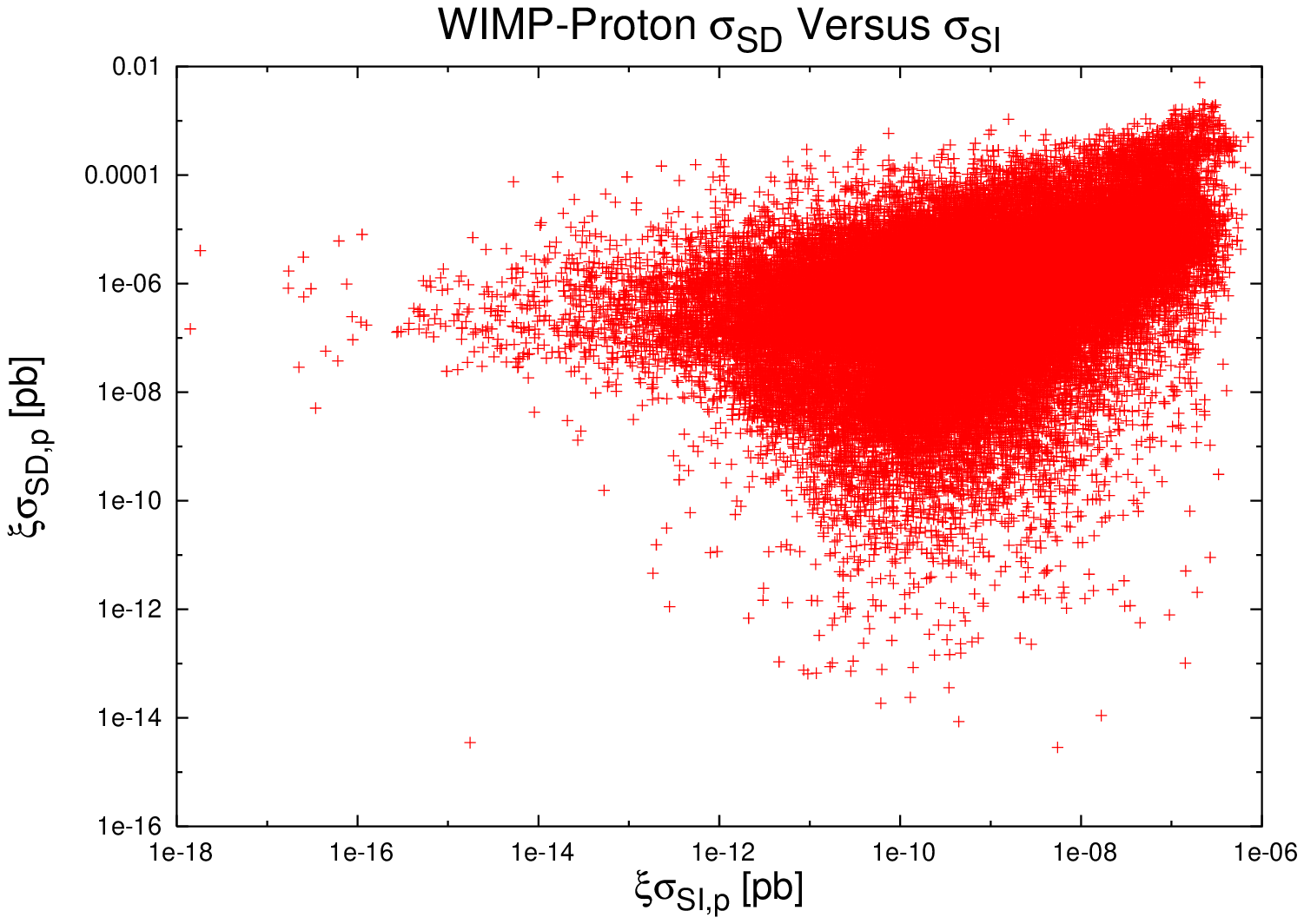}
\hspace*{0.1cm}
\includegraphics[width=6.0cm,angle=-90]{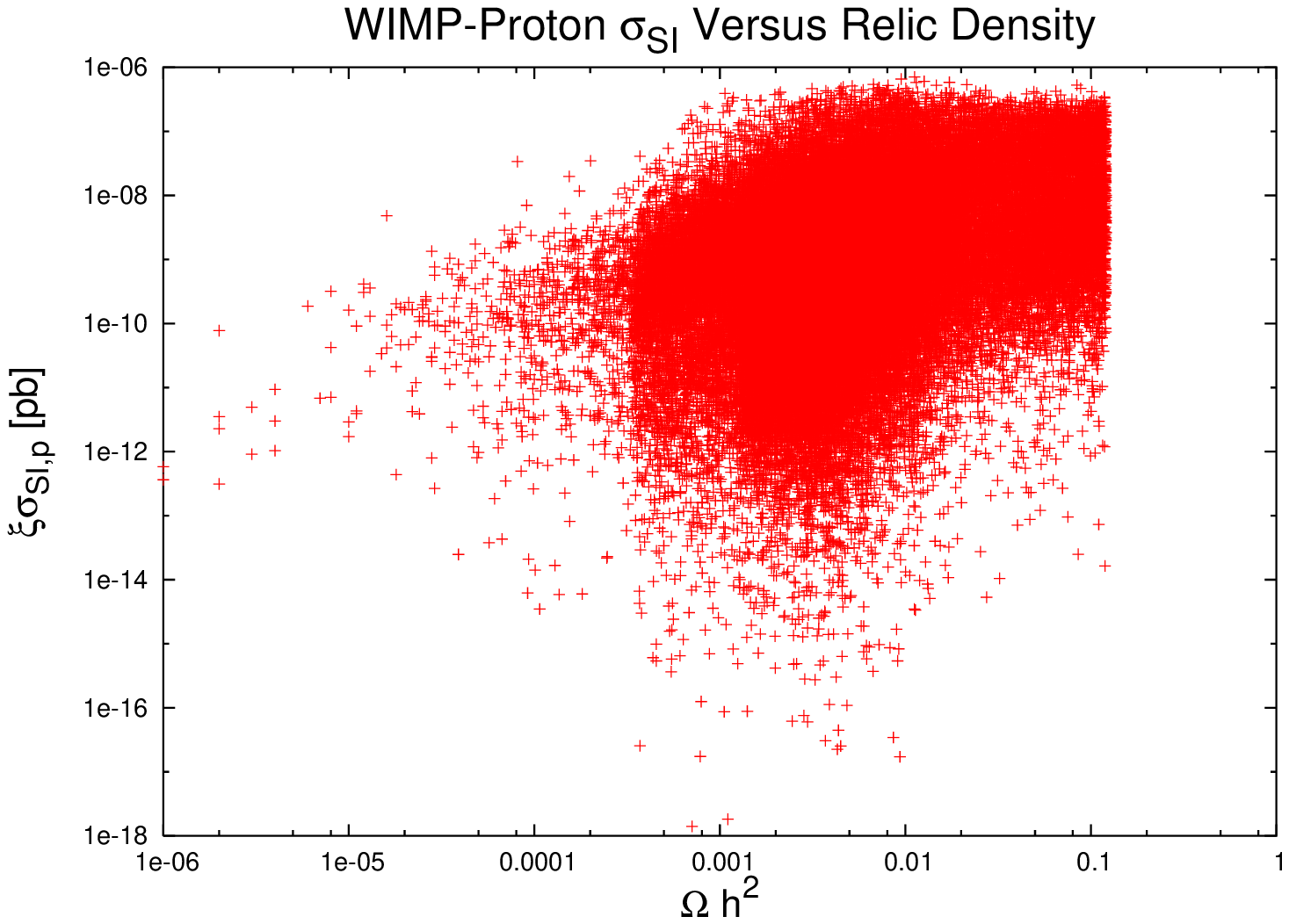}}
\caption{Distributions of relevant dark matter predictions from our model sample subject to the constraints discussed in the text. All 
cross sections are in pb.}
\label{fig10}
\end{figure}

We show a number of relevant predictions for dark matter in our  flat prior 
model sample Fig.~\ref{fig10}. In the top panels we see the prediction for $\Omega h^2$ as a 
function of both the LSP mass as well as the LSP-nLSP mass splitting, $\Delta m$. Note that the predicted range for $\Omega h^2$ is quite large but is found to peak 
near $\sim 0.004$, which is about 1/30 of the observed value obtained by WMAP as discussed above. 
These models then would require that there be a substantial amount of dark matter 
from other sources such as axions. The range of predictions presented here for the relic density is found to be much larger than those obtained by other analyses{\cite {big}} that 
were restricted to specific SUSY breaking scenarios. We also show the scaled spin-independent as well as spin-dependent LSP-proton scattering 
cross sections relevant for direct detection experiments. Note that again the predicted ranges that we find for these scattering cross sections are far 
larger than those based on mSUGRA, as obtained by, \eg, Ref.{\cite{Barger:2008qd}}.  In fact, we find that the range of values we obtain from our pMSSM model 
set essentially covers the entire theoretical anticipated region found in this reference for all different types of theories beyond the SM.  Thus, this observable can not 
be used as a model discriminator, \eg, to distinguish between SUSY versus Universal Extra Dimensions or Little Higgs thermal relics.  Clearly these results show that the range 
of expectations from the pMSSM is {\it far larger} than those predicted within any specific SUSY breaking scenario. 

\begin{figure}[htbp]
\centerline{
\includegraphics[width=9.0cm,angle=0]{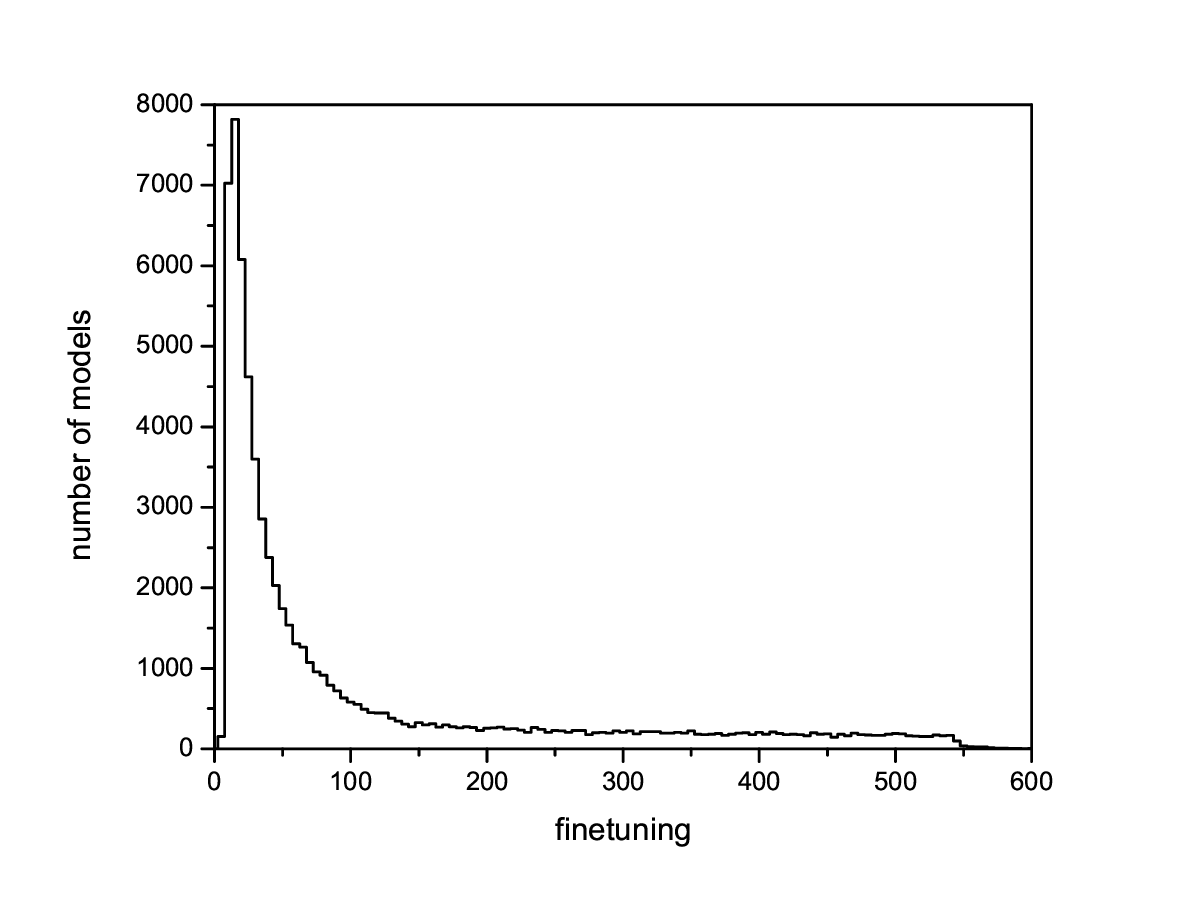}
\hspace*{-0.1cm}
\includegraphics[width=8.0cm,angle=0]{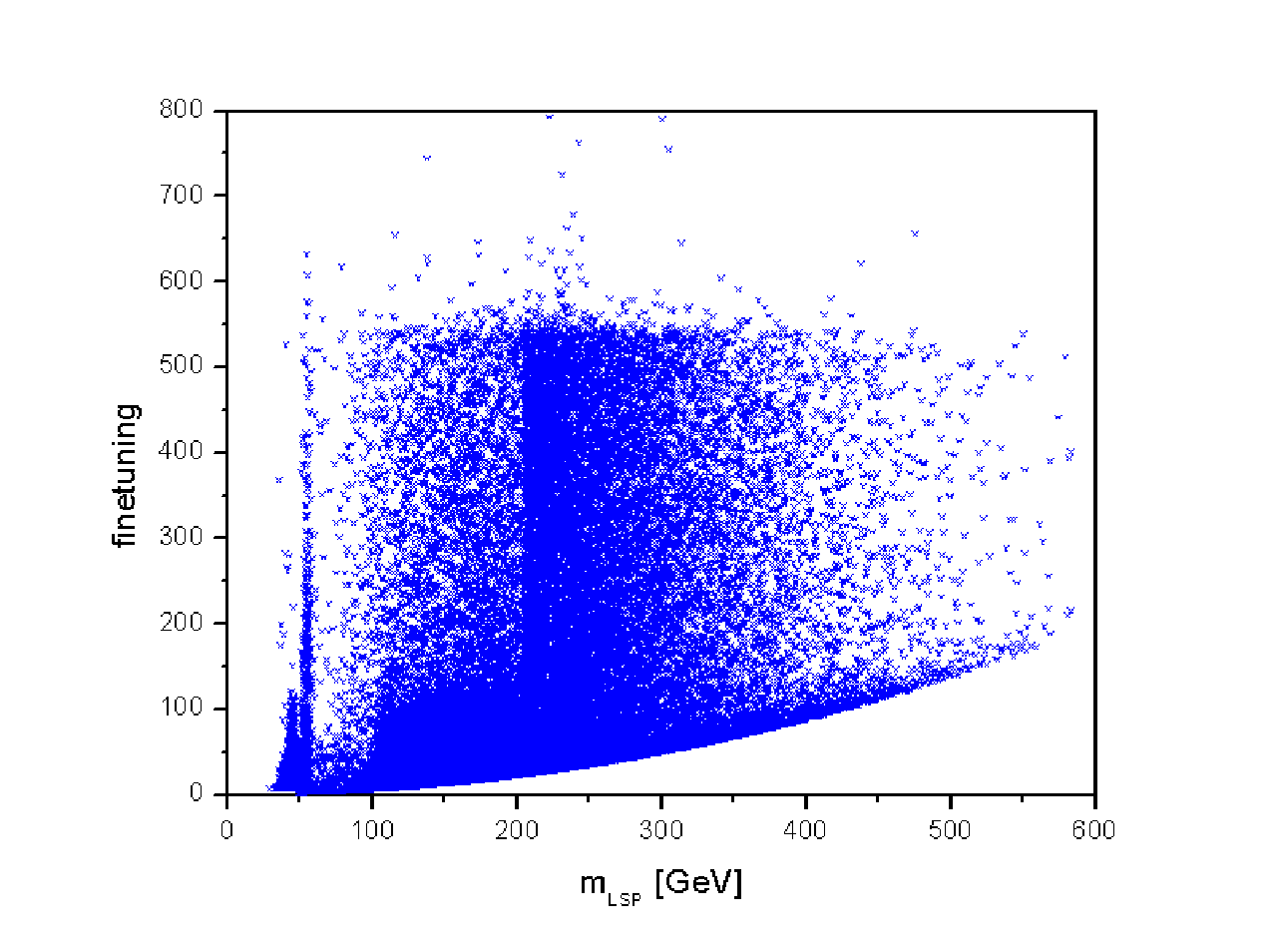}}
\vspace*{0.1cm}
\centerline{
\includegraphics[width=9.0cm,angle=0]{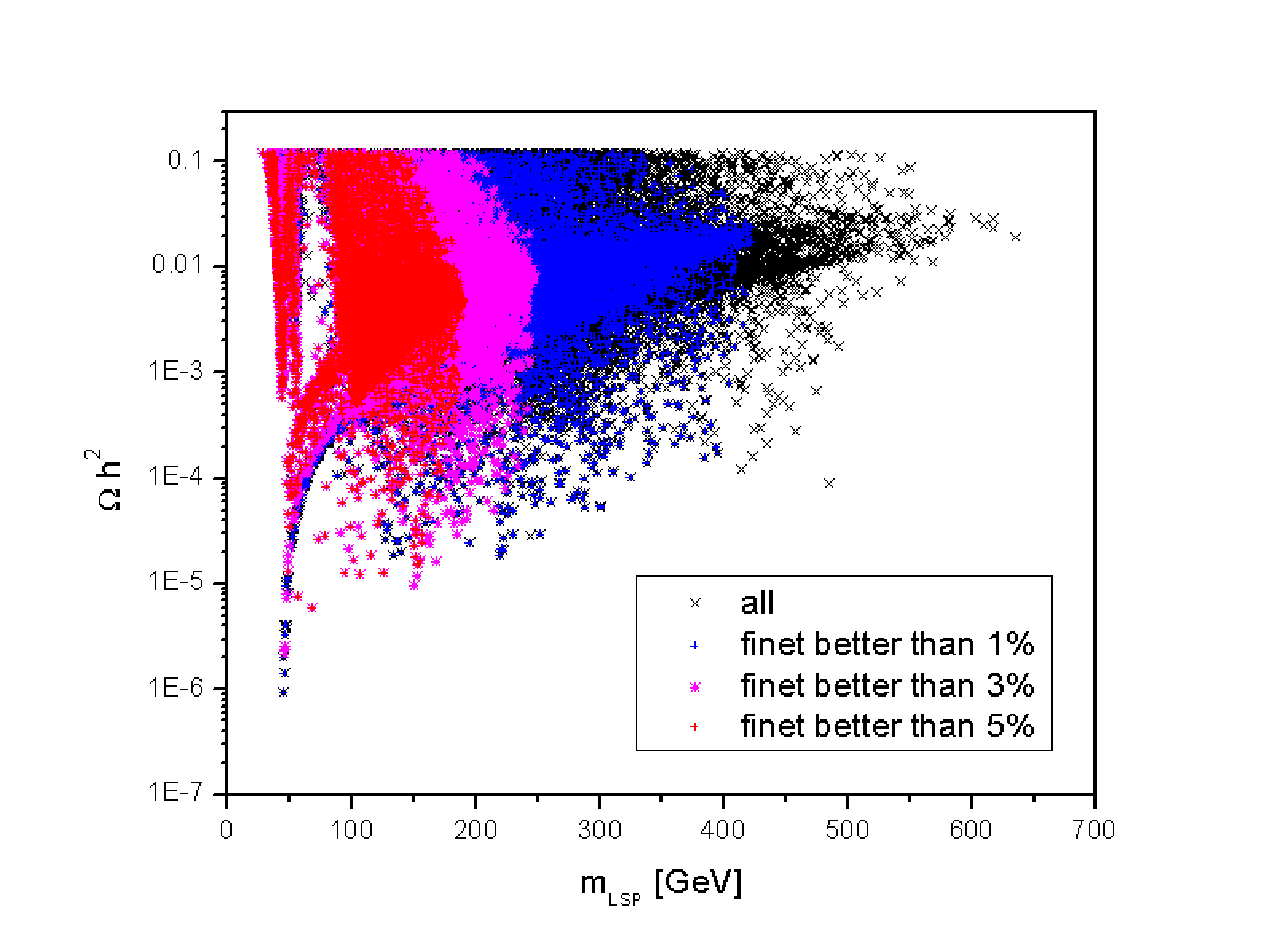}}
\caption{The distribution of fine-tuning ($=\Delta$) in our set of models and as a function of the LSP mass. Also shown is the correlation of the amount of fine-tuning 
with the predicted dark matter density.}
\label{fig11}
\end{figure}

The amount of fine-tuning in our model sample is described in Fig.~\ref{fig11}.  Here we see that  there is a clear set of pMSSM 
models with a rather reasonable degree of fine-tuning, unlike the current situation within  mSUGRA{\cite {big}}. Following the analysis presented in 
Ref. {\cite {finetuning}}, we define the quantities 
\begin{equation}
A(\xi)\,=\,\left| \frac{\partial\log
 m_Z^2}{\partial\log \xi}\right| \,
\end{equation}
where $\xi$ are the set of relevant Higgs Lagrangian parameters within the pMSSM. Then, for example 
\begin{eqnarray}
A(\mu) &=&
\frac{4\mu^2}{m_Z^2}\,\left(1+\frac{m_A^2+m_Z^2}{m_A^2} \tan^2 2\beta
\right), \nonumber \\
A(b) &=& \left( 1+\frac{m_A^2}{m_Z^2}\right)\tan^2
2\beta, \\
A(m_u^2) &=& \left| \frac{1}{2}\cos2\beta
+\frac{m_A^2}{m_Z^2}\cos^2\beta
-\frac{\mu^2}{m_Z^2}\right|\times\left(1-\frac{1}{\cos2\beta}+
\frac{m_A^2+m_Z^2}{m_A^2} \tan^2 2\beta \right), \nonumber \\
A(m_d^2) &=&
\left| -\frac{1}{2}\cos2\beta +\frac{m_A^2}{m_Z^2}\sin^2\beta
-\frac{\mu^2}{m_Z^2}\right|\times\left|1+\frac{1}{\cos2\beta}+
\frac{m_A^2+m_Z^2}{m_A^2} \tan^2 2\beta \right|, \nonumber 
\end{eqnarray}
where it is has been assumed that $\tan\beta>1$. Here, $m_{u,d}^2$ are the effective mass parameters for the two Higgs doublets generating masses for the $u,d$-type quarks, 
respectively, and $b$ is the parameter in the Higgs potential that can be rewritten in terms of $\tan \beta$ and $m_A$. The overall fine-tuning $\Delta$ is defined by 
adding these four quantities in quadrature. Values of $\Delta$ far above unity then indicate large amounts of fine-tuning. In this figure we 
see that a reasonable set of our surviving models can have very low values of finetuning. This figure also shows that the overall amount of fine-tuning indeed peaks at 
rather low values in our set of models, far smaller than that obtained in the case of mSUGRA. Note that the fine-tuning percentage given in the bottom panel of this figure is simply the 
value of $1/\Delta$ expressed as a percent.

\subsection{Log Priors}

In the case where we employed log priors, a far smaller number (and fraction) of models survive the confrontation with the experimental constraints.  In this case, we randomly sampled  $2 \cdot 10^6$ models and found that only $\sim 3000$ models remained after imposing the theoretical and experimental constraints.  The main reason for this is rather 
straightforward to understand. In the log prior scan, although the range of the SUSY masses has been expanded, most of the generated parameter space points correspond 
to models with very light SUSY particles which are more highly constrained by existing experimental data and, consequently, far fewer of the models survive.  Recall that we employed a smaller model sample in this case as our primary goal in undertaking the log prior study is to 
examine any differences with the flat prior analysis and consequently our presentation of the corresponding 
results will be somewhat brief.  

\begin{figure}[htbp]
\centerline{
\includegraphics[width=9.0cm,angle=0]{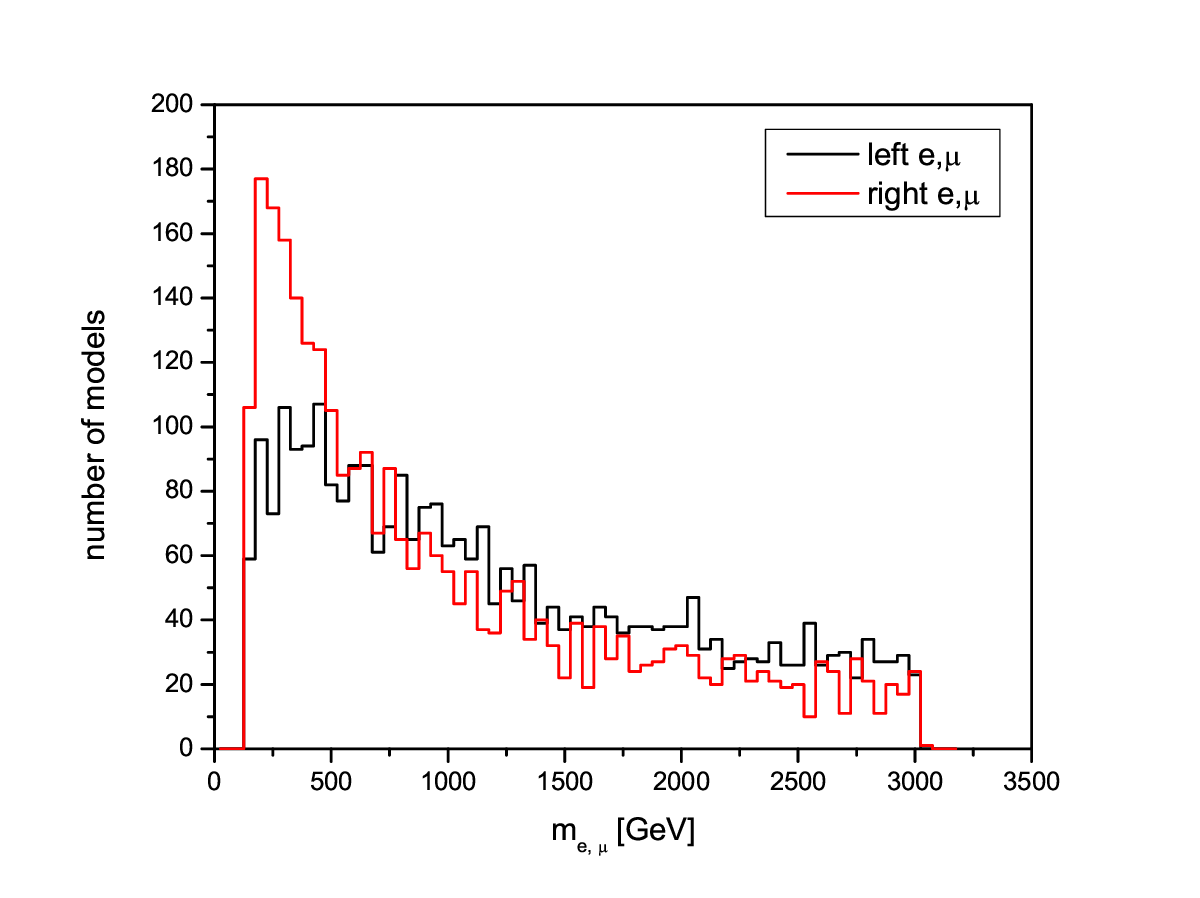}
\hspace*{0.1cm}
\includegraphics[width=9.0cm,angle=0]{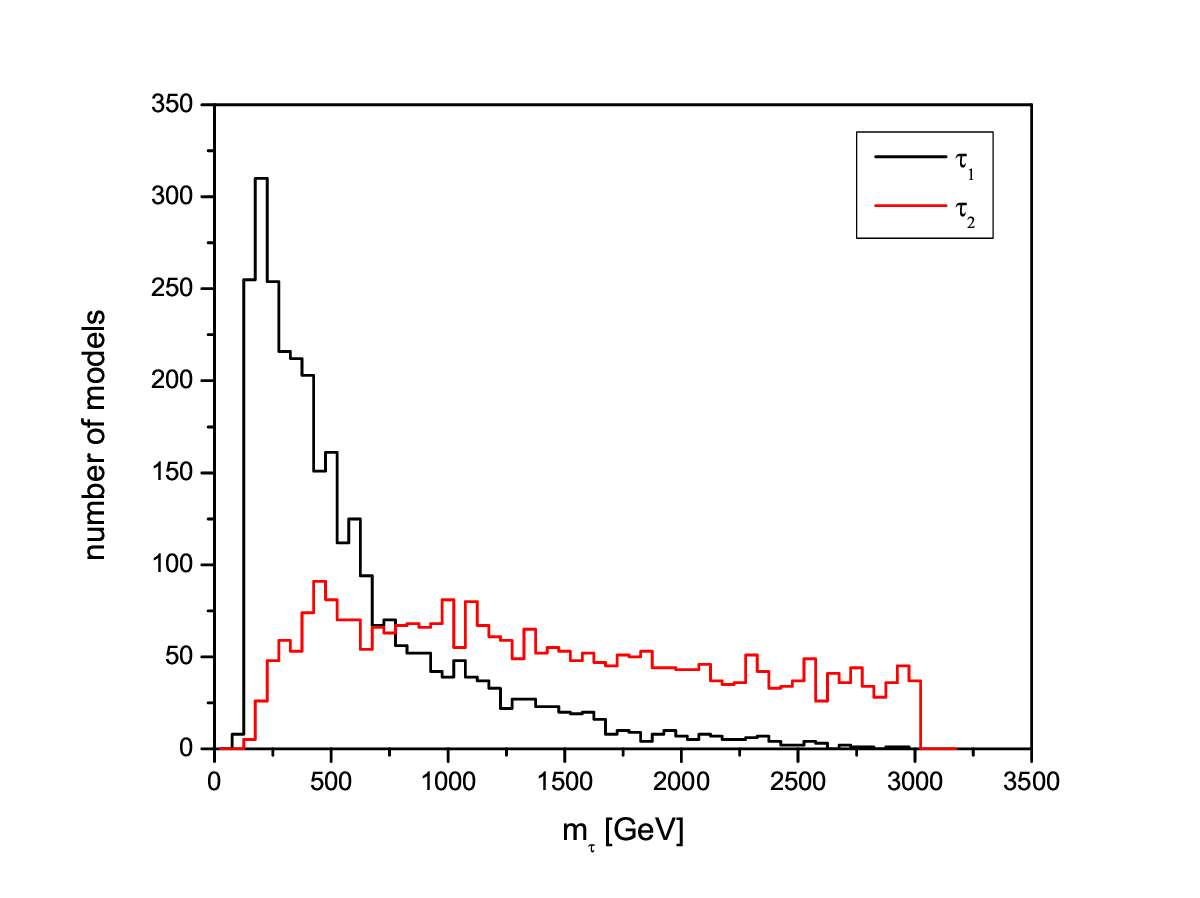}}
\vspace*{0.1cm}
\centerline{
\includegraphics[width=9.0cm,angle=0]{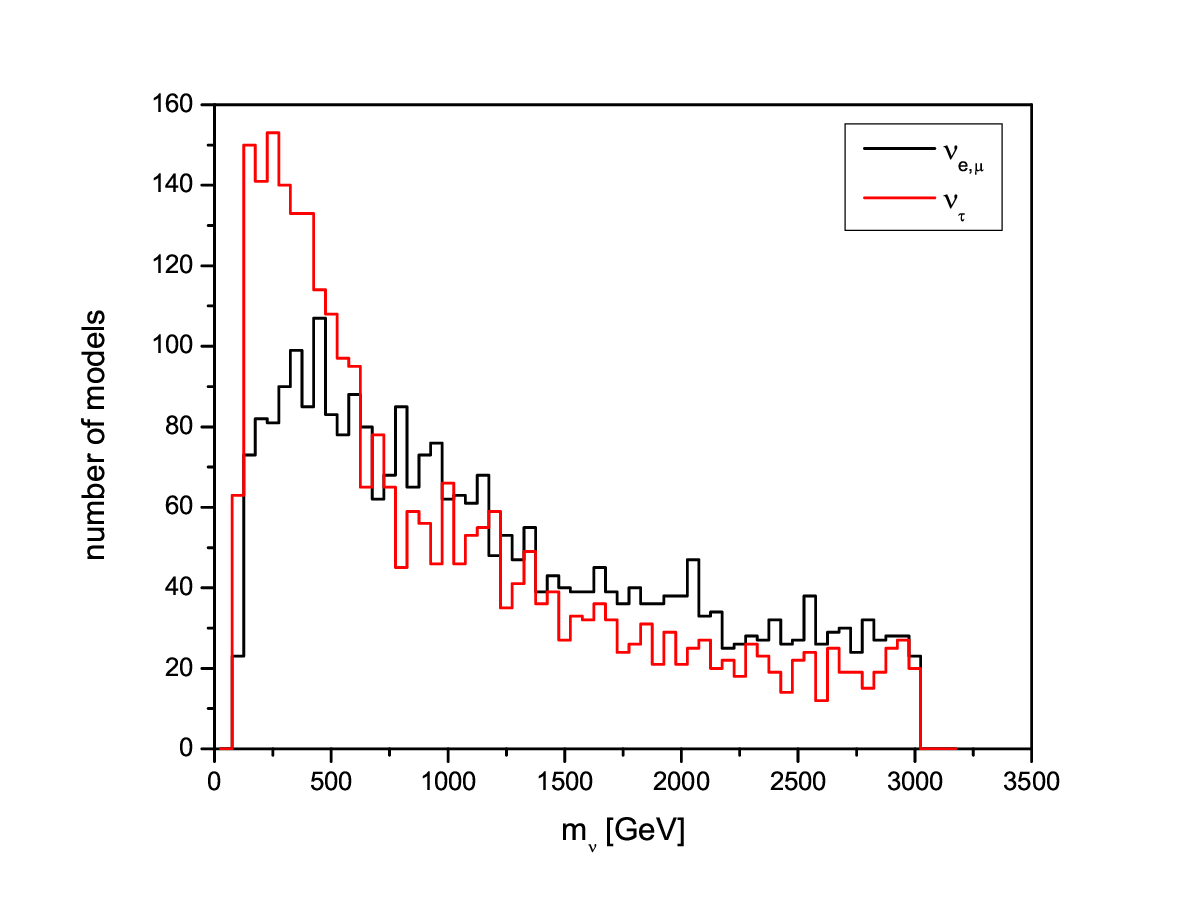}}
\caption{Distribution of slepton masses for the set of log prior models satisfying all of our constraints: selectrons/smuons(top left), staus(top right) and 
sneutrinos(bottom).}
\label{fig12}
\end{figure}

\begin{figure}[htbp]
\centerline{
\includegraphics[width=9.0cm,angle=0]{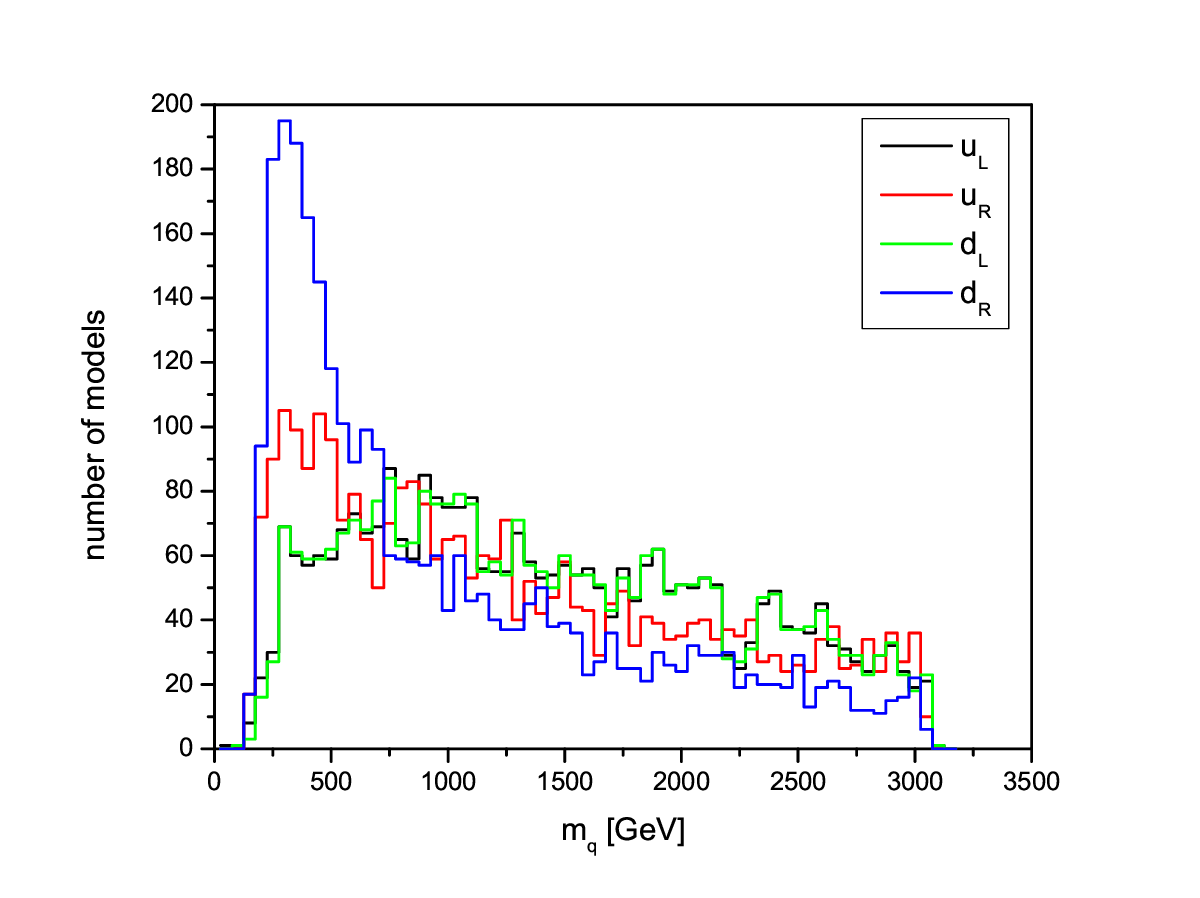}
\hspace*{0.1cm}
\includegraphics[width=9.0cm,angle=0]{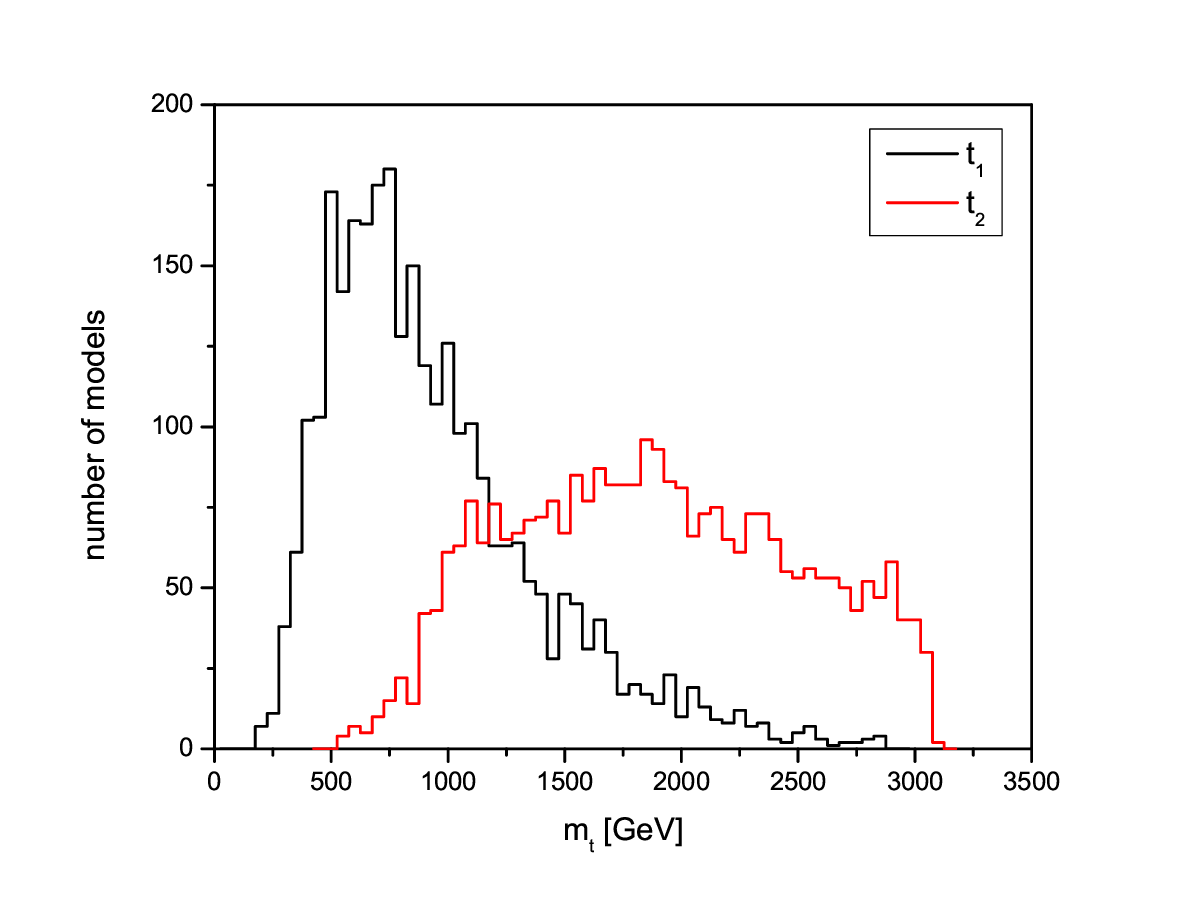}}
\vspace*{0.1cm}
\centerline{
\includegraphics[width=9.0cm,angle=0]{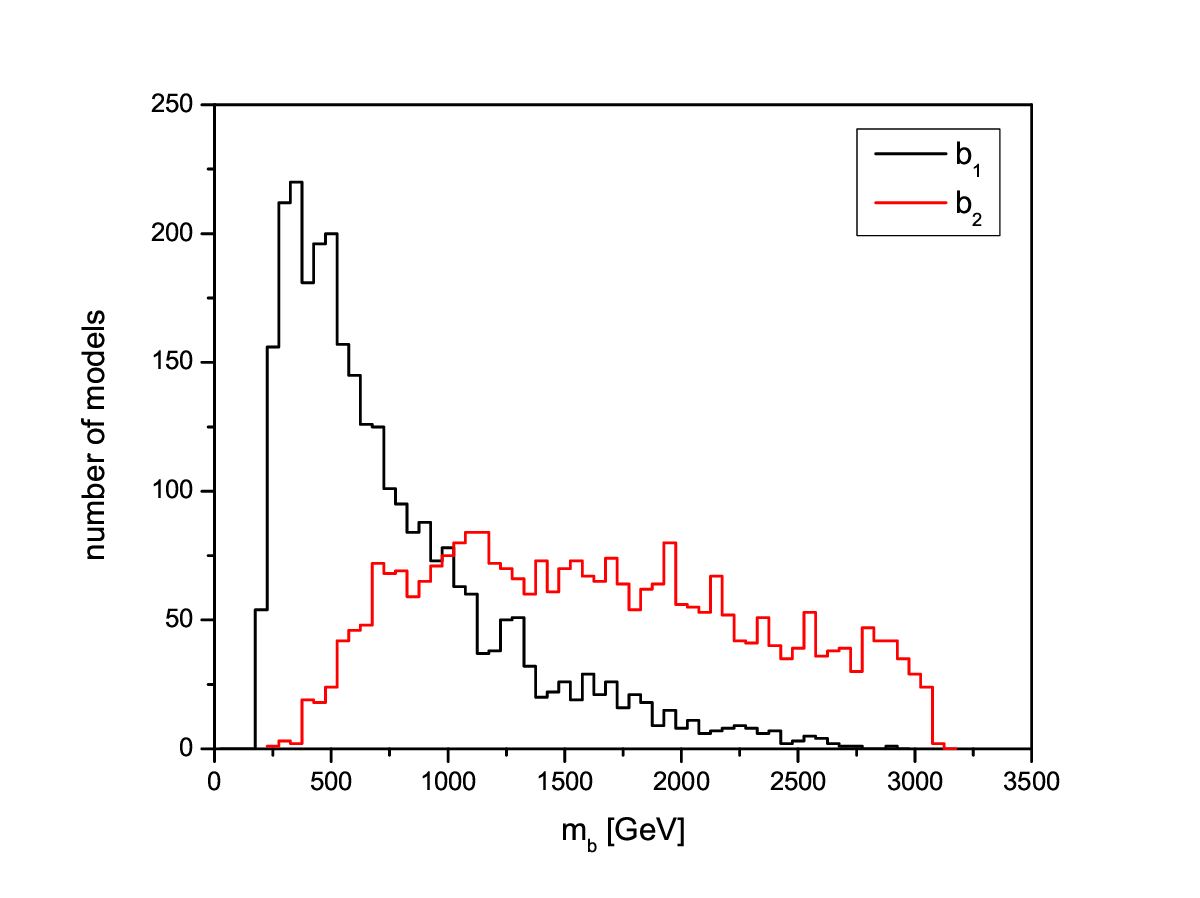}}
\caption{Distribution of squark masses for the set of log prior models satisfying all of our constraints: `light' (first/second generation) squarks are in the upper left panel, stops in 
the upper right panel and sbottoms are in the lower panel.}
\label{fig13}
\end{figure}

\begin{figure}[htbp]
\centerline{
\includegraphics[width=13.0cm,angle=0]{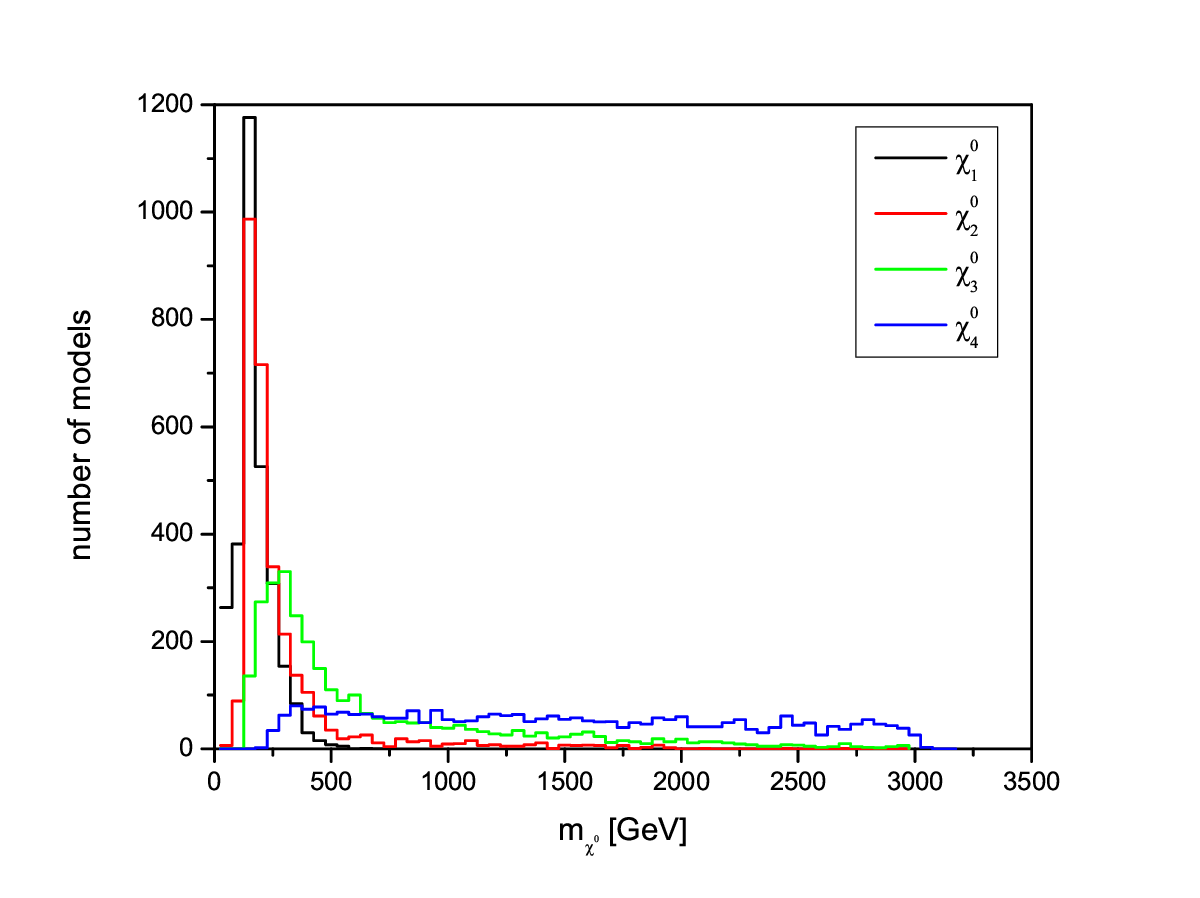}}
\vspace*{0.1cm}
\centerline{
\includegraphics[width=13.0cm,angle=0]{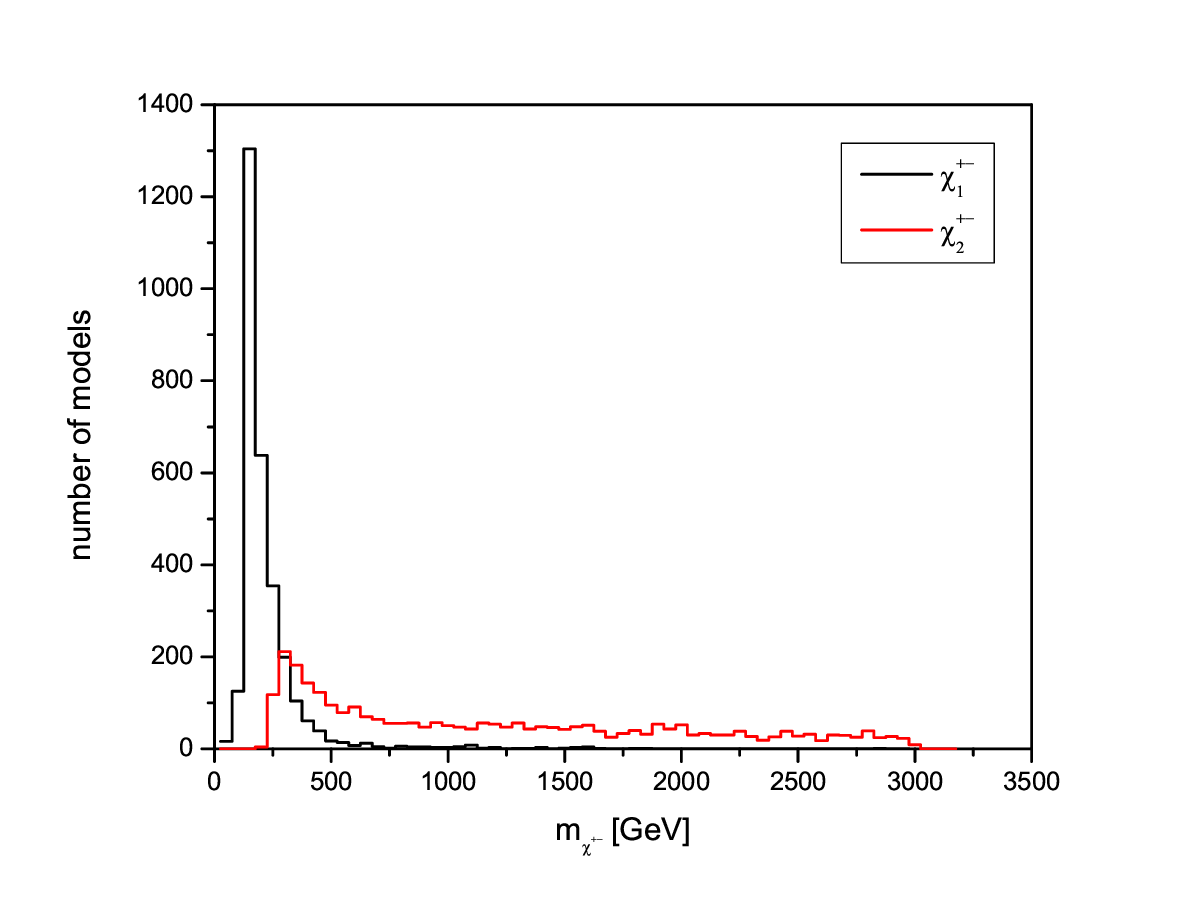}}
\caption{Distribution of neutralino and chargino masses for the set of log prior models satisfying all of our constraints.}
\label{fig14}
\end{figure}

\begin{figure}[htbp]
\centerline{
\includegraphics[width=13.0cm,angle=0]{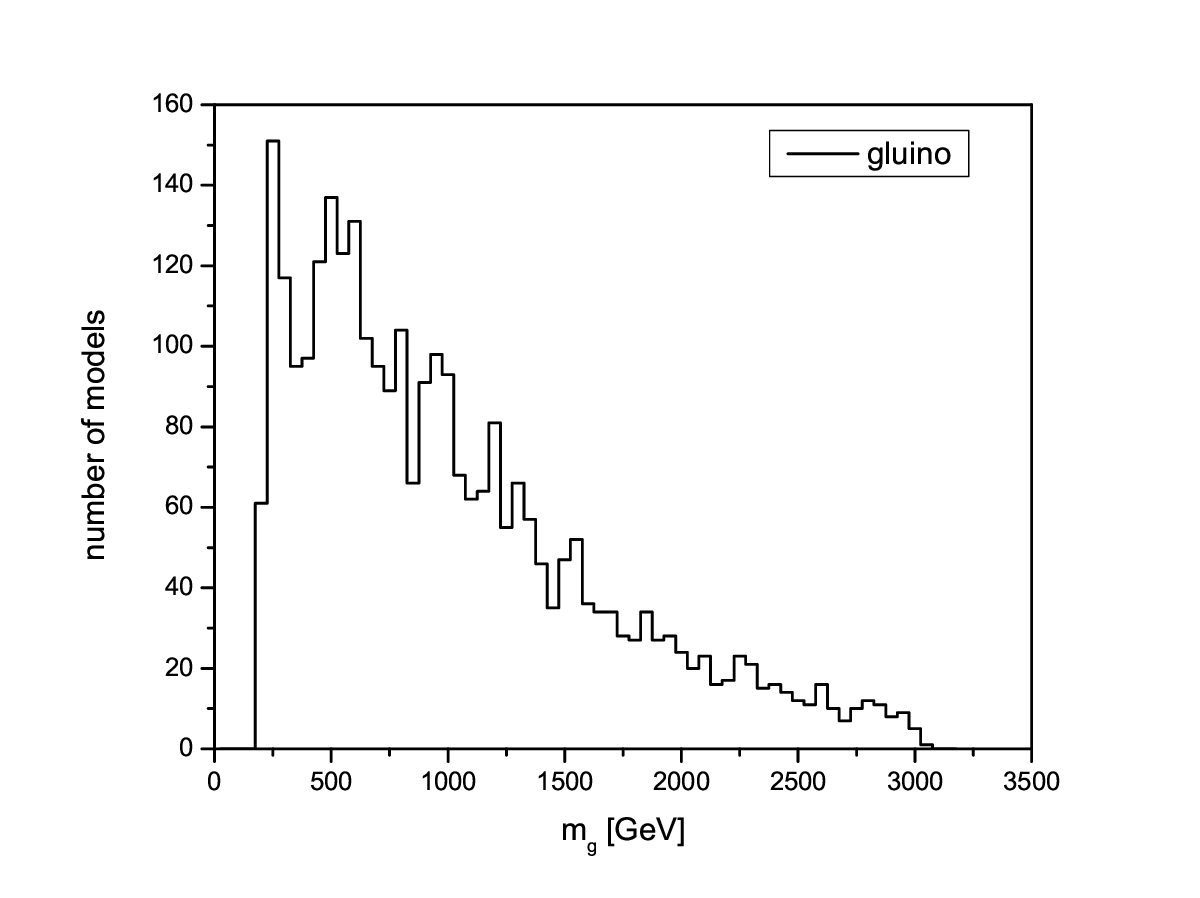}}
\vspace*{0.2cm}
\centerline{
\includegraphics[width=13.0cm,angle=0]{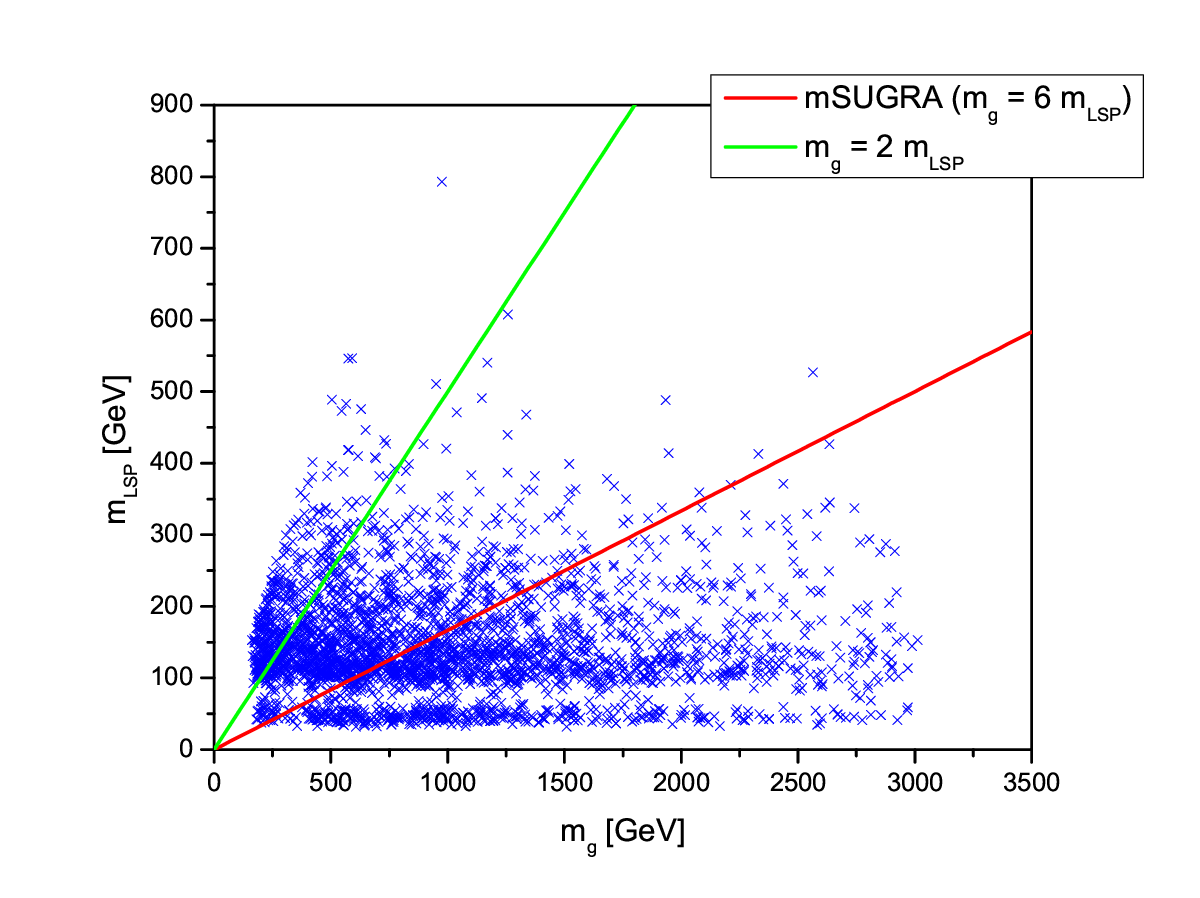}}
\caption{Distribution of gluino masses and a comparison of the gluino and LSP masses for the set of log prior models satisfying all of our constraints.}
\label{fig15}
\end{figure}

\begin{figure}[htbp]
\centerline{
\includegraphics[width=13.0cm,angle=0]{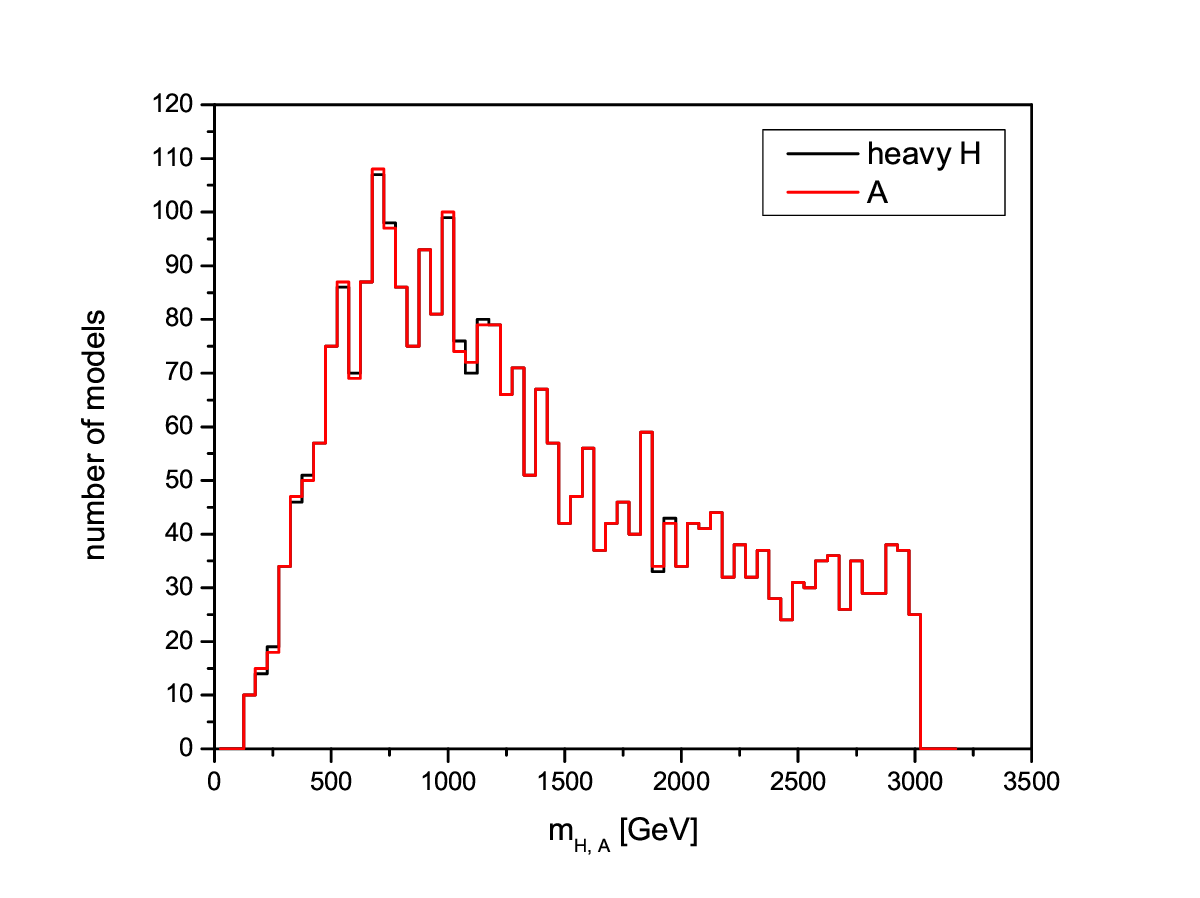}}
\vspace*{0.1cm}
\centerline{
\includegraphics[width=13.0cm,angle=0]{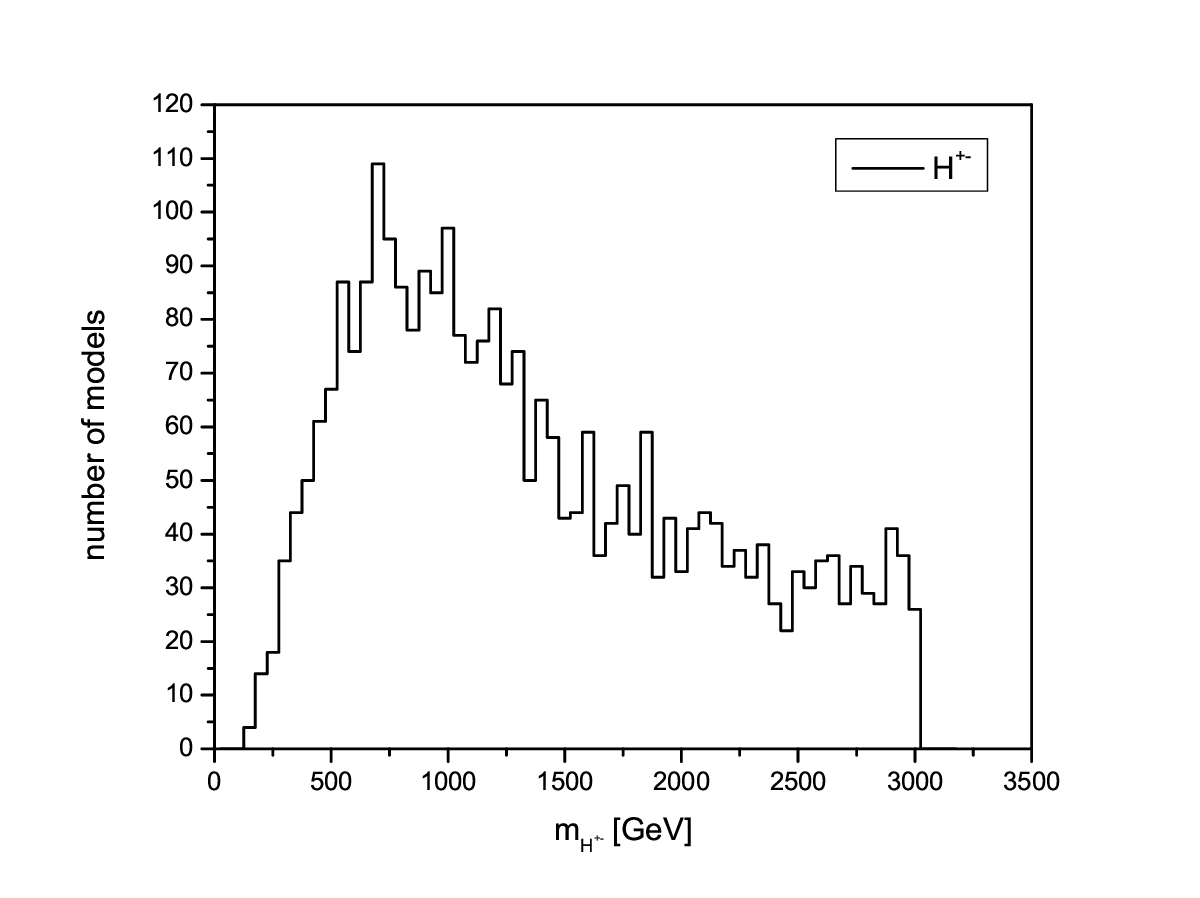}}
\caption{Distribution of heavy and charged Higgs masses for the set of log prior models satisfying all of our constraints.}
\label{fig16}
\end{figure}

\begin{figure}[htbp]
\centerline{
\includegraphics[width=13.0cm,angle=0]{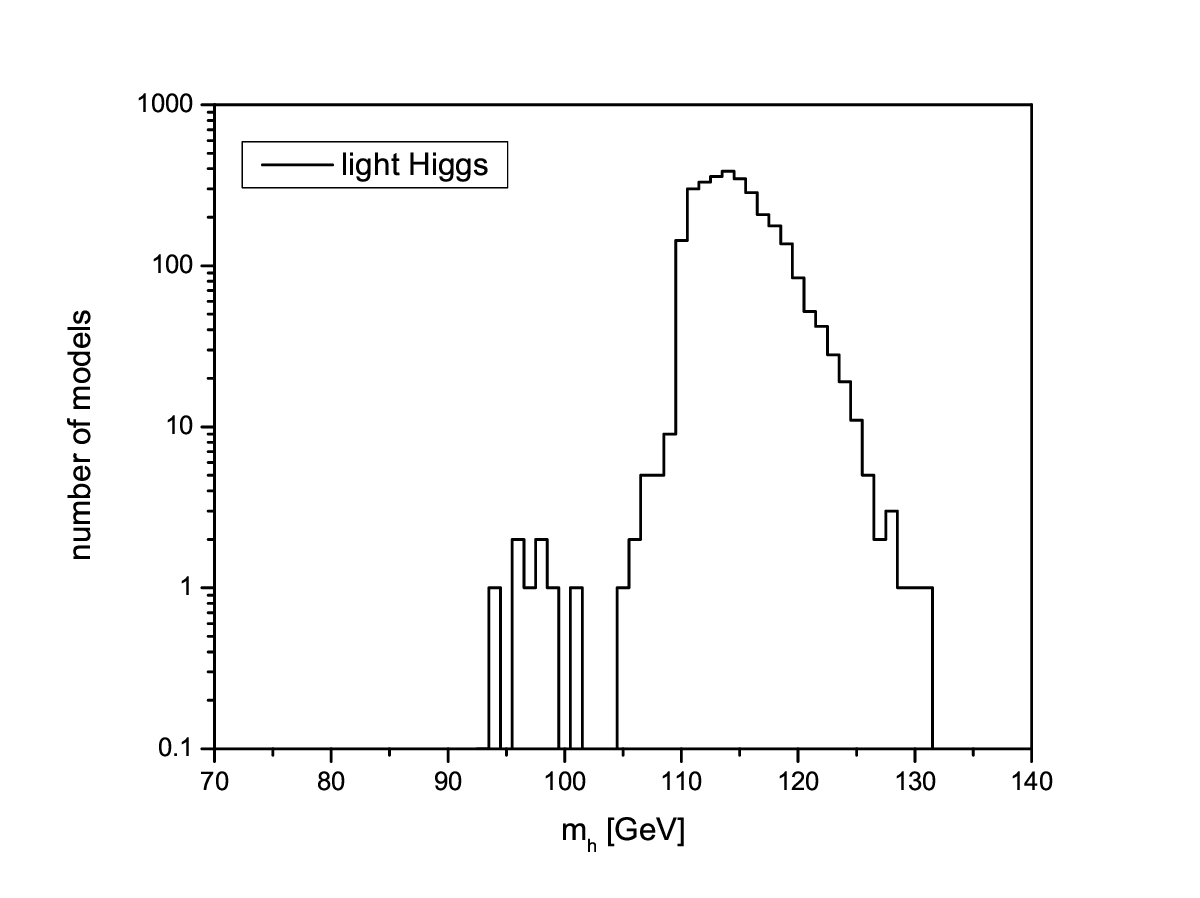}}
\caption{Distribution of the light Higgs mass on a log scale for the set of log prior models satisfying all of our constraints.}
\label{fig17}
\end{figure}

We begin our comparison with the flat prior results by examining the various Higgs and sparticle spectra.  
Figures~\ref{fig12}~--~\ref{fig17} show the set of spectrum results  for the set of log prior models which survive all of the constraints.  These should be 
compared to Figs.~\ref{fig1}-~\ref{fig6} above in the case of flat priors. As can be seen 
these spectra are qualitatively very similar to what we have seen in the case of flat priors above, except for the obvious two effects already mentioned: the 
ranges of the allowed mass values have been extended so that larger masses are reached and the sample size is significantly smaller.  However, the broad features discussed above for the flat prior sample also hold true for the log prior sample.
The main feature that we observed 
in both log and flat prior samples is the apparent preference for relatively light spartners with this 
preference being somewhat stronger in the log prior case as would be expected. While the low end of the log prior distributions are similar to those in the flat 
prior analysis, naturally, the log prior spectra have longish tails which stretch out to significantly 
larger mass values.  Larger $A$-terms, for example, can lead to greater stop mixing effects which then produce larger values for the light Higgs mass; this also 
leads to a slight increase in the maximum value of the Higgs mass in comparison to the flat prior case, as is observed in Fig.~\ref{fig17}.

\begin{figure}[htbp]
\centerline{
\includegraphics[width=13.0cm,angle=0]{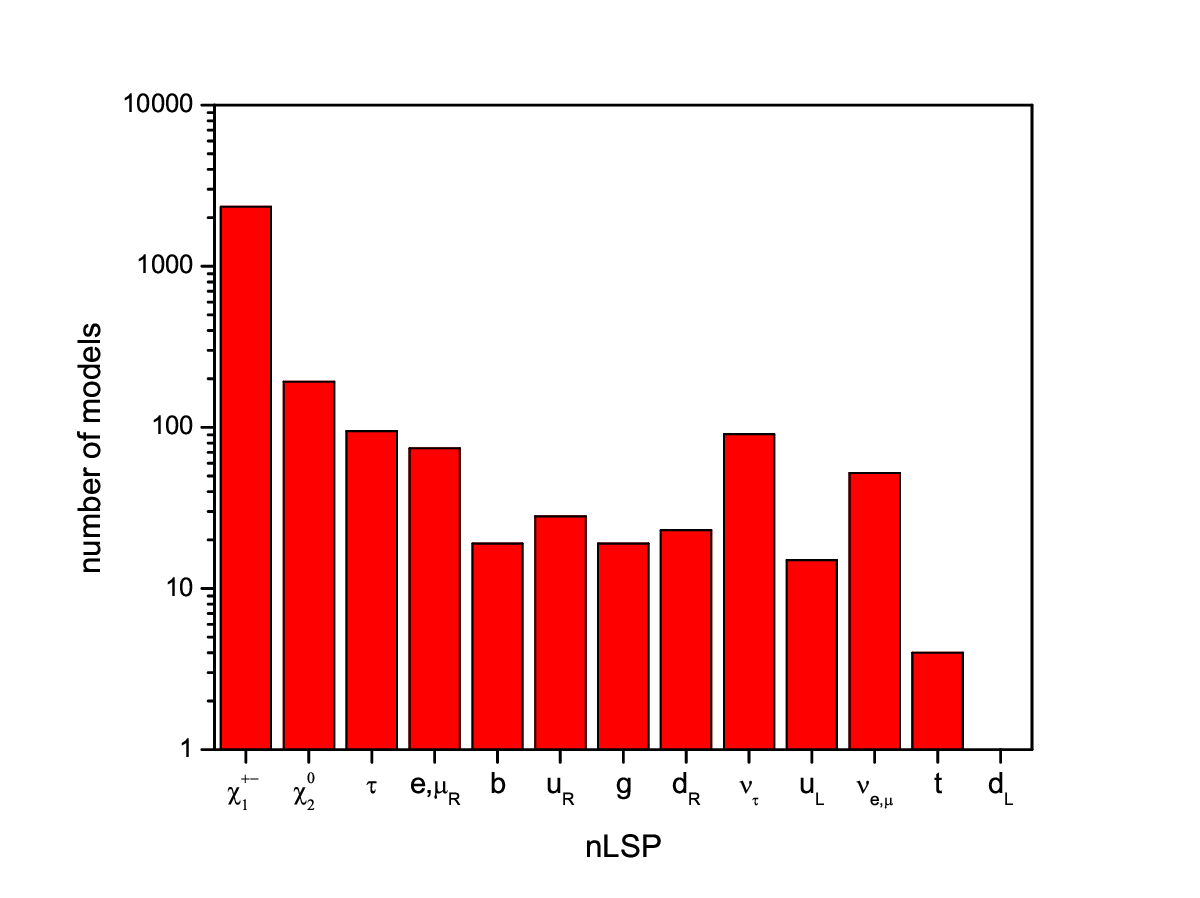}}
\vspace*{0.2cm}
\centerline{
\includegraphics[width=15.0cm,angle=0]{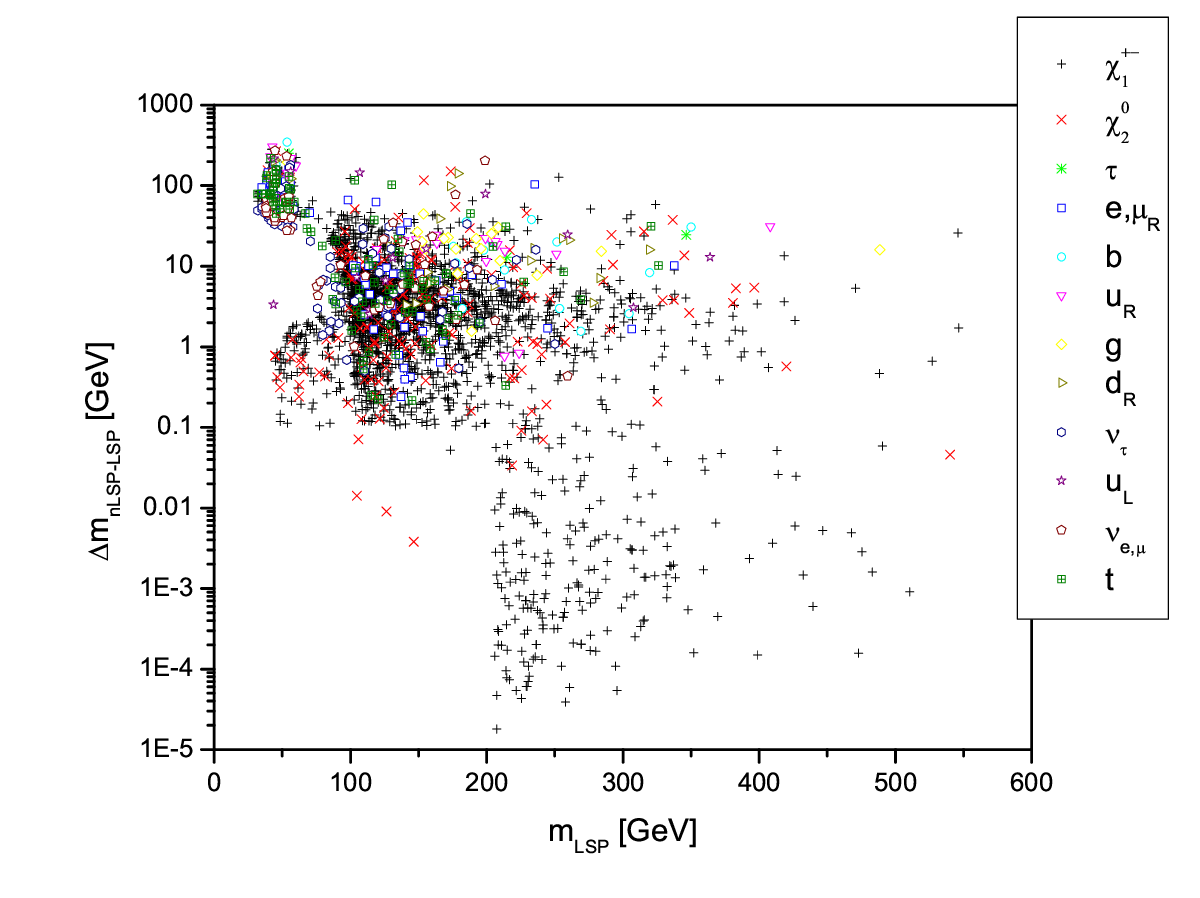}}
\caption{(Top) Identity of the nLSP. (Bottom) nLSP-LSP mass splitting as a function of the LSP mass. Both apply in the case of log priors.}
\label{fig18}
\end{figure}
\begin{figure}[htbp]
\centerline{
\includegraphics[width=13.0cm,angle=0]{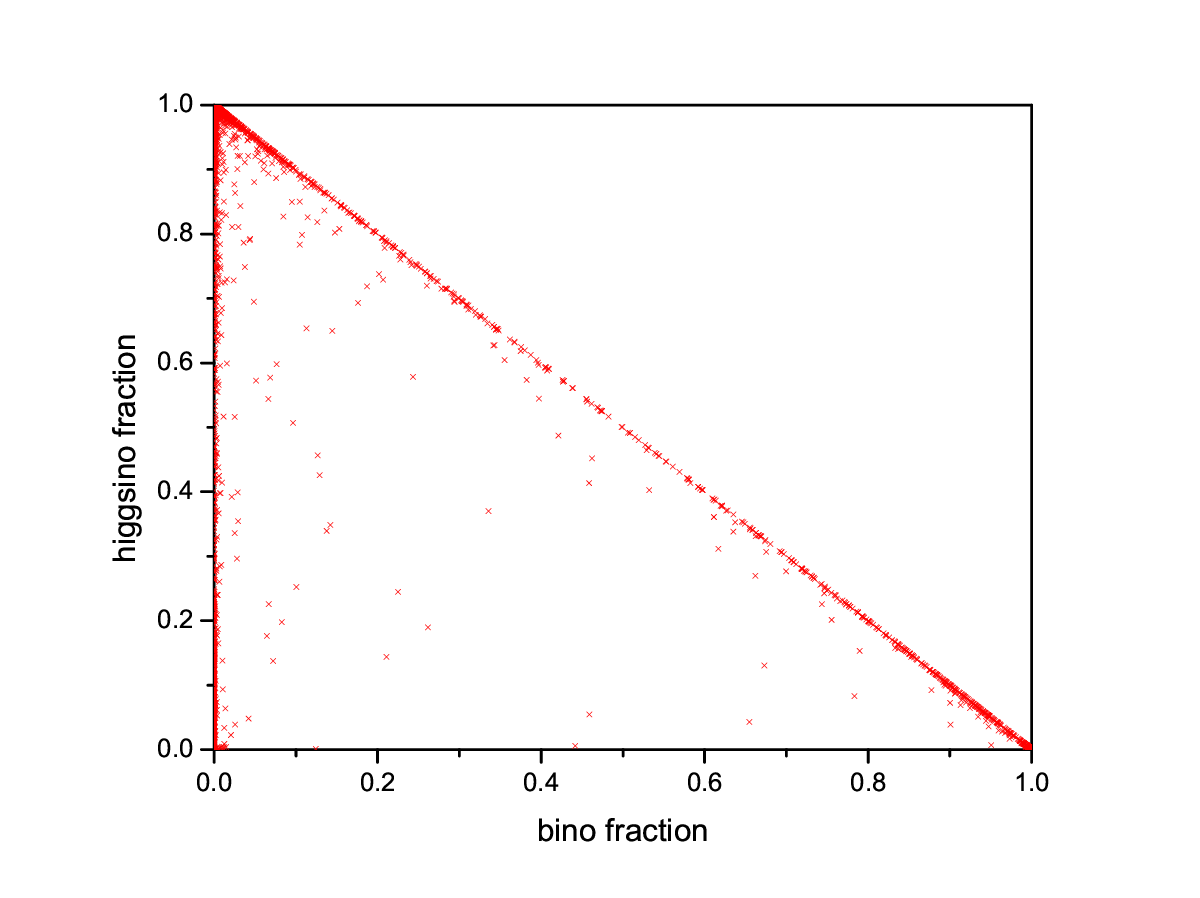}}
\vspace*{0.1cm}
\centerline{
\includegraphics[width=13.0cm,angle=0]{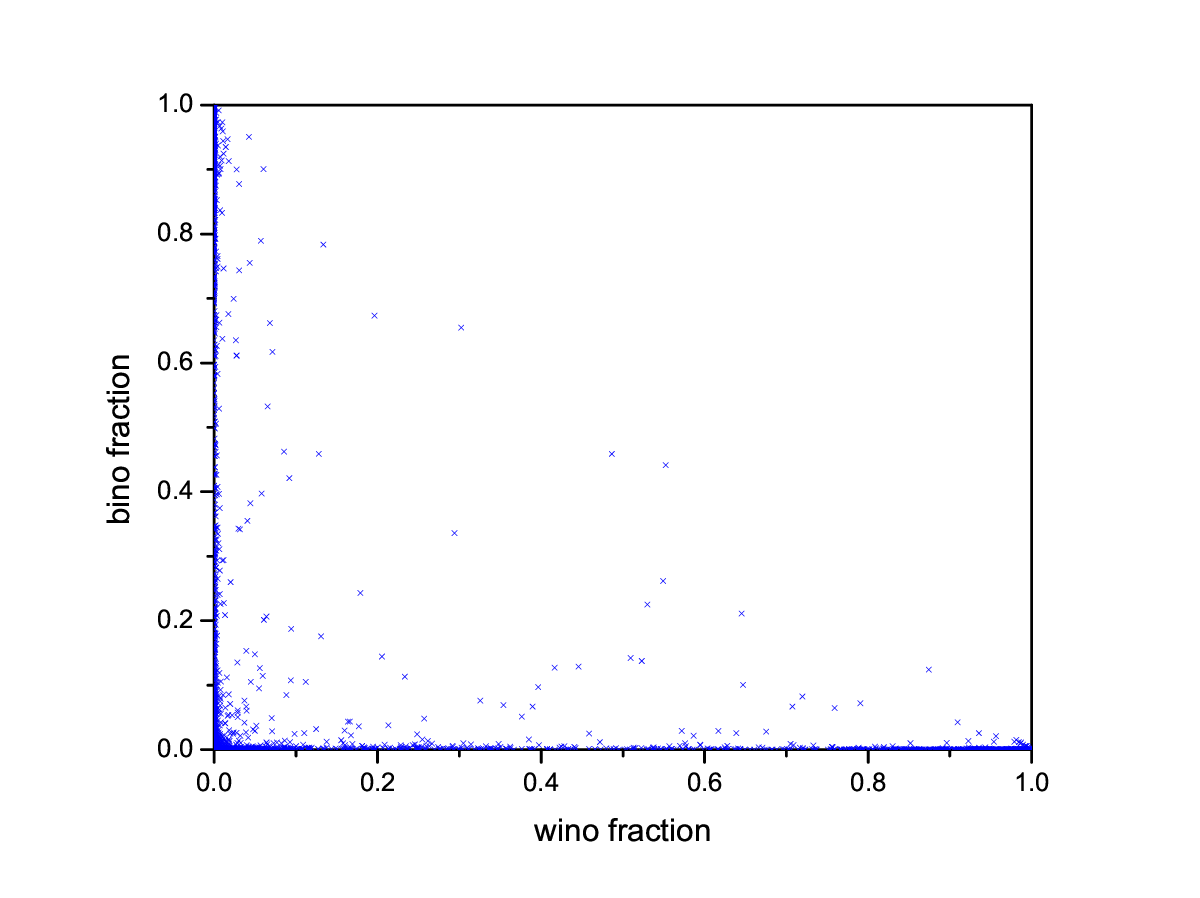}}
\caption{Wino/Higgsino/Bino content of the LSP in the case of log priors.}
\label{fig19}
\end{figure}
\begin{figure}[htbp]
\centerline{
\includegraphics[width=13.0cm,angle=0]{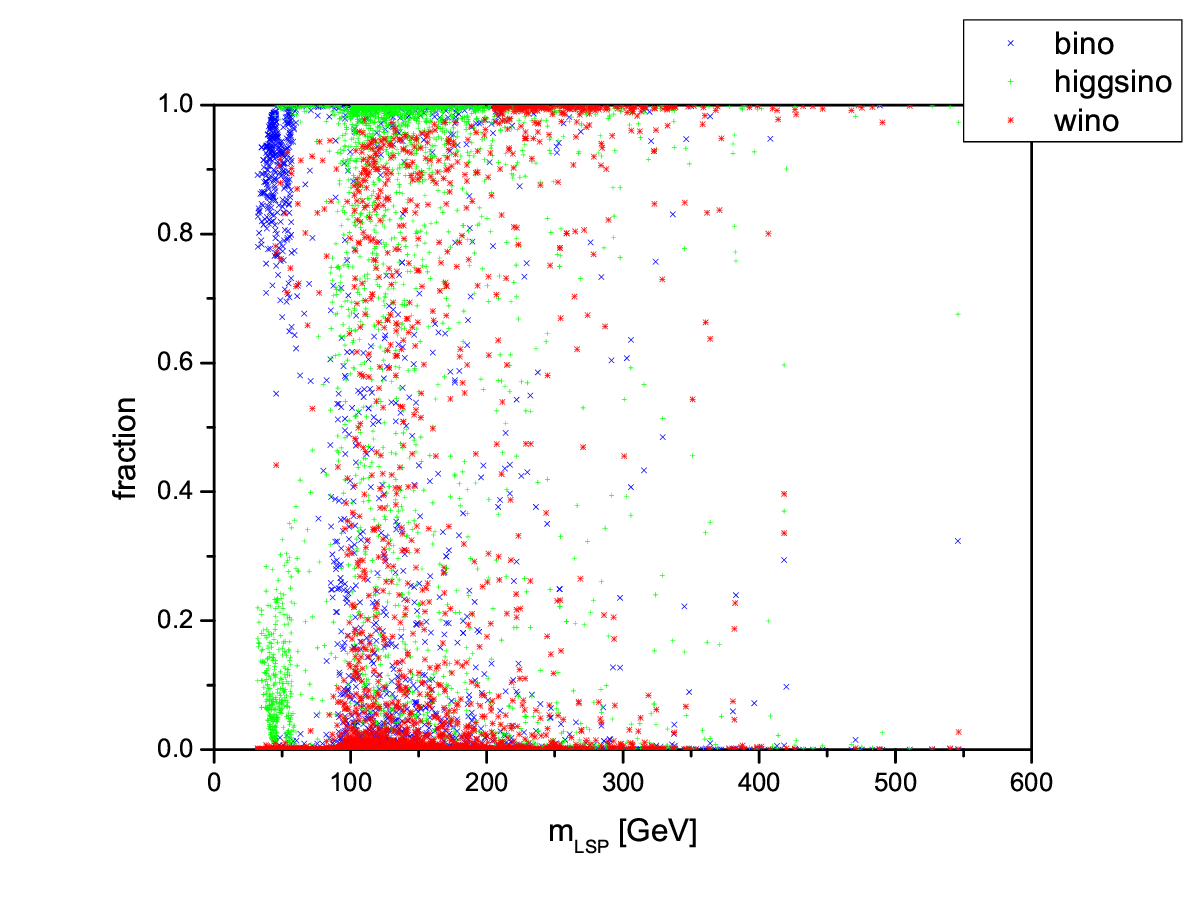}}
\vspace*{0.1cm}
\centerline{
\includegraphics[width=13.0cm,angle=0]{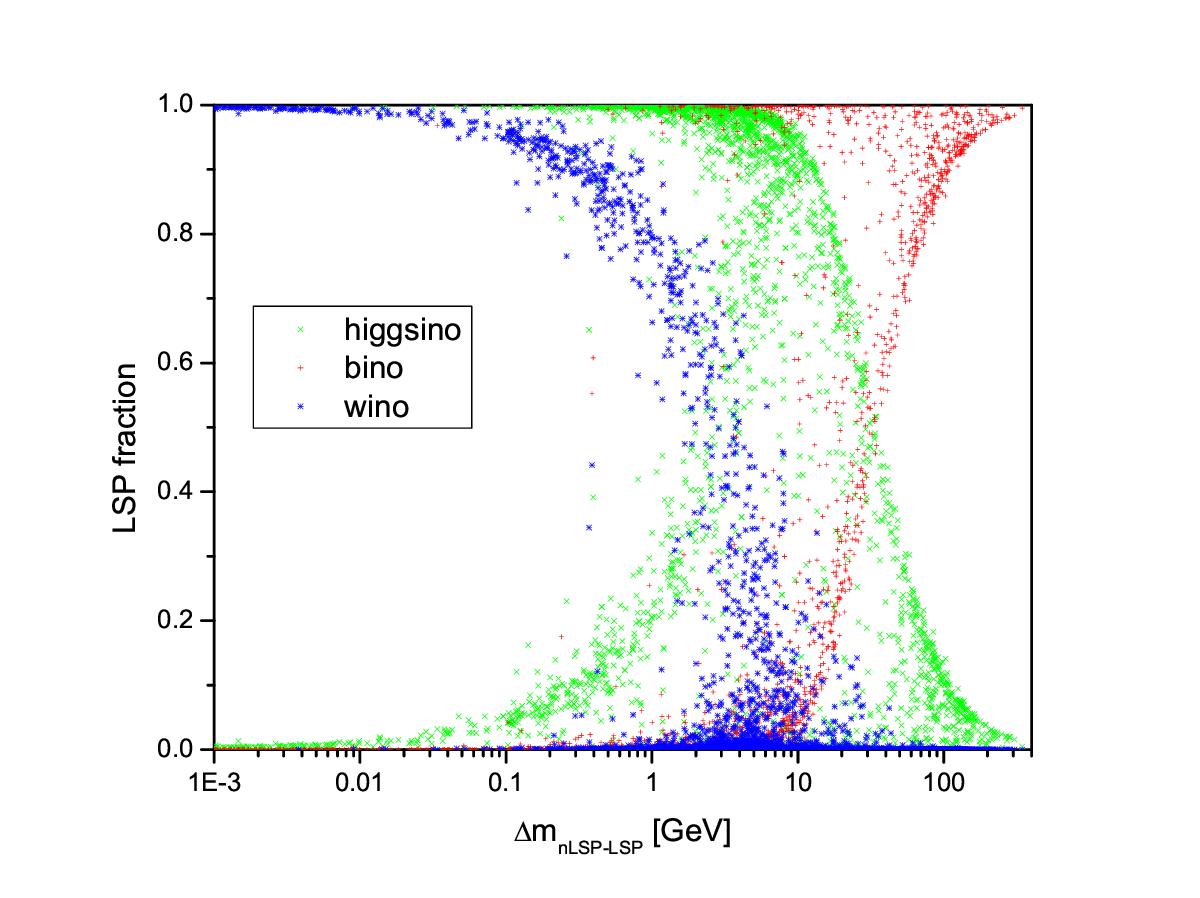}}
\caption{Fractional Wino/Higgsino/bino content of the LSP in the case of log priors as functions of the LSP mass(top) and the nLSP-LSP mass splitting(bottom). 
Note that each of the log prior models has three entries on these figures.}
\label{fig19p}
\end{figure}

Figures.~\ref{fig18} and ~\ref{fig19} describe the nature of the nLSP and the corresponding mass 
splitting with the LSP for the case of log priors, as well as the associated  
electroweak origin of the LSP.  While different in some details from the flat prior results, here we see that the properties of the LSP and the possible 
identities of the nLSP are essentially the same in the two cases.   The slight difference in the identity of the nLSP between the log and flat prior scenarios is 
due to the difference in the statistics of the two sample sizes.
Overall, the main distinguishing feature that we observe arises again due to the preference for lighter sparticle masses 
in the case of log priors and the reduced statistical power of the surviving sample. 
Complementary information on the Wino/Higgsino/Bino content of the LSP in the case of log priors is provided in Figure~\ref{fig19p}. Here we see results which are 
qualitatively similar to those found in the case of flat priors. 

\begin{figure}[htbp]
\centerline{
\includegraphics[width=6.0cm,angle=-90]{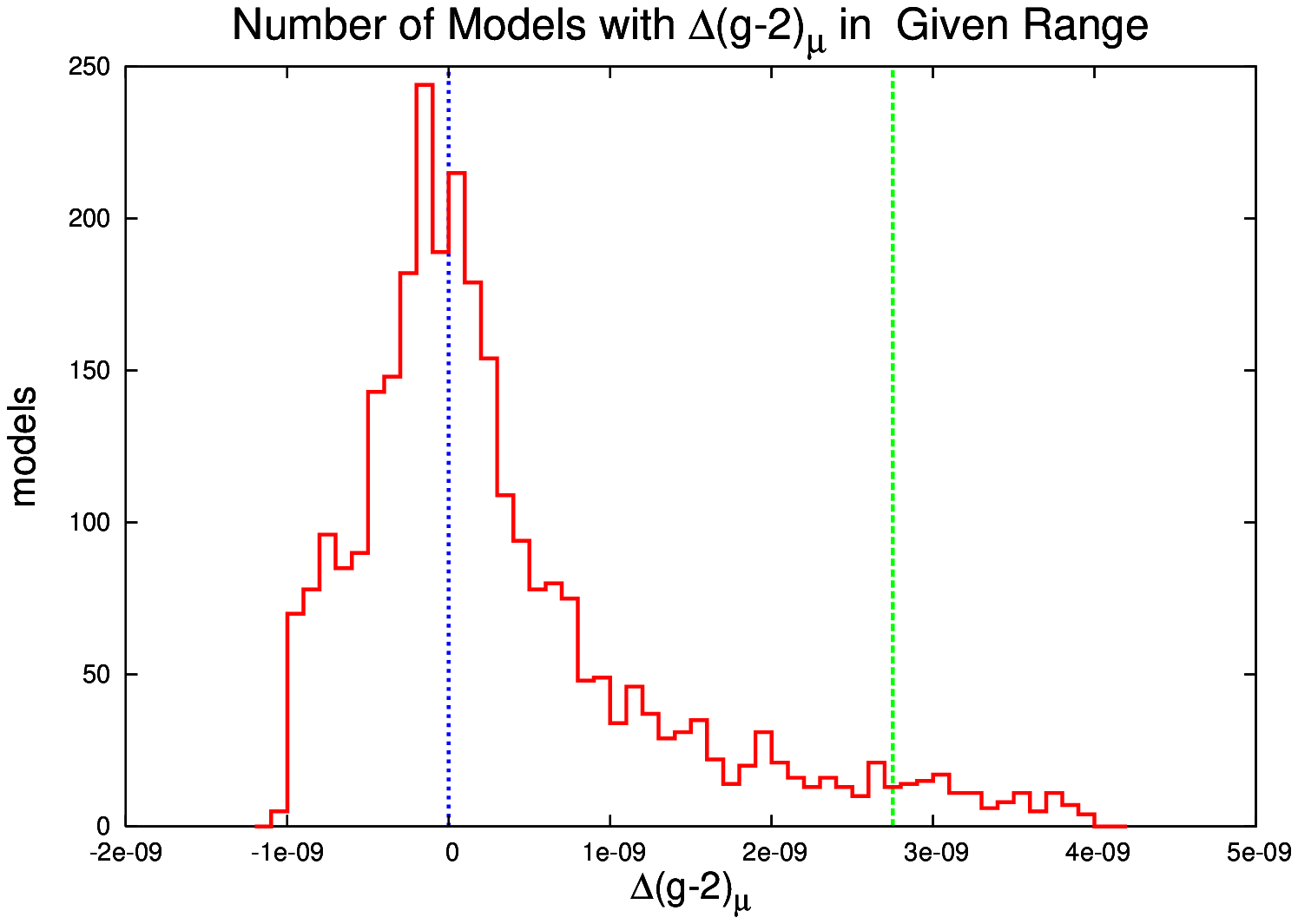}
\hspace*{0.1cm}
\includegraphics[width=6.0cm,angle=-90]{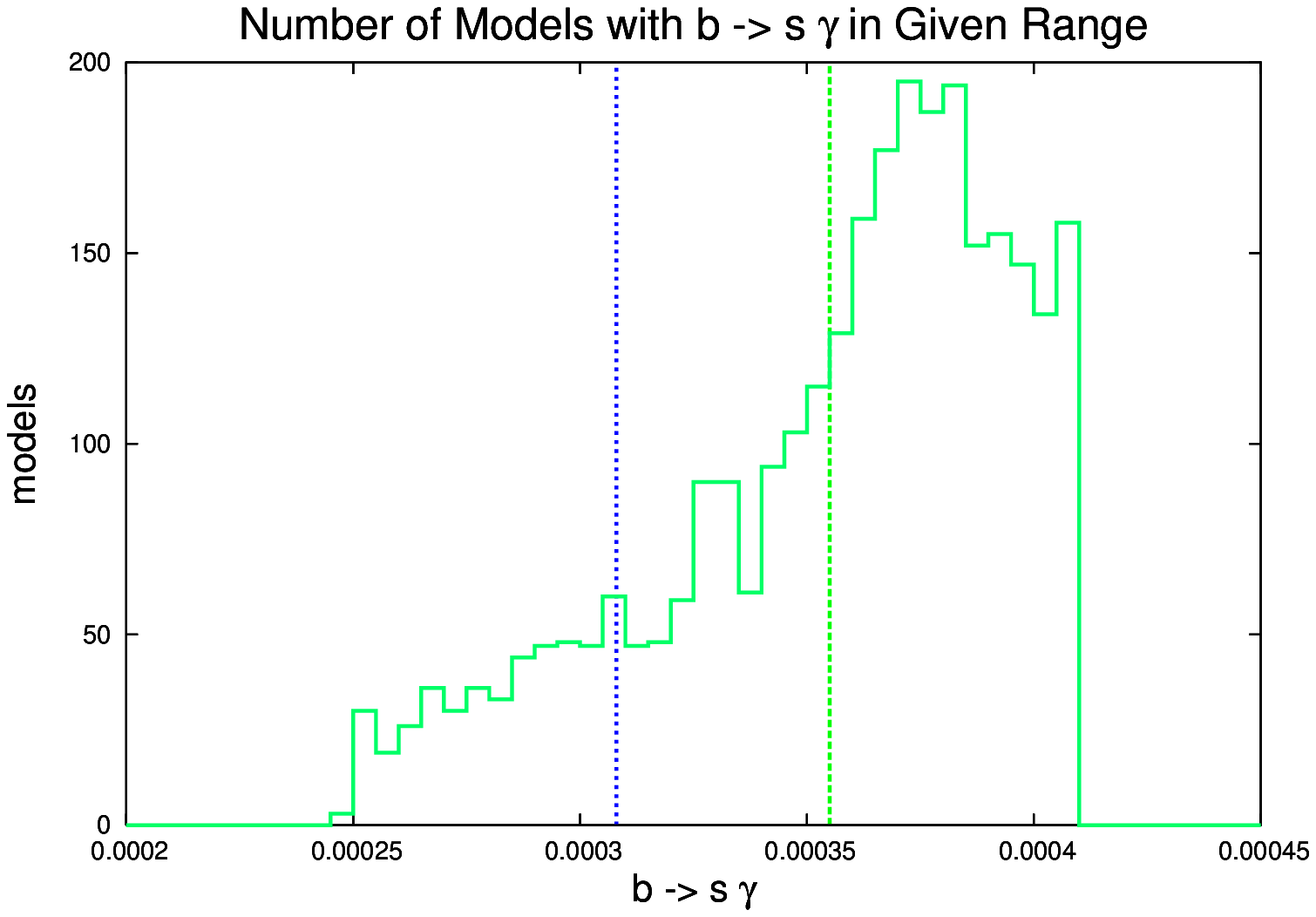}}
\vspace*{0.1cm}
\centerline{
\includegraphics[width=6.0cm,angle=-90]{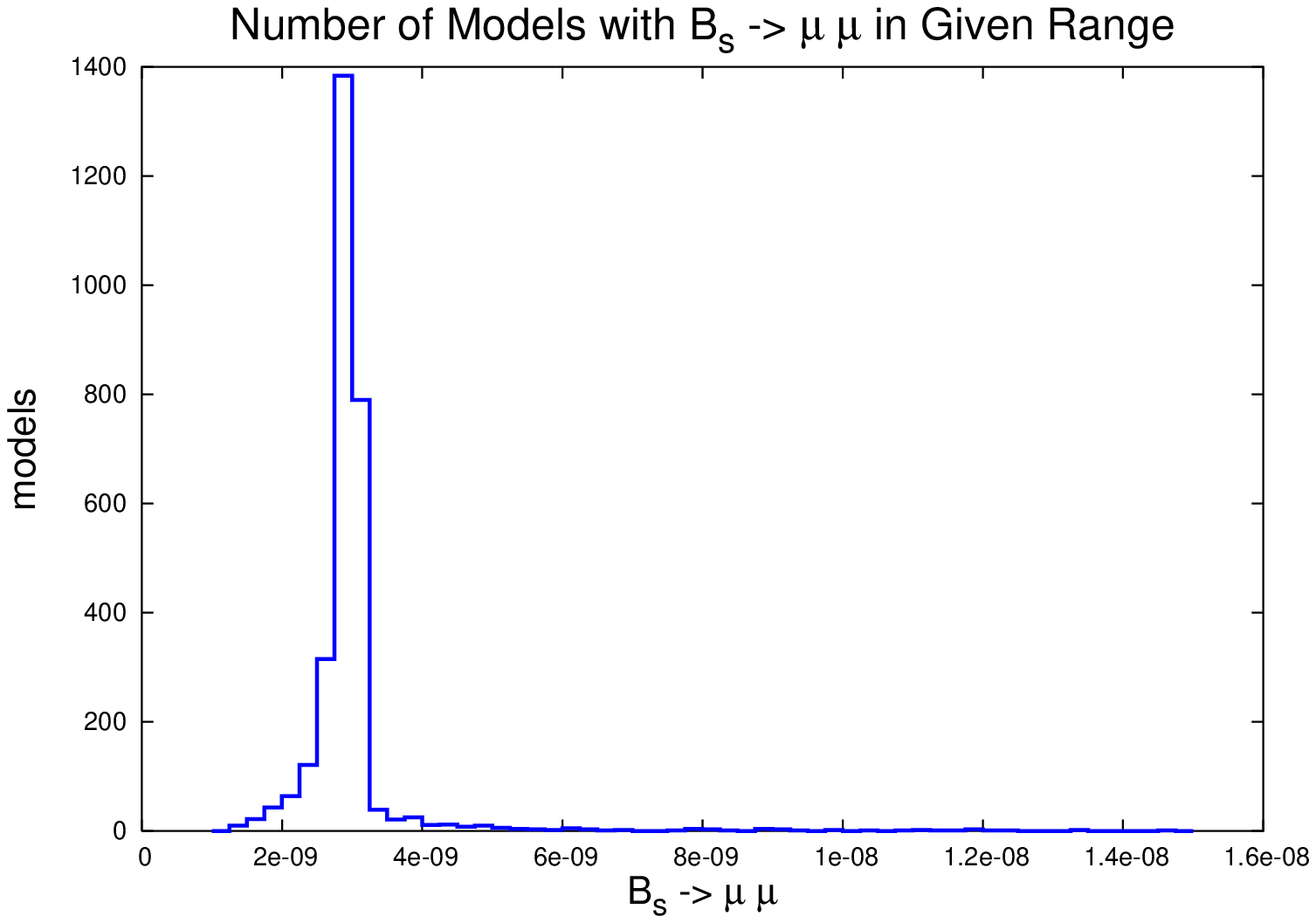}
\hspace*{0.1cm}
\includegraphics[width=6.0cm,angle=-90]{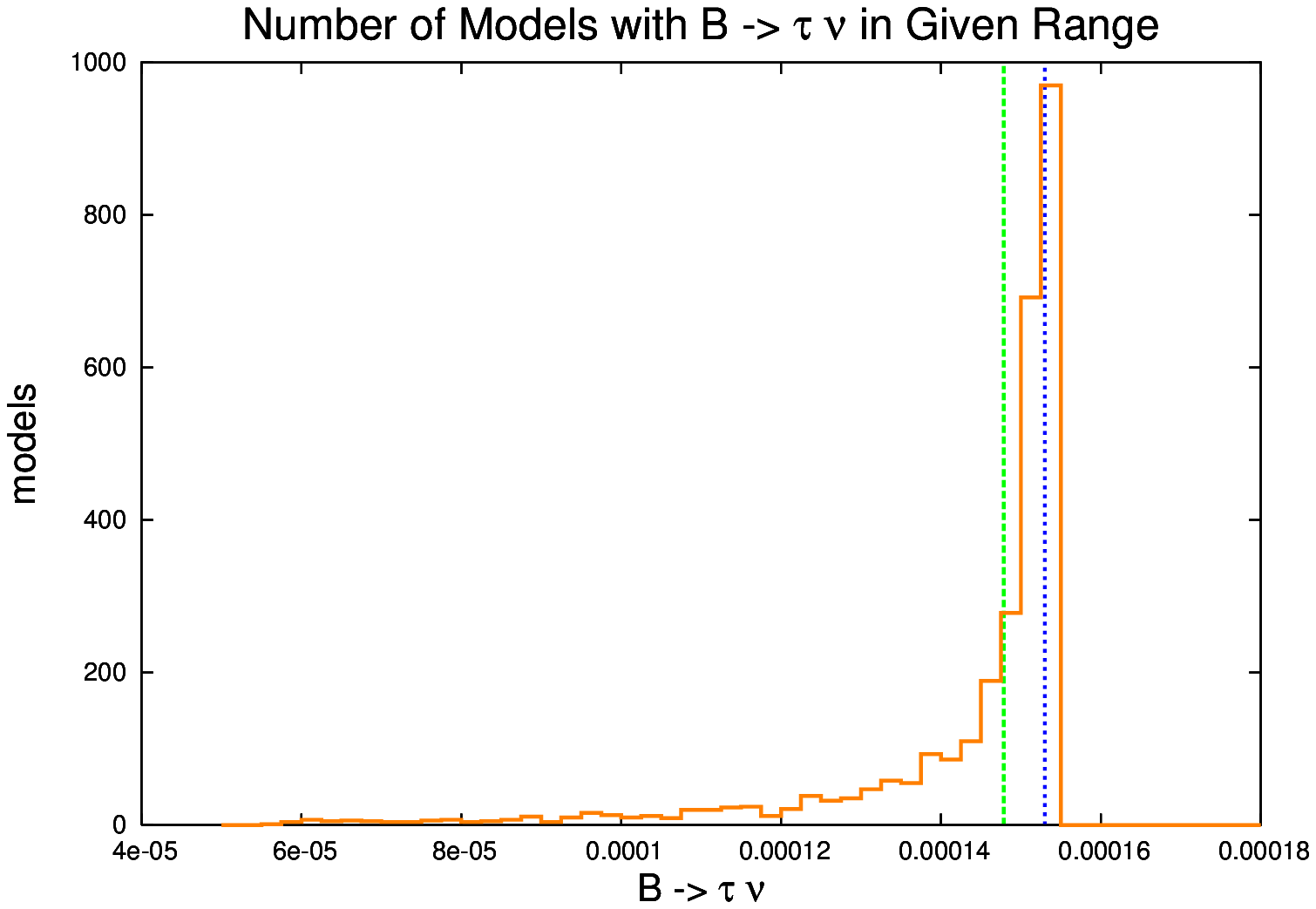}}
\centerline{
\includegraphics[width=6.0cm,angle=-90]{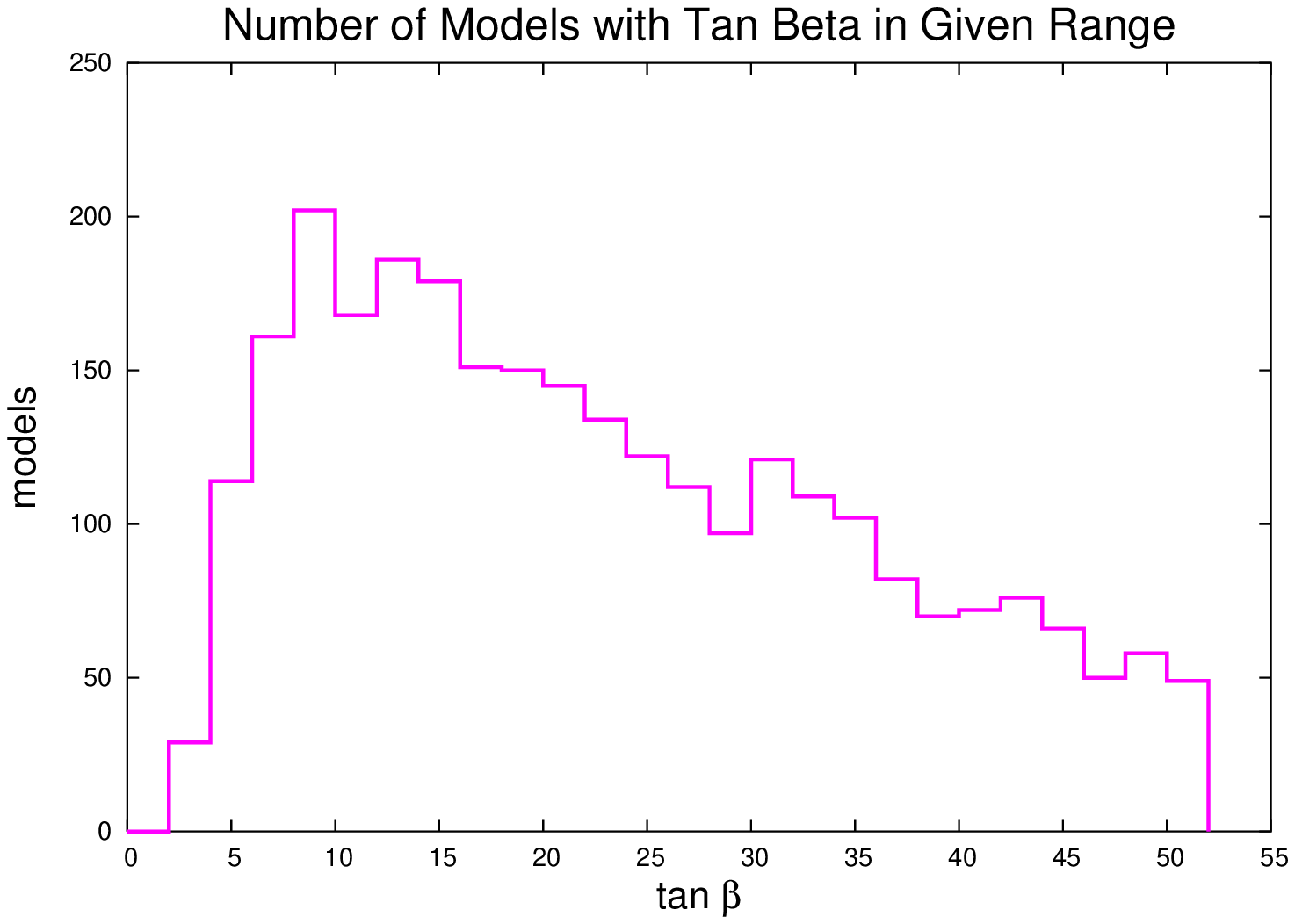}
\hspace*{0.1cm}
\includegraphics[width=6.0cm,angle=-90]{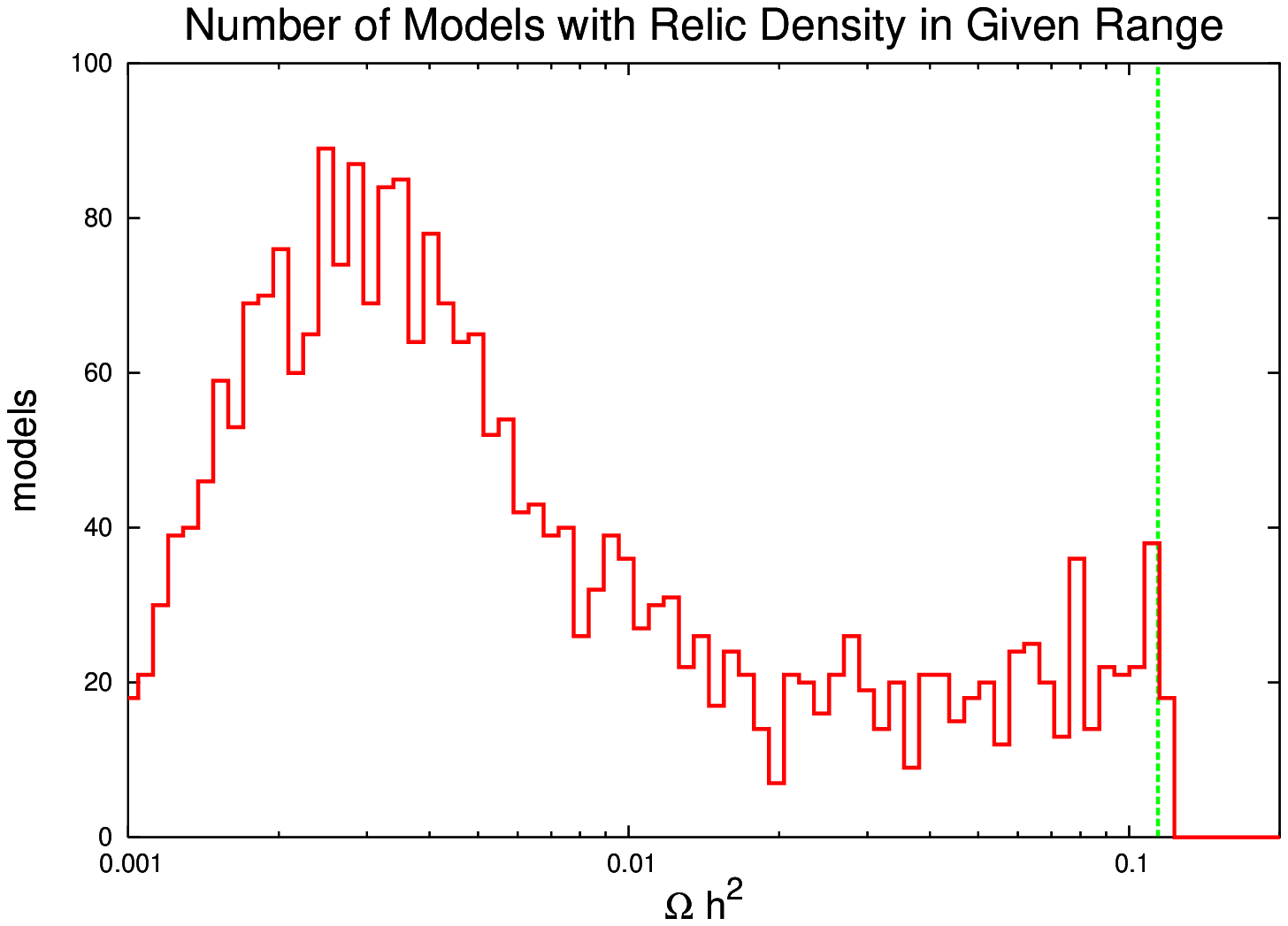}}
\caption{Distributions of predictions for several observables as well as $\tan \beta$ for our model sample subject to the constraints discussed in the text in 
the case of log priors. 
The blue and green dashed lines show the SM predictions as well as the current central values obtained by experiment, respectively. }
\label{fig20}
\end{figure}

\begin{figure}[htbp]
\centerline{
\includegraphics[width=6.0cm,angle=-90]{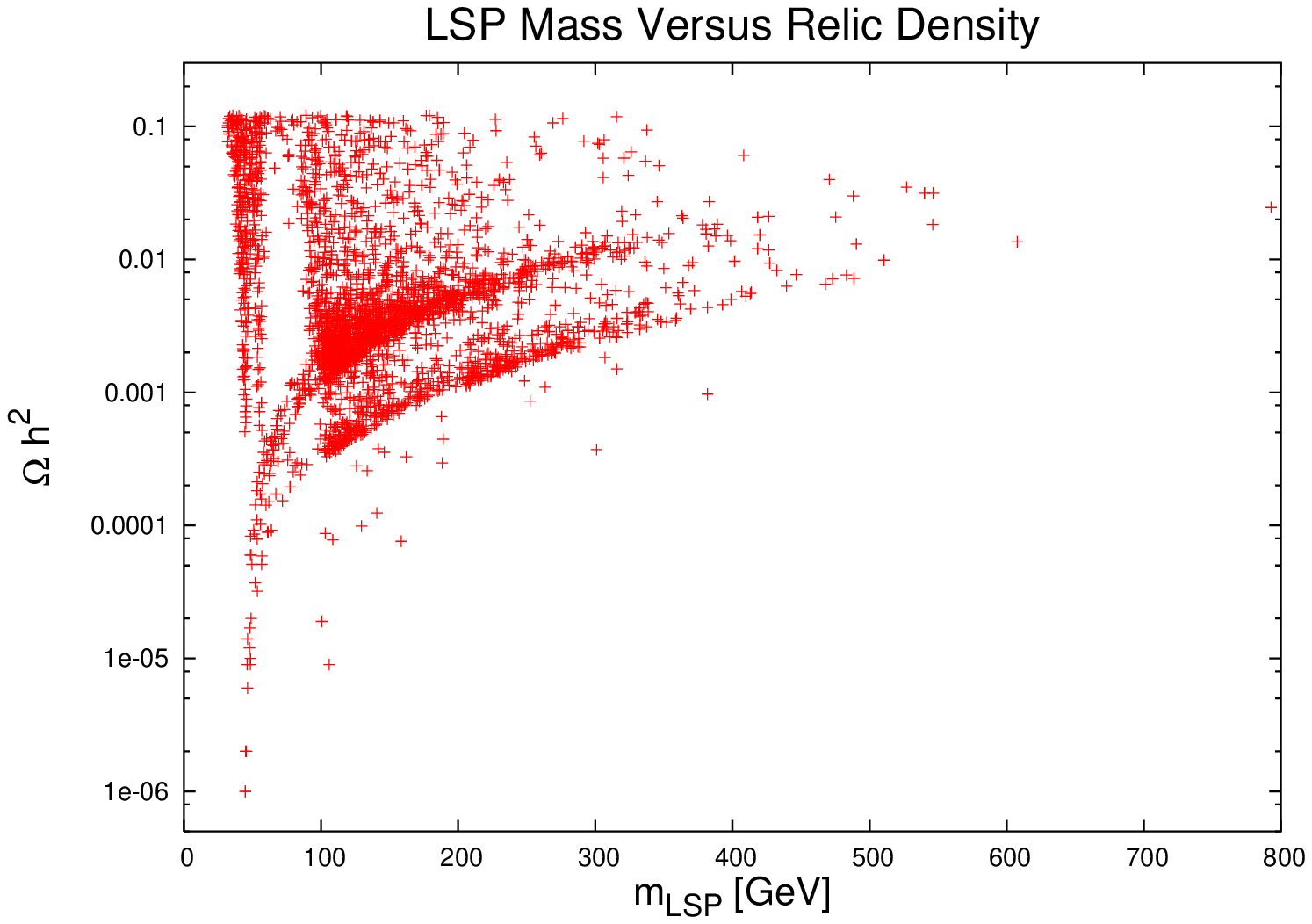}
\hspace*{0.1cm}
\includegraphics[width=6.0cm,angle=-90]{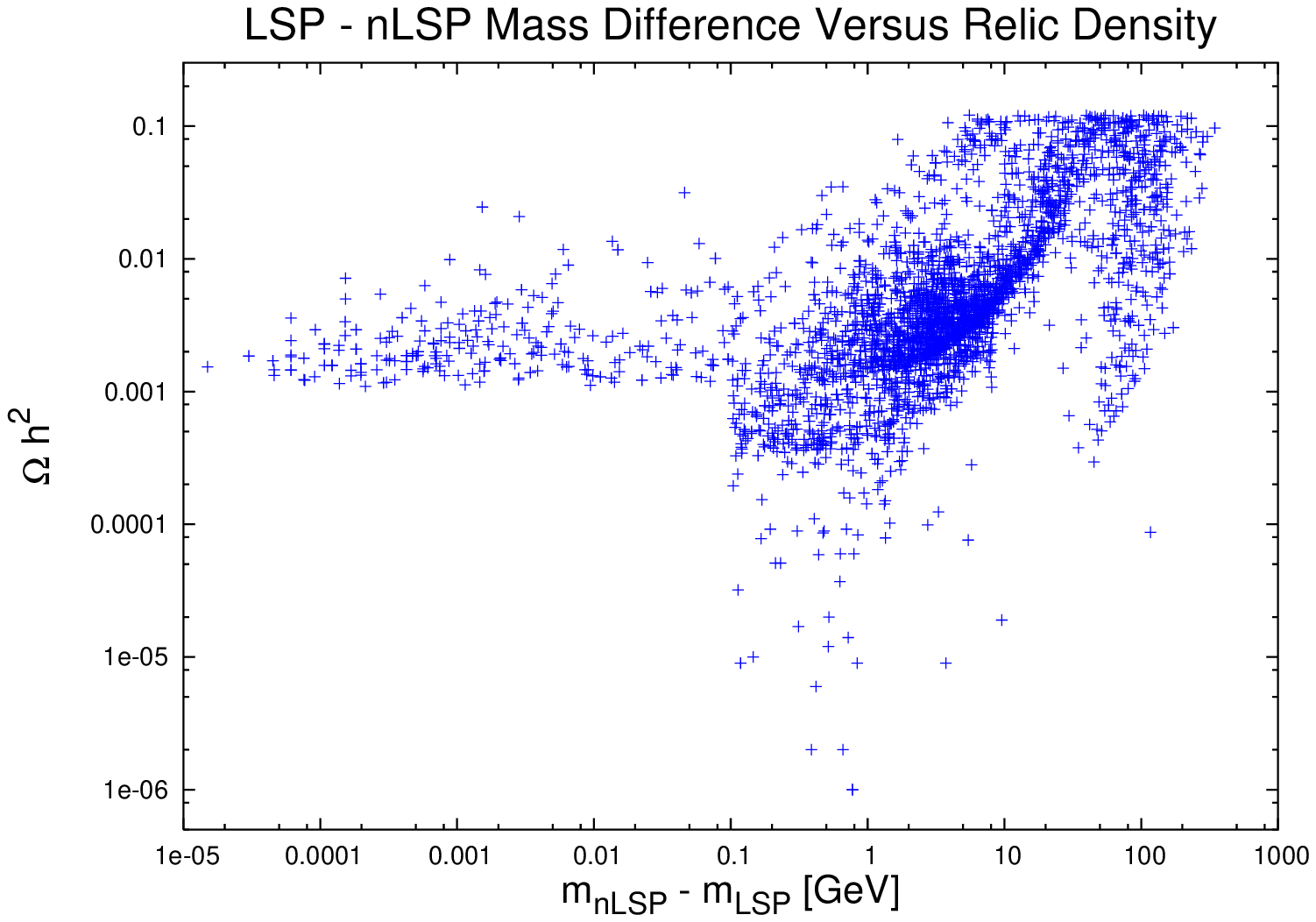}}
\vspace*{0.1cm}
\centerline{
\includegraphics[width=6.0cm,angle=-90]{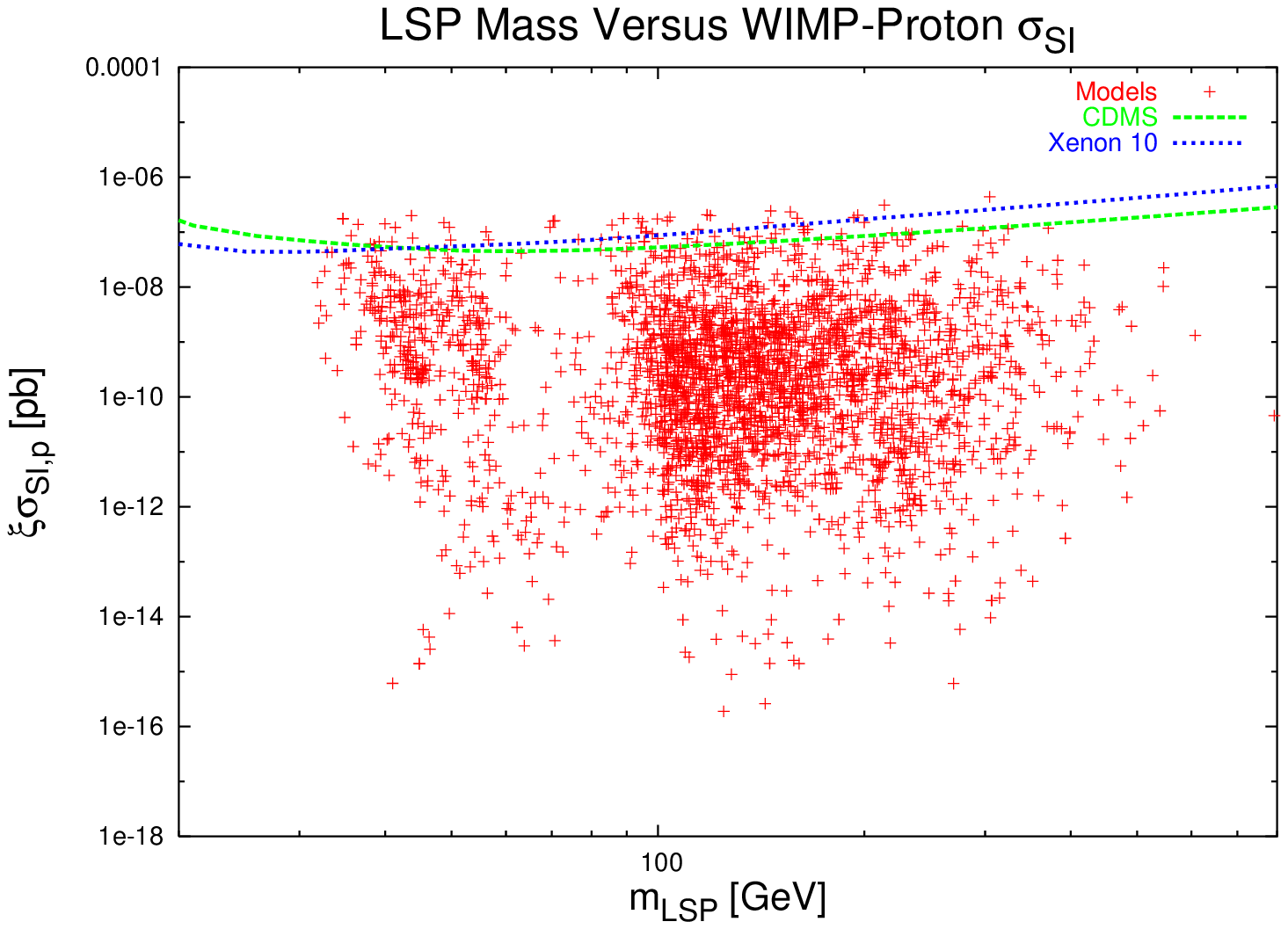}
\hspace*{0.1cm}
\includegraphics[width=6.0cm,angle=-90]{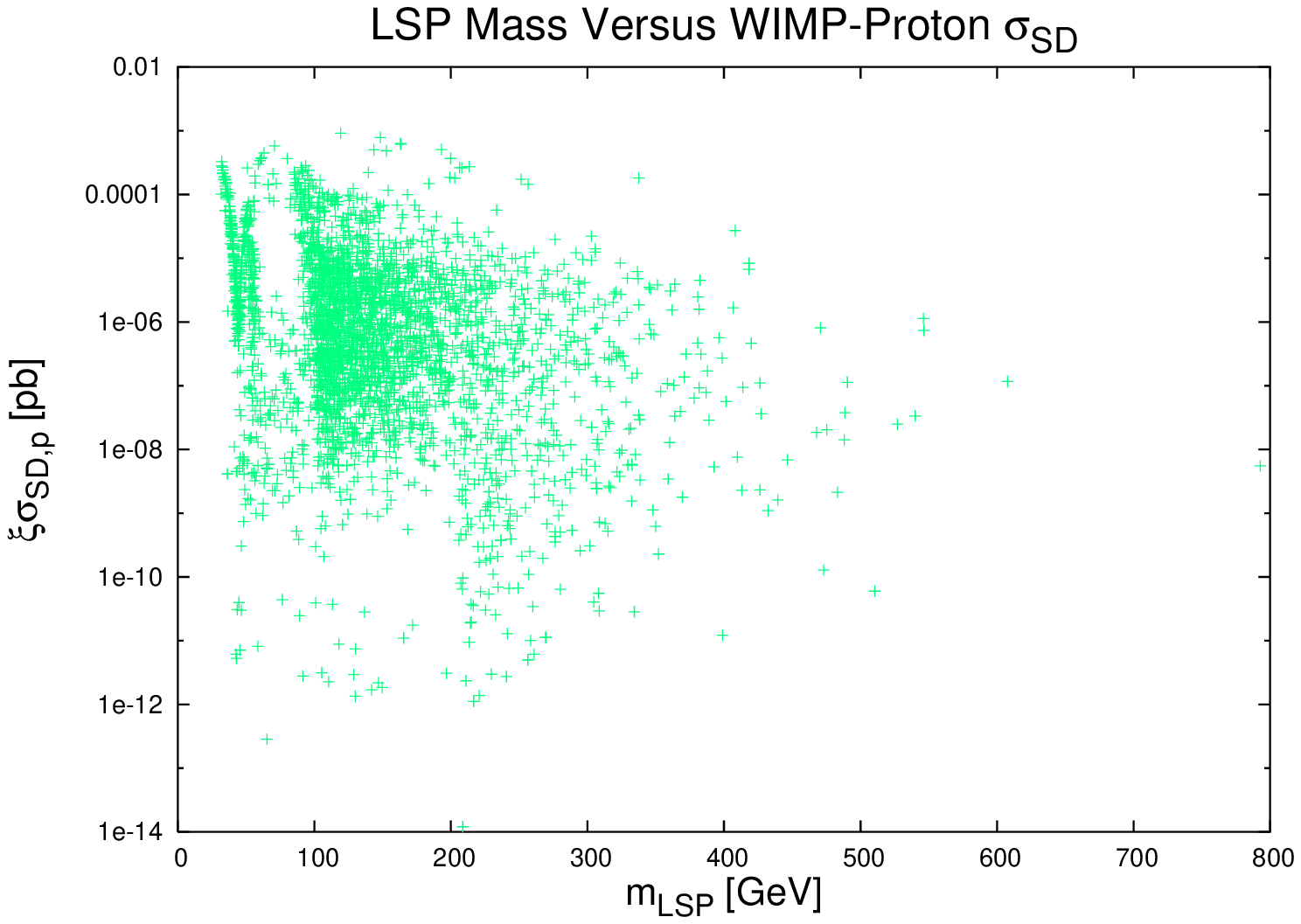}}
\centerline{
\includegraphics[width=6.0cm,angle=-90]{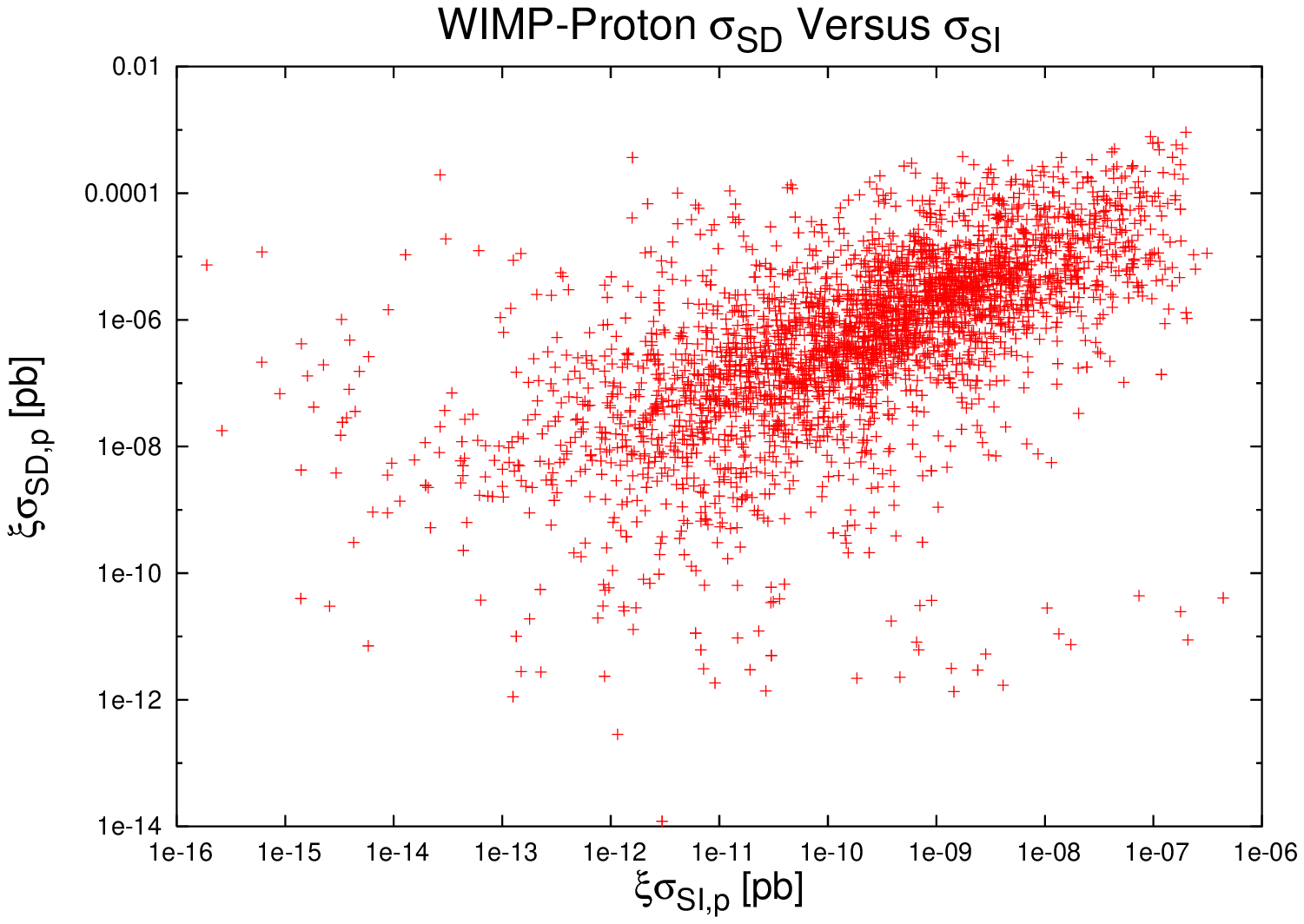}
\hspace*{0.1cm}
\includegraphics[width=6.0cm,angle=-90]{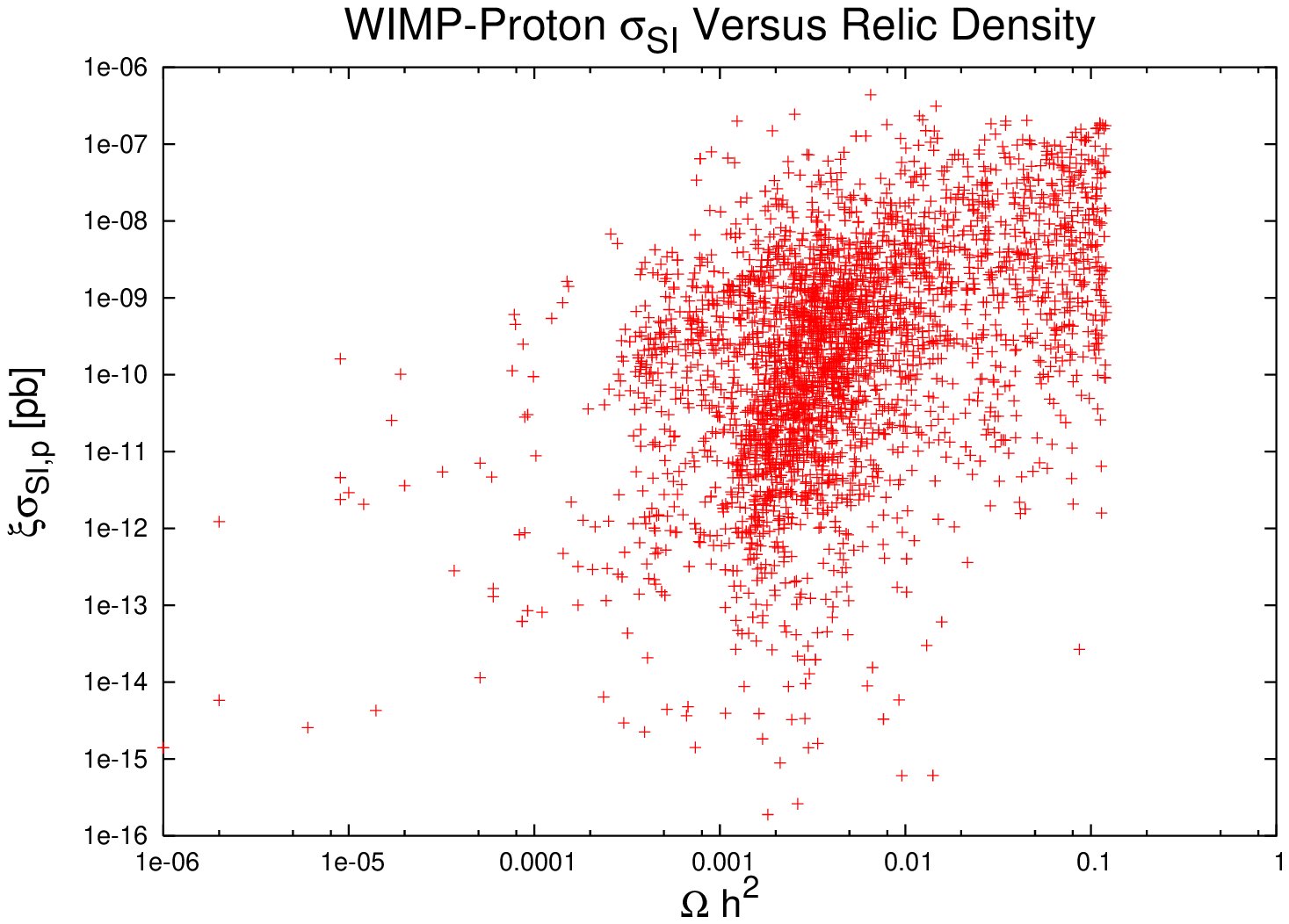}}
\caption{Distributions of relevant dark matter predictions for our model sample in the case of log priors subject to the constraints discussed in the text. All 
cross sections are in pb.}
\label{fig21}
\end{figure}

\begin{figure}[htbp]
\centerline{
\includegraphics[width=9.0cm,angle=0]{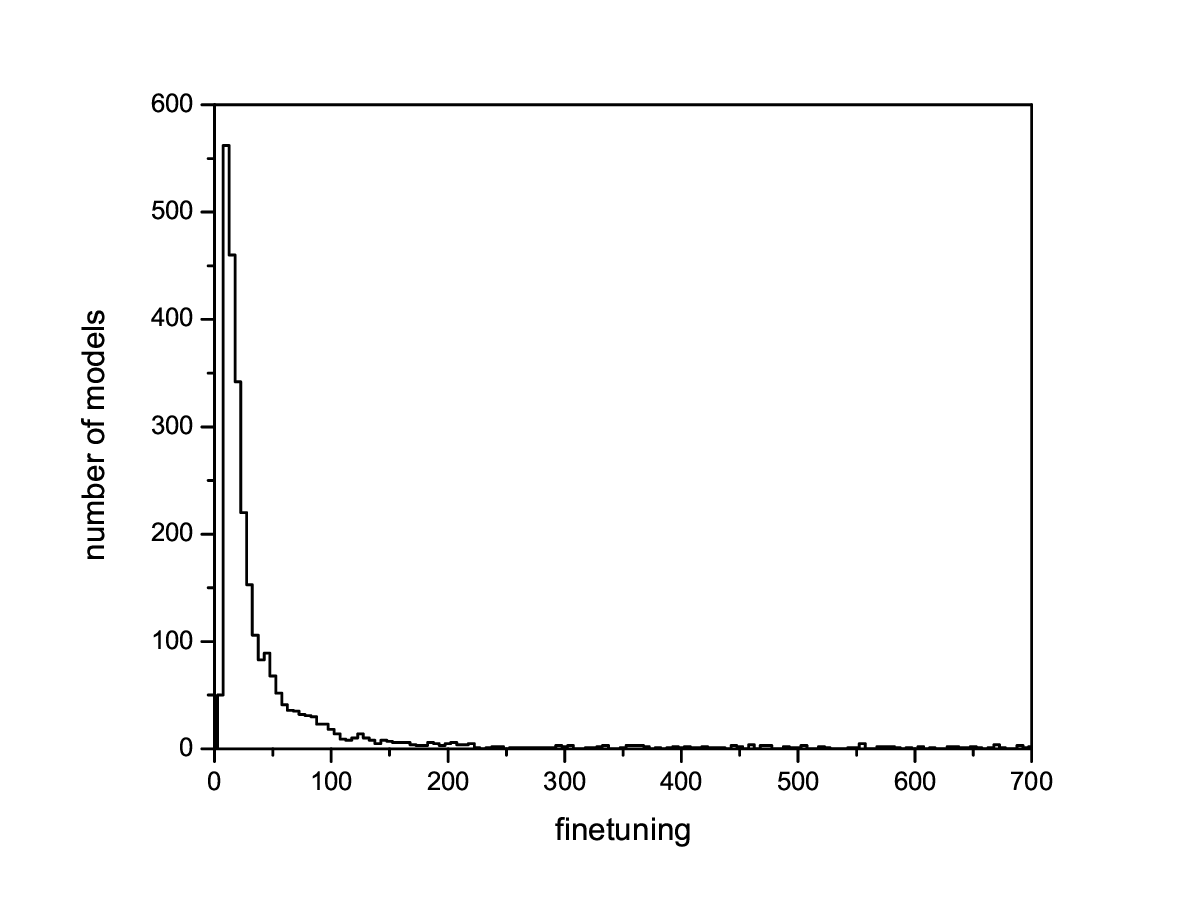}
\hspace*{0.1cm}
\includegraphics[width=9.0cm,angle=0]{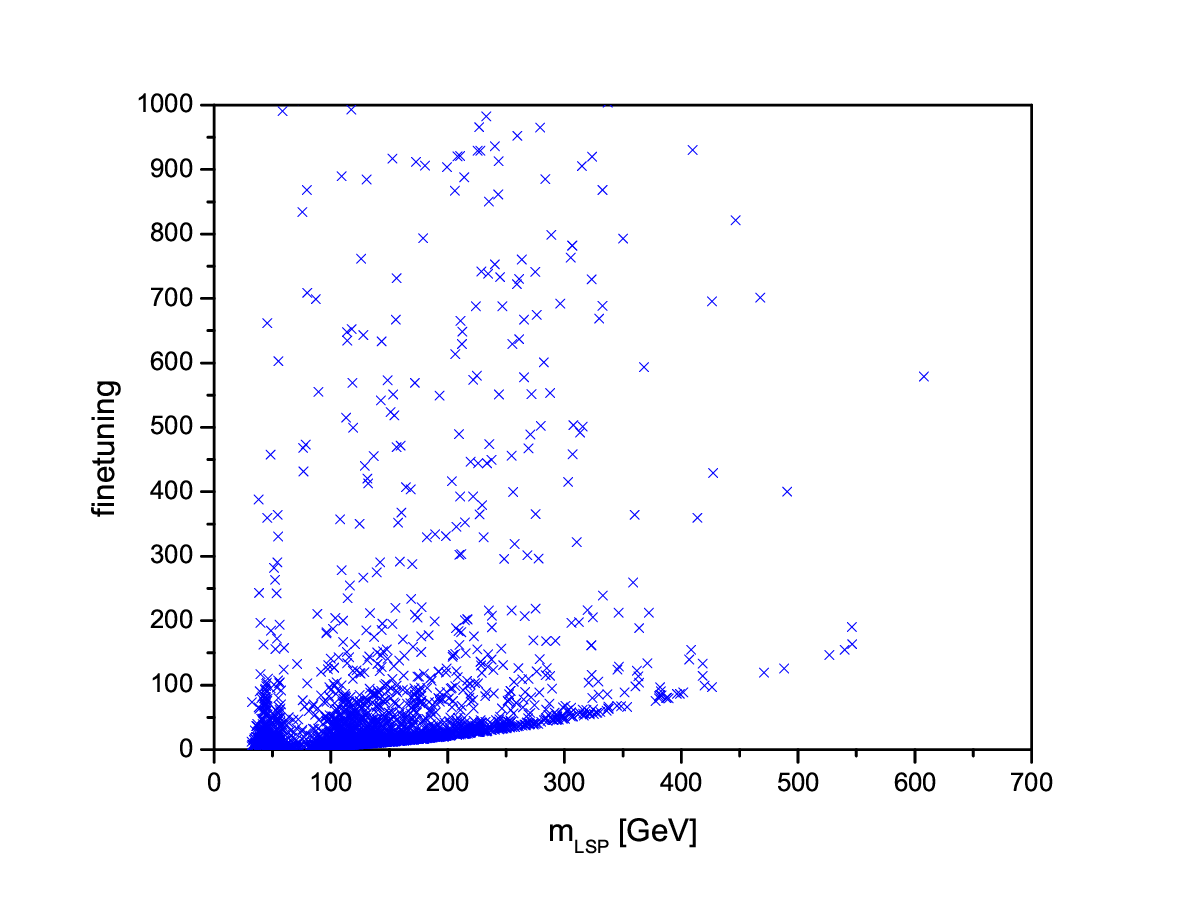}}
\vspace*{0.1cm}
\centerline{
\includegraphics[width=9.0cm,angle=0]{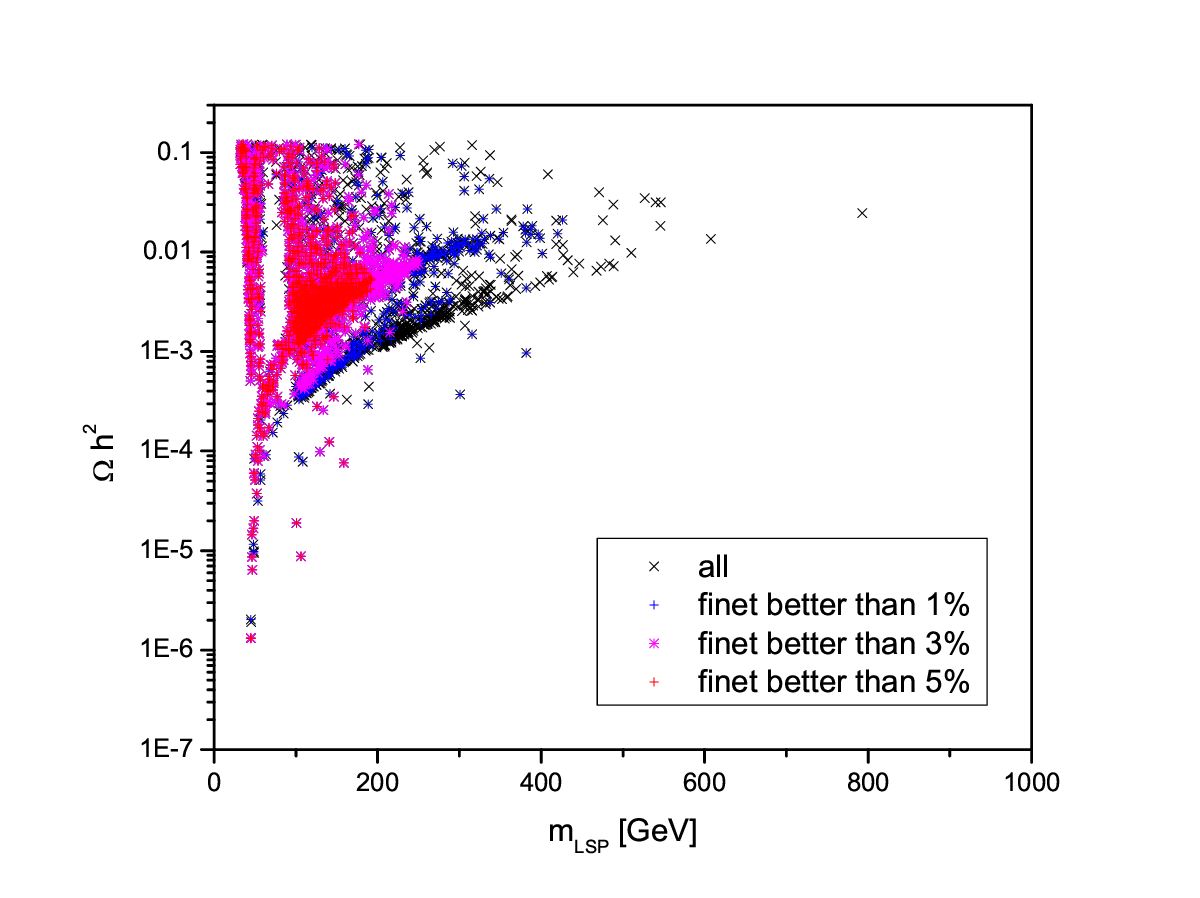}}
\caption{The distribution of fine-tuning ($=\Delta$) in our set of models with log priors 
and as a function of the LSP mass. Also shown is the correlation of the amount of fine-tuning with the predicted dark matter density.}
\label{fig22}
\end{figure}

The log prior predictions for a number of experimental observables are displayed in Figs.~\ref{fig20} and ~\ref{fig21}.  These are seen to be quite similar 
qualitatively to those obtained 
in the case of flat priors. Some minor differences are: ($i$) the slight preference for a larger branching fraction for the $b\to s\gamma$ transition in the case of log 
priors; ($ii$) the lack of a bimodal peaking structure for $\Delta (g-2)_\mu$ in the log prior case, which is directly linked to the lighter spartner masses and ($iii$) 
though peaked near the same value as in the flat prior 
sample, the relative number of models in the log prior sample that predict larger values of $\Omega h^2$ is found to be somewhat larger. 
All of these small effects can be traced to the log tilted mass spectra and the corresponding extended mass range. 
In Fig.~\ref{fig22} we see that the log prior models are perhaps even less fine tuned than those obtained in the flat prior case which is not a surprise since 
lighter spartner masses are now favored.


From the analysis presented above, we conclude that on the whole it would seem safe to say that the flat and log priors lead to very similar results for almost 
all of the quantities examined with no significantly large differences.

\section{Discussion}

There are many ways to begin a detailed collider study of the varied signatures arising from the very large set of models that we have generated. Such studies are 
beyond the scope of the present paper but 
there are some observations that we can make in addition to those above. 

One of the best ways to probe the phenomenological implications of our set of models for the LHC is to examine the mass spectrum beyond the LSP and nLSP. Feldman, Liu and 
Nath (FLN){\cite {nath}} have performed a previous study within mSUGRA where they categorize and examine the 22 possible orderings for the 4 lightest new particles in the 
spectrum.  These are rank ordered by the number of times they occur while scanning the parameter space. This mass spectrum ordering is important as it influences the 
various cascade decay chains for heavy colored SUSY particles produced at the LHC. To get a feeling for our results, we have repeated this analysis using 
both our flat and log prior model sets. The first question we addressed was how frequently do we obtain a model which can be identified as one of the 22 found 
by FLN for either set of pMSSM priors. This is answered in Table~\ref{nathy}. Interestingly, we see that one of the FLN orderings, mSP10, does {\it not} occur 
in either of our flat or log prior model sample for the pMSSM at this level of statistics indicating that they occur very rarely. 
Many of the other FLN models are seen to occur rather rarely in our two model pMSSM sets, whereas the most common FLN 
models are also common in our two model sets.

\begin{table}
\centering
\begin{tabular}{|l|l|c|c|} \hline\hline
\multicolumn{2}{|c|}{}&\multicolumn{2}{|c|}{\% of Models}\\ \hline
mSP & Mass Pattern  & Linear Priors & Log Priors \\ \hline

mSP1 &  $\tilde\chi_1^0<\tilde\chi_1^\pm<\tilde\chi_2^0<\tilde\chi_3^0$ &9.82 & 18.59\\
mSP2 &  $\tilde\chi_1^0<\tilde\chi_1^\pm<\tilde\chi_2^0<A/H$ & 2.08 & 0.68\\
mSP3 &  $\tilde\chi_1^0<\tilde\chi_1^\pm<\tilde\chi_2^0<\tilde \tau_1$ & 5.31 & 6.64\\
mSP4 &  $\tilde\chi_1^0<\tilde\chi_1^\pm<\tilde\chi_2^0<\tilde g$ &2.96 & 3.73\\ \hline

mSP5 &  $\tilde\chi_1^0<\tilde\tau_1<\tilde\ell_R<\tilde\nu_\tau$ &0.02 & 0.14\\
mSP6 &  $\tilde\chi_1^0<\tilde\tau_1<\tilde\chi_1^\pm<\tilde\chi_2^0$ &0.46 & 1.22\\
mSP7 &  $\tilde\chi_1^0<\tilde\tau_1<\tilde\ell_R<\tilde\chi_1^\pm$ &0.02 & 0.03\\
mSP8 &  $\tilde\chi_1^0<\tilde\tau_1<A\sim H$ &0.10 & 0\\
mSP9 &  $\tilde\chi_1^0<\tilde\tau_1<\tilde\ell_R<A/H$ &0.01 & 0\\
mSP10 &  $\tilde\chi_1^0<\tilde\tau_1<\tilde t_1<\tilde\ell_R$ &0 & 0\\ \hline

mSP11 &  $\tilde\chi_1^0<\tilde t_1<\tilde\chi_1^\pm<\tilde\chi_2^0$ &0.09 & 0\\
mSP12 &  $\tilde\chi_1^0<\tilde t_1<\tilde\tau_1<\tilde\chi_1^\pm$ &0.01 & 0\\
mSP13 &  $\tilde\chi_1^0<\tilde t_1<\tilde\tau_1<\tilde\ell_R$ &0.01 & 0 \\ \hline

mSP14 &  $\tilde\chi_1^0<A\sim H<H^\pm$ &0.35 & 0.10\\
mSP15 &  $\tilde\chi_1^0<A\sim H<\tilde\chi_1^\pm$ &0.08 &0\\
mSP16 &  $\tilde\chi_1^0<A\sim H<\tilde\tau_1$ &0.01 & 0.03\\ \hline

mSP17 &  $\tilde\chi_1^0<\tilde\tau_1<\tilde\chi_2^0<\tilde\chi_1^\pm$ &0.18 &0.41 \\
mSP18 &  $\tilde\chi_1^0<\tilde\tau_1<\tilde\ell_R<\tilde t_1$ &0.01 & 0 \\
mSP19 &  $\tilde\chi_1^0<\tilde\tau_1<\tilde t_1<\tilde\chi_1^\pm$ & 0.01 & 0\\ \hline

mSP20 &  $\tilde\chi_1^0<\tilde t_1<\tilde\chi_2^0<\tilde\chi_1^\pm$ &0.06 & 0\\
mSP21 &  $\tilde\chi_1^0<\tilde t_1<\tilde\tau_1<\tilde\chi_2^0$ &0.01 & 0\\ \hline

mSP22 &  $\tilde\chi_1^0<\tilde\chi_2^0<\tilde\chi_1^\pm<\tilde g$ &0.27 & 0.51\\

\hline\hline
\end{tabular}
\caption{Frequency of occurrence for the FLN{\cite {nath}} model sequences in our our flat and log prior pMSSM model samples.}
\label{nathy}
\end{table}

The next question we addressed was how many mass patterns do we find for the 4 lightest new particles for both sets of priors and how common are they. This is a 
generalization of the FLN analysis to the pMSSM. In the case of 
flat (log) priors we find $1109 (267)$ different mass patterns for the four lightest sparticles; the difference in these two numbers is largely due to the 
much greater statistics in the the flat prior case. However, in either scenario this number is many times larger than that found by FLN and it is likely that even more 
patterns may emerge if greater statistics were to become available. 
This result should not be surprising since we are imposing no constraints on the SUSY breaking mechanism and hence
the sparticle mass ordering above the LSP is in general completely arbitrary.  Table~\ref{ourres} shows 
the set of 25 particle mass orderings for the 4 lightest sparticles
that occur most frequently amongst our surviving models for both flat and log priors. Note that the sets for the flat and log priors 
are generically similar but are different in detail. Note that the most frequently occurring FLN orderings are also among the 25 that we obtain for either set of priors. 
Conversely, however, most of the FLN orderings do not appear amongst our top 25. We note, however, that once the assumptions of minimal SUGRA are relaxed more of the 
new patterns that we find in the pMSSM begin to emerge. Four of these are given by the NUSP1-3 and DBSP6 models as found in the last two papers of Ref.~{\cite {nath}}.

\begin{table}
\centering
\begin{tabular}{|l|c|l|c|} \hline\hline
\multicolumn{2}{|c|}{Linear Priors}&\multicolumn{2}{|c|}{Log Priors}\\ \hline
Mass Pattern  & \% of Models & Mass Pattern & \% of Models \\ \hline

$\tilde\chi_1^0<\tilde\chi_1^\pm<\tilde\chi_2^0<\tilde\chi_3^0$ & 9.82 & 
  $\tilde\chi_1^0<\tilde\chi_1^\pm<\tilde\chi_2^0<\tilde\chi_3^0$ &18.59\\
$\tilde\chi_1^0<\tilde\chi_1^\pm<\tilde\chi_2^0<\tilde \ell_R$ & 5.39& 
  $\tilde\chi_1^0<\tilde\chi_1^\pm<\tilde\chi_2^0<\tilde\nu_\tau$ & 7.72\\
$\tilde\chi_1^0<\tilde\chi_1^\pm<\tilde\chi_2^0<\tilde\tau_1$ & 5.31& 
  $\tilde\chi_1^0<\tilde\chi_1^\pm<\tilde\chi_2^0<\tilde \ell_R$ & 6.67\\
$\tilde\chi_1^0<\tilde\chi_1^\pm<\tilde\chi_2^0<\tilde\nu_\tau$ & 5.02& 
  $\tilde\chi_1^0<\tilde\chi_1^\pm<\tilde\chi_2^0<\tilde\tau_1$ &6.64\\
$\tilde\chi_1^0<\tilde\chi_1^\pm<\tilde\chi_2^0<\tilde b_1$ & 4.89& 
  $\tilde\chi_1^0<\tilde\chi_1^\pm<\tilde\chi_2^0<\tilde d_R$ &5.18\\
$\tilde\chi_1^0<\tilde\chi_1^\pm<\tilde\chi_2^0<\tilde d_R$ & 4.49& 
  $\tilde\chi_1^0<\tilde\chi_1^\pm<\tilde\chi_2^0<\tilde\nu_\ell$ &4.50\\
$\tilde\chi_1^0<\tilde\chi_1^\pm<\tilde\chi_2^0<\tilde u_R$ & 3.82& 
  $\tilde\chi_1^0<\tilde\chi_1^\pm<\tilde\chi_2^0<\tilde b_1$ &3.76\\
$\tilde\chi_1^0<\tilde\chi_1^\pm<\tilde\chi_2^0<\tilde g$ & 2.96& 
  $\tilde\chi_1^0<\tilde\chi_1^\pm<\tilde\chi_2^0<\tilde g$ &3.73\\
$\tilde\chi_1^0<\tilde\chi_1^\pm<\tilde\chi_2^0<\tilde \nu_\ell$ & 2.67& 
  $\tilde\chi_1^0<\tilde\chi_1^\pm<\tilde\chi_2^0<\tilde u_R$ &2.74\\
$\tilde\chi_1^0<\tilde\chi_1^\pm<\tilde\chi_2^0<\tilde u_L$ & 2.35& 
  $\tilde\chi_1^0<\tilde\chi_1^\pm<\tilde\nu_\tau<\tilde\tau_1$ &2.27\\

$\tilde\chi_1^0<\tilde\chi_1^\pm<\tilde\nu_\tau<\tilde\tau_1$ &2.19& 
  $\tilde\chi_1^0<\tilde\chi_2^0<\tilde\chi_1^\pm<\tilde\chi_3^0$ &2.24\\
$\tilde\chi_1^0<\tilde\chi_2^0<\tilde\chi_1^\pm<\tilde\chi_3^0$ & 2.15& 
  $\tilde\chi_1^0<\tilde\chi_1^\pm<\tilde \ell_R<\tilde\chi_2^0$ &1.42\\
$\tilde\chi_1^0<\tilde\chi_1^\pm<\tilde\chi_2^0< A$  &2.00& 
  $\tilde\chi_1^0<\tilde\chi_1^\pm<\tilde\chi_2^0<\tilde u_L$ &1.32 \\
$\tilde\chi_1^0<\tilde\chi_1^\pm<\tilde\chi_2^0<\tilde t_1$ &1.40& 
  $\tilde\chi_1^0<\tilde\tau_1<\tilde\chi_1^\pm<\tilde\chi_2^0$ &1.22\\
$\tilde\chi_1^0<\tilde\chi_1^\pm<\tilde\nu_\ell<\tilde \ell_L$ &1.37& 
  $\tilde\chi_1^0<\tilde\chi_1^\pm<\tilde\tau_1<\tilde\chi_2^0$ &1.19\\

$\tilde\chi_1^0<\tilde\chi_1^\pm<\tilde\tau_1<\tilde\chi_2^0$ &1.35& 
  $\tilde\chi_1^0<\tilde\chi_2^0<\tilde\chi_1^\pm<\tilde\nu_\tau$ &1.15\\
$\tilde\chi_1^0<\tilde\chi_1^\pm<\tilde \ell_R<\tilde\chi_2^0$ &1.32& 
  $\tilde\chi_1^0<\tilde \ell_R<\tilde\chi_1^\pm<\tilde\chi_2^0$ &1.05 \\
$A<H<H^\pm<\tilde\chi_1^0$ &1.24& 
  $\tilde\chi_1^0<\tilde\nu_\tau<\tilde\tau_1<\tilde\chi_1^\pm$ &1.02 \\
$\tilde\chi_1^0<\tilde\chi_1^\pm<\tilde d_R<\tilde\chi_2^0$ &1.03& 
  $\tilde\chi_1^0<\tilde\chi_1^\pm<\tilde\nu_\ell<\tilde \ell_L$ & 0.95\\
$\tilde\chi_1^0<\tilde\chi_1^\pm<\tilde u_L<\tilde d_L$ &0.95& 
  $\tilde\chi_1^0<\tilde\chi_1^\pm<\tilde d_R<\tilde\chi_2^0$ &0.71\\

$\tilde\chi_1^0<\tilde\chi_1^\pm<\tilde b_1<\tilde\chi_2^0$ &0.89& 
  $\tilde\chi_1^0<\tilde\nu_\tau<\tilde\chi_1^\pm<\tilde\chi_2^0$ &0.68\\
$\tilde\chi_1^0<\tilde\chi_1^\pm<\tilde u_R<\tilde\chi_2^0$ &0.84& 
  $\tilde\chi_1^0<\tilde\chi_1^\pm<\tilde\chi_2^0< A$ &0.64\\
$\tilde\chi_1^0<\tilde\chi_1^\pm<A<H$ &0.74& 
  $\tilde\chi_1^0<\tilde\chi_1^\pm<\tilde\nu_\tau<\tilde\chi_2^0$ &0.61\\
$\tilde\chi_1^0<\tilde\chi_1^\pm<\tilde g<\tilde\chi_2^0$ &0.65& 
  $\tilde\chi_1^0<\tilde\chi_2^0<\tilde\chi_1^\pm<\tilde d_R$ &0.54\\
$\tilde\chi_1^0<\tilde\chi_1^\pm<\tilde\tau_1<\tilde\nu_\tau$ &0.51& 
  $\tilde\chi_1^0<\tilde\chi_1^\pm<\tilde\tau_1<\tilde\nu_\tau$ &0.54\\

\hline\hline
\end{tabular}
\caption{The 25 most common mass orderings for the four lightest sparticles amongst our pMSSM model sets assuming either flat (left) or log(right) priors.}
\label{ourres}
\end{table}

In conclusion, we see that the predictions of mSUGRA are greatly extended when examining the full pMSSM.

\section{Conclusions}

In this paper we begin a detailed study of the 19 parameter, CP-conserving pMSSM under the assumptions of ($i$) minimal flavor violation, ($ii$) the LSP being 
identified with the lightest neutralino, ($iii$) only the Yukawa couplings of the third generation are important and ($iv$) the WMAP measurement of the dark matter 
relic density is taken only as an upper limit to that arising from the LSP. Carefully subjecting numerous combinations of parameter 
space points, with mass terms randomly generated assuming either flat or log priors, to a large number of theoretical and experimental constraints 
we arrive at two sets of models that survive all of the restrictions. 

With these two sets of models, we first examined the resulting distributions of expected Higgs and sparticle masses; 
qualitatively and even semi-quantitatively either choice of prior was found to return quite similar results except for trivially obvious differences due to how 
the parameters were generated. In both cases, we found some instances where surprisingly
light sparticles are still allowed by the data.
Next, we studied the identity of the nLSP and its mass 
splitting with the LSP. We found that for either choice of priors, over a dozen sparticles can 
play the role of the nLSP with more or less the same probabilities in the two model samples. 
The most common nLSP identities were found to be $\tilde \chi_1^\pm$ and $\tilde \chi_2^0$ with other possibilities being $\tilde e_R, \tilde \tau_1$ 
and $\tilde t_1$. Less familiar nLSP identities were also found to occur such as the gluino or the first/second generation squarks. We also showed that 
the mass splitting between these two 
states, $\Delta m\equiv m_{nLSP}-m_{LSP}$, which is a critical parameter for collider signals and studies, was found to vary over a very large 
range of six orders of magnitude with values 
in excess of $\sim 100$ GeV and as small as $\sim 100$ keV or less.  The greatest range was found in the case where the lightest chargino was the nLSP. Small values of 
$\Delta m$ were found to be quite common particularly in the case of the lightest chargino or the second lightest neutralino being the nLSP. 
This would imply that searches for long-lived and (essentially) stable 
particles at the LHC will be of particular importance. In some cases, squarks and gluinos were found to be relatively light and not widely separated in mass from the 
corresponding LSP. The decays of such states would likely lead to soft jets in the final state which may be difficult to observe above SM backgrounds at the LHC.  
The mass spectra of the models we obtained were such as to substantially diminish the needed amount of fine tuning in comparison to, \eg, mSUGRA-based scenarios. 

Subsequently, we examined the postdictions arising from both model sets for a number of observables such as the branching fraction for $b\to s\gamma$  
and the shift in the value of the $g-2$ of 
the muon; these were found to be reasonably similar for both cases of priors. Since the predicted dark matter density was allowed to float to any value at or below 
that obtained by WMAP, we found that the average contribution of the LSP to dark matter peaked at $\sim 4\%$, with long tails, implying that most of the time some 
other source of dark matter must typically dominate. However, we do see that there is a reasonable fraction of models where the LSP contribution to the dark matter 
density is, in fact, quite appreciable in comparison to the WMAP measurement. 
These results were found to hold for either choice of priors. We then went further and examined the implications of these models for the direct 
detection of dark matter; we found that the range of possible cross sections is far larger than that obtained in any of the well-known SUSY breaking frameworks, such 
as the cMSSM or mSUGRA. Furthermore these ranges extended into the regions one would have expected to only arise from either the Universal Extra Dimensions or 
Little Higgs models. This implies that 
these observables alone cannot be used to discriminate between different Terascale models. 

It is important to note that while the two sets of priors have led to qualitatively very similar pictures, some {\it quantitative} differences do appear 
as can be seen from the figures above. For example, the mass spectra of the SUSY particles in both cases, while qualitatively similar, are different in detail 
since the log prior case generally leads to smaller masses.  

Clearly the detailed study of the pMSSM models is quite exciting as well as complex; we are just at the initial stages of this analysis and there are many things 
remaining to be done which we will address in subsequent publications.  In addition, we
plan to make our model sets public in the near future, in order to encourage a 
wider set of possible analyses. Hopefully we will not have to wait too long until we actually have real data from the LHC to examine.

\section{Acknowledgments}
The authors would like to thank J.~Conley, R.~Cotta and D.~Maitre for computational aid and S.~Kraml and T.~Tait for discussions related to this work. 
They would also like to 
thank M.~Peskin for the suggested title of the paper and J.~Baglio for his input during the initial stages of this analysis. CFB thanks the Institut f\"ur Physik, 
Fachbereich Theoretische Physik, Karl-Franzens-Universit\"at Graz and the LPSC
Grenoble for hospitality. CFB's work is supported in part by funds provided by the U.S. Department of Energy (D.O.E.) under cooperative 
research agreement DE-FC02-94ER40818.

\clearpage

%
%%%%%%%%%%%%%%%%%%--- References
%%%%%%%%%%%%%%%%%%%%%%%%%%%%%%%%%%%%%%%%%%%%%%%%%%%%%%%
\def\MPL #1 #2 #3 {Mod. Phys. Lett. {\bf#1},\ #2 (#3)}
\def\NPB #1 #2 #3 {Nucl. Phys. {\bf#1},\ #2 (#3)}
\def\PLB #1 #2 #3 {Phys. Lett. {\bf#1},\ #2 (#3)}
\def\PR #1 #2 #3 {Phys. Rep. {\bf#1},\ #2 (#3)}
\def\PRD #1 #2 #3 {Phys. Rev. {\bf#1},\ #2 (#3)}
\def\PRL #1 #2 #3 {Phys. Rev. Lett. {\bf#1},\ #2 (#3)}
\def\RMP #1 #2 #3 {Rev. Mod. Phys. {\bf#1},\ #2 (#3)}
\def\NIM #1 #2 #3 {Nuc. Inst. Meth. {\bf#1},\ #2 (#3)}
\def\ZPC #1 #2 #3 {Z. Phys. {\bf#1},\ #2 (#3)}
\def\EJPC #1 #2 #3 {E. Phys. J. {\bf#1},\ #2 (#3)}
\def\IJMP #1 #2 #3 {Int. J. Mod. Phys. {\bf#1},\ #2 (#3)}
\def\JHEP #1 #2 #3 {J. High En. Phys. {\bf#1},\ #2 (#3)}


\begin{thebibliography}{99}
\bibitem{discuss}
We would like to thank the authors of this code for communications on this point and note that they have recently corrected this feature in their code.
\end{thebibliography}
\end{document}